\documentclass[a4paper,12pt]{book}
\usepackage[utf8]{inputenc}
\pdfoutput=1
\usepackage{etoolbox}
\usepackage{float}
\usepackage{graphicx}%
\usepackage[lofdepth,lotdepth]{subfig}
\usepackage{epstopdf}
\usepackage[hyperfootnotes=false]{hyperref} 
\usepackage[titletoc]{appendix}
\usepackage{setspace}
\usepackage[square,sort,comma,numbers]{natbib}
\usepackage{amsmath}
\usepackage{amssymb}
\usepackage{epsfig}
\usepackage{caption}
\usepackage{textcomp}
  
\usepackage{slashed}
\usepackage{pdfpages} 
\usepackage{tikz,pgfplots}

\usepackage[top=3.5cm,bottom=2.5cm,left=3.5cm,right=3cm,includeheadfoot,twoside]{geometry}

\usepackage{fancyhdr}                          
  \fancypagestyle{plain}{
    \fancyhead{}                               
  }
  
\usepackage{xpatch}

\usepackage[nottoc]{tocbibind}
\usepackage{cleveref}
\makeatletter

\xpatchcmd{\@makechapterhead}{%
  \huge\bfseries \@chapapp\space \thechapter
}{%
  \Huge\bfseries\centering \@chapapp\space \thechapter
}{\typeout{Patched @makechapterhead}}{\typeout{Patching of @makechapterhead failed}}

\makeatother
  
 \makeatletter
 \let\l@chapter\l@section
 \let\l@section\l@subsection
 \let\l@subsection\l@subsubsection
 \makeatother
  
\begin{document}
\pagenumbering{roman}

 \begin{titlepage}

\setlength{\voffset}{0pt}
\setlength{\hoffset}{0pt}

\centering

\Large{University of São Paulo\\
Physics Institute}

\vspace{\stretch{2}}

\LARGE{\bf
Phenomenology of Vector-like Fermions in Physics Beyond the Standard Model}
\vspace{\stretch{1}}

\Large{Victor Manuel Peralta Cano}

\vspace{\stretch{1}}

\vspace{\stretch{0.025}}

\begin{flushright}
\begin{minipage}{0.56\textwidth}
\small{Dissertation submitted to the Physics Institute of the University of São Paulo in partial fulfillment of the requirements 
for the degree of Doctor of Science}
\end{minipage}
\end{flushright}



\vspace{\stretch{0.7}}


\begin{flushleft}
\normalsize
\end{flushleft}

\vspace{\stretch{0.5}}



\vspace{\stretch{1}}


\vspace{\stretch{1}}
\normalsize
São Paulo\\
2017

\end{titlepage}

\cleardoublepage
\phantomsection
\addcontentsline{toc}{chapter}{Acknowledgments}
\chapter*{\centerline{\bf Acknowledgments}}
\pagestyle{fancy}

First of all, I would like to thank my supervisor, Gustavo Alberto Burdman, for his encouragement and guidance throughout 
this project. I am deeply indebted to Department of Mathematical Physics at the University of Sao Paulo's 
Physics Institute. I am also grateful to the elementary particle physics group, for
its helpful tips, seminars and enjoyable discussions.\newline

\noindent I would like to thank to my thesis committee members, Gustavo Burdman, Enrico Bertuzzo, Oscar José Pinto Éboli, Ricardo D'Elia Matheus and 
Sérgio Ferraz Novaes for their for
their important comments.\newline

\noindent In particular, thanks to my friends and colleagues, Leonardo Lima, Lindber Salas, Yuber, Denis, Boris, Gabi, 
Leila, Leonardo Duarte, Hugo Camacho for great discussions on elementary particle physics.\newline

\noindent Thanks too to my undergraduate professors Jorge Abel Espichán Carrillo and 
Jaime Francisco Vento Flores and Jesús Félix Sánchez Ortiz who encouraged me in 
the area of physics.\newline

\noindent I am very grateful to my parents, Sebastián Peralta, Cirila Cano for their unconditional support all along.\newline

\noindent I would like to Acknowledge the financial support of the Committee for the Advancement of Higher Education 
(CAPES).
\cleardoublepage
\phantomsection
\addcontentsline{toc}{chapter}{Abstract} 
\chapter*{\centerline{\bf Abstract}}

The Standard model of particle physics provides a successful theory to understand the experimental results of 
the electroweak and strong interactions. However, it does not have a satisfactory explanation for the 
hierarchy problem. Many extensions of the Standard Model that solve the hierarchy problem result in new particles. We will study the phenomenology 
of vector-like fermions resulting in theories where the Higgs boson is typically a pseudo-Nambu-Goldstone
boson. In these theories we study the case where a heavy fermion 
will be heavier than a heavy gluon, and then the channel of a heavy fermion decaying into a color octet is considered. We study this 
phenomenology at high energy colliders, both the LHC as well as future machines.


 
   
\tableofcontents
 

\pagenumbering{arabic}
\chapter{\bf Introduction}

  The Standard Model (SM) of elementary particle physics is a quantum field theory that describes the 
  strong, weak and electromagnetic interactions between elementary particles. The gauge sector of the SM is 
  $SU(3)_{C}\times SU(2)_{L}\times U(1)_{Y}$, where $SU(3)_{C}$ and $SU(2)_{L}\times U(1)_{Y}$ indicate the strong 
  and electroweak interactions, respectively. In the SM the electroweak symmetry breaking (EWSB) $SU(2)_{L}\times U(1)_{Y}\rightarrow U(1)_{EM}$ 
  (here $U(1)_{EM}$ corresponds to the electromagnetic interaction) is due to the Higgs sector. As a result, after the EWSB
  the weak vector bosons and the fermions obtain masses through the Higgs mechanism. In this mechanism the 
  Higgs scalar field is a doublet of $SU(2)_{L}$ and it acquires a vacuum expectation value (VEV) such that the symmetry 
  $SU(2)_{L}\times U(1)_{Y}$ is spontaneously broken.\newline
  
\noindent  The recent discovery of a scalar boson at the Large Hadron Collider (LHC)~\citep{Aad:2012tfa,Chatrchyan:2012ufa}
 seems to indicate that it is the SM Higgs boson. If this particle is the Higgs boson of the SM then this 
  discovery confirms that the SM is consistent. As a consequence of the measurements related to the SM ~\citep{Beringer:1900zz}, 
  we now have direct evidence of all the SM spectrum.\newline 
  
  Despite the success of the SM when compared with experiment ~\citep{Beringer:1900zz,Montanet:1994xu} we have many 
  reasons to believe that the SM is not complete since, to say the least, gravity is not included. The SM does not 
 provide a satisfactory explanation to the hierarchy problem ~\citep{Burdman:2007ck,Tavares:2013dga}. Here we focus on extensions of the SM that solve 
 the hierarchy problem. In particular, we study quiver 
 theories~\citep{Bai:2009ij,Burdman:2012sb,Burdman:2013qfa,Burdman:2014ixa,Nayara-Burdman,Lima-Burdman} and their phenomenology. Using quiver theories 
 ~\cite{Burdman:2007ck,Gherghetta:2000qt,Lillie:2007yh} as well as other similar theories where the Higgs is a pseudo-Nambu-Goldstone
boson (pNGB).\newline

We study the fermion excitations in these models by computing their masses and wave functions. 
Considering both cases left- and right-handed zero mode fermions, we will study the more relevant phenomenology. 

To reproduce the phenomenology of these fermion excitations, we compute all their couplings. 
The one to the Higgs sector will be crucial to do the phenomenology. 
For instance, the coupling to the first excitations of the gauge bosons will be neglected, 
but only because we computed first and now we know they will play no role in the pair production.\newline 

We study the phenomenology of vector-like quarks at high energy Colliders. Prompted by our results in quiver theories, the vector-like quark 
can be heavier than the excitedgluons, we study the phenomenology of production and decay of vector-like quark at the LHC and beyond taking into 
account the decay channel $T\rightarrow Gt$, with $Q$ the vector-like quark, $G$ the heavy gluon and $q$ a SM quark.\newline

We start in chapter 2 by presenting the SM, jointly with the motivations to study physics beyond SM. In chapter 3
we briefly study one simple Little Higgs model~\citep{Schmaltz:2005ky} and introduce the quiver 
theories, for both gauge bosons and fermions. The Higgs is induced as a pNGB, by 
considering the quiver theory of EWSB. And then the couplings of fermions excited states were computed. In chapter 4 we study the phenomenology 
at the LHC of the excited heavy quarks in quiver theories, by using the most relevant couplings computed. 
In chapter 5 we begin 
to study the phenomenology of the heavy quarks at a future high energy Collider in a general vector-like theories. Finally we conclude and present the future studies in chapter 6.
\chapter{\bf The Standard Model}

The Standard Model (SM) of elementary particle physics is a quantum field theory that describes the electromagnetic, 
weak and strong interactions. The theoretical and experimental research in elementary particle physics in the 60s 
gave evidence of a possible unification of the weak and electromagnetic interactions, 
due to the fact that both are of vectorial nature and have universal couplings. In other words, 
they are both described by a gauge theory. Finally, between the years 60 and 70, 
the SM was first developed by Glashow, Weinberg and Salam, setting the foundations of our modern understanding 
of elementary particles.\newline
\newline\noindent The four basic ingredients necessary to the SM are: 
quarks, leptons, gauge bosons and the Higgs boson. All the electroweak and strong interactions 
are explained by gauge theories, namely the SM Lagrangian is invariant under the gauge transformations 
of $SU(3)_C\times SU(2)_L\times U(1)_Y$, for the strong and electroweak interactions respectively. 
We can study the strong interactions separately of the electroweak interactions, 
since both gauge sectors do not mix. The SM has a domain of applicability of at least several hundred of GeV. 
It is worth mentioning that not only works splendidly in theory, but it has also passed every experiment test so far. 
In addition, the model presents important symmetries in describing such interactions. 
In general, the symmetries have a central role in physics, namely, 
they protect some physical quantity and determine the dynamic structure of the fields. In this chapter, 
besides a brief introduction to the SM, we will present the gauge hierarchy problem and the 
problem of the hierarchy of SM fermion masses.

\newpage\section{Quantum Chromodynamics}

The sector of strong interactions of the SM better known as QCD is a non-Abelian gauge 
theory where generators belong to the local symmetry group $SU(3)_C$, and the internal degree of freedom 
is the named color. One of the properties of QCD is asymptotic freedom, that makes possible to 
use perturbative methods at very small distances~\citep{Dono2002}. The Lagrangian of QCD is

\begin{equation}\label{qcd-th} \mathcal{L} =
-\frac{1}{4} F_{\mu\nu}^i F^{\mu\nu i} + \sum_r ~\bar{q}_{r\alpha}~i
\displaystyle{\not} D^\alpha_{\mu \beta}~ q_r^\beta.
\end{equation}

Notice that in (\ref{qcd-th}), 
$r$ is the flavor index of quarks; the index of the adjoint representation is $i= 1,2,\ldots,8$ of $8$ generators  
of $SU(3)_C$; $~\alpha,\beta=1,2,3~$ are the indexes of the fundamental representation, these are the indexes of color, 
$q_r^\beta$ is a the quark field
and $F_{\mu\nu}^i $ is the gauge field strength tensor  
which is given by
\begin{equation}\label{FmunuSU3} 
F_{\mu\nu}^i = \partial_\mu G_\nu^i -\partial_\nu
 G_\mu^i - g_s f_{ijk} G_\mu^j G_\nu^k.
\end{equation}

\noindent In (\ref{FmunuSU3}) the fields $G_\mu^i$ are the gluons fields and the structure constants $f_{ijk}~~ (i,j,k = 1,2,\ldots,8)$ 
are defined by
\begin{equation}\label{structureconstants} [\lambda^i,\lambda^j]=2if_{ijk}\lambda^k.
\end{equation}

\noindent The matrices $\lambda^k$ are the Gell-Mann matrices for $SU(3)$. The covariant derivative is given by
\begin{equation}\label{covariant-SU(3)_c} 
D^\alpha_{\mu \beta}= \partial_\mu~
 \delta_\beta^\alpha~ + ~ \frac {1}{2}ig_s~ G_\mu^i~ \lambda_\beta^{\alpha i},
\end{equation}
\noindent with $g_s$ the gauge coupling constant in $SU(3)_c$ local. We should point out that 
the mass terms for the quarks fields will appear after the EW symmetry breaking. 

\section{Electroweak Model}\label{E-W}

The Glashow-Weinberg-Salam theory (GWS) where fermions are chiral is a gauge theory with symmetry 
$SU(2)_L\times U(1)_Y$, where $L$ and $Y$ represent the left-handed chirality with symmetry of weak 
isospin, and weak hypercharge, respectively. The left-handed leptons are
\begin{equation} \label{f:lL}
  \textsf{L}_e= \left (
\begin{array}{c}
\nu_e\\
e^-
\end{array} \right )_L ~~~~~~~~
 \textsf{L}_\mu= \left (
\begin{array}{c}
\nu_\mu\\
\mu^-
\end{array} \right )_L ~~~~~~~~
 \textsf{L}_\tau= \left (
\begin{array}{c}
\nu_\tau\\
\tau^-
\end{array} \right )_L, ~~~~~~~~
\end{equation}

\noindent with weak isospin \begin{math} I_\ell= 1/2 \end{math} and hypercharge \begin{math} Y_{L_\ell}=-1 \end{math}, 
and the right-handed leptons are
\begin{equation}\label{leptons-right}  
\textsf{R}_{e,\mu,\tau}= e_R,\mu_R,\tau_R ,
\end{equation}
with hypercharge \begin{math} Y_{R_\ell}=-2 \end{math}, where the hypercharge is given by the 
Gell-Mann and Nishijima relation \begin{math} Q=I_3+ (1/2)Y \end{math}. The SM does not include the right-handed 
neutrinos because these have no SM gauge quantum numbers. The left-handed quarks are given by 

\begin{equation} \label{f:qL}
  \textsf{L}_q^1= \left (
\begin{array}{c}
u\\
d
\end{array} \right )_L ~~~~~~~~
\textsf{L}_q^2= \left (
\begin{array}{c}
c\\
s
\end{array} \right )_L ~~~~~~~~
\textsf{L}_q^3= \left (
\begin{array}{c}
t\\
b
\end{array} \right )_L, ~~~~~~~~
\end{equation}

\noindent with weak isospin \begin{math} I_q= 1/2 \end{math} and hypercharge \begin{math} Y_{L_q}=1/3 \end{math}. The right-handed quarks are
\begin{equation} \label{f:qRu} \textsf{R}_u^{(1,2,3)}= u_R, c_R, t_R 
\end{equation}
and
\begin{equation}\label{f:qRd} \textsf{R}_d^{(1,2,3)}= d_R, s_R, b_R 
\end{equation}
with \begin{math} Y_{R_u}= 4/3 \end{math} and \begin{math} Y_{R_d}=- 2/3 \end{math}, respectively.\newline
\newline Now, we write the Lagrangian $\mathcal{L}$ that is invariant under the gauge symmetry $SU(2)_{L}\times U(1)_{Y}$ 
in three parts, as follows

\begin{equation}\label{L-EW} 
\mathcal{L} = \mathcal{L}_{G} + \mathcal{L}_{F} + \mathcal{L}_{H},
\end{equation}

\noindent with 

\begin{equation}\label{f:Lgauge}
\mathcal{L}_{G} = -\frac{1}{4} W_{\mu\nu}^i W^{\mu\nu i} - \frac{1}{4} B_{\mu\nu}B^{\mu\nu},
\end{equation}

\noindent where the field strength tensor $W_{\mu\nu}^i$ corresponds to the gauge symmetry $SU(2)_L$, with coupling $g$ 
and is given by
\begin{equation}\label{tensorSU(2)}
W_{\mu\nu}^i = \partial_\nu W_\mu^i - \partial_\mu W_\nu^i + g \varepsilon_{ijk}W_\mu^j W_\nu^k,
\end{equation}

\noindent and the field strength tensor $B_{\mu\nu}$ corresponds to the $U(1)_Y$ field with coupling $g'$ and is 
given by
\begin{equation}\label{tensorU(1)}
B_{\mu\nu} = \partial_\nu B_\mu - \partial_\mu B_\nu.
\end{equation}

\noindent The kinetic term for the fermions is written in two parts
\begin{equation}\label{fermionsSM} 
\mathcal{L}_{F} =  \mathcal{L}_{\mathrm{leptons}}+ \mathcal{L}_{\mathrm{quarks}},
\end{equation}
that are given by

\begin{align}\label{f:Lleptons}
\mathcal{L}_{\mathrm{leptons}} &=\overline{\textsf{R}}_\ell^n ~ i \gamma^\mu \left( \partial_\mu+ i \frac{g'}{2} B_\mu Y \right) \textsf{R}_\ell^n \nonumber \\
 & + \overline{\textsf{L}}_\ell^n ~ i \gamma^\mu \left( \partial_\mu + i \frac{g'}{2} B_\mu Y  + i\frac{g}{2} \vec{\sigma} \cdot \overrightarrow{W}_\mu\right) \textsf{L}_\ell^n,
\end{align}
where the index $\ell$ represents the leptons ($ \ell= e,\mu,\tau$); 
and the second term in (\ref{fermionsSM}) is

\begin{align}\label{f:Lquarks}
\mathcal{L}_{\mathrm{quarks}} &= \overline{\textsf{R}}_u^n ~i \gamma^\mu \left( \partial_\mu+ i \frac{g'}{2} B_\mu Y \right) \textsf{R}_u^n + \overline{\textsf{R}}_d^{n} ~i \gamma^\mu\left( \partial_\mu + i \frac{g'}{2} B_\mu Y \right) \textsf{R}_d^n  \nonumber\\
&+\overline{\textsf{L}}_q^n ~i \gamma^\mu \left( \partial_\mu + i \frac{g'}{2} B_\mu Y  + i\frac{g}{2} \vec{\sigma} \cdot \overrightarrow{W}_\mu\right) \textsf{L}_q^n,
\end{align}
where the index $n$ is the index of generations $(n = 1,2,3)$. In (\ref{f:Lleptons}) and (\ref{f:Lquarks}) 
$\vec{\sigma}$ represents the Pauli matrices: 
\begin{align} \label{Pauli}
&\sigma^1=\left( \begin{array}{cc}0 & ~~1\\1 & ~~0\end{array} \right),~\sigma^2=\left( \begin{array}{cc}0 & -i\\i & ~~0\\\end{array} \right),~\sigma^3=\left( \begin{array}{cc}0 & -1\\0 & ~~1\\\end{array} \right).
\end{align}
The Lagrangian (\ref{f:Lgauge}) has four massless gauge bosons $W_\mu^1$, $W_\mu^2$, $W_\mu^3$ and $B_\mu $ 
because their mass terms are not invariant under gauge transformations. In addition, the gauge symmetry 
$SU(2)_L\times U(1)_Y$ prohibits mass 
terms for fermions, since the left- and right- handed components of the fermionic fields transform 
in different ways under the $SU(2)_L \times U(1)_Y$ gauge symmetry.\newline
\newline In the SM, the electroweak symmetry is spontaneously broken through the Higgs 
mechanism. This is done by introducing a scalar that is a doublet of $SU(2)_L$, such that it includes 
two complex scalar fields as follows
\begin{equation} \label{f:dHiggs}
 \Phi= \left (
\begin{array}{c}
\phi^+\\
\phi^0
\end{array} \right ),
\end{equation}

\noindent and with hypercharge $Y_\Phi=1$. The Lagrangian for $\Phi$ is given by

\begin{equation}\label{f:Lescalar}
 \mathcal{L}_{\Phi} = (D^\mu\Phi)^\dag(D_\mu\Phi)- V(\Phi^\dag\Phi),
\end{equation}
where 
\begin{equation}\label{covariantderivative-EW}
 D_\mu=\partial_\mu + i\frac{g'}{2}B_\mu Y + i\frac{g}{2}\vec{ \sigma} \cdot \overrightarrow{W}_\mu
\end{equation} 
and the potential in $\mathcal{L}_{\Phi}$ is 

\begin{equation}\label{f:potencialH}
 V(\Phi^\dag\Phi) = \mu^2 (\Phi^\dag\Phi) + \lambda(\Phi^\dag\Phi)^2.
\end{equation}

In the following, we will work with the ground state that minimizes the potential (\ref{f:potencialH}), 
considering the case for $\mu^2 < 0$. Then we can write the VEV for $\Phi$ as follows: 

\begin{equation} \label{f:vacuo}
 \langle\Phi\rangle_0= \frac{1}{\sqrt{2}}\left (
\begin{array}{c}
0\\
v
\end{array} \right ),
\end{equation}
where $v=\sqrt{-\mu^2/\lambda}$. Using the nonlinear sigma model~\citep{peskin}, jointly with (\ref{f:vacuo}), 
we parameterize the field $\Phi$ as
\begin{equation}\label{Higgsdoublet-NLsigma}
 \Phi=  \text{exp}(\frac{i\xi^j(x) \sigma^j}{2v}) \left (
\begin{array}{c}
0\\
(v + h(x))/\sqrt{2}
\end{array} \right ),
\end{equation}
where $\sigma^j$ are Pauli matrices (\ref{Pauli}). But we will work in the unitary gauge, such that 
the Nambu-Goldstone Bosons (NGB's) $\xi^j$ can be removed by the following gauge transformation:
\begin{equation}\label{f:phigaugeunitario}
 \Phi \rightarrow  \text{exp}(\frac{-i\xi^j(x) \sigma^j}{2v})\Phi = \frac{1}{\sqrt{2}} \left (
\begin{array}{c}
0\\
v + h(x)
\end{array} \right ),
\end{equation}
so $\mathcal{L}_{\Phi}$ in (\ref{f:Lescalar}) can be written in terms of physical fields as
\begin{align}\label{L_Phi_after_EWBS_GU}
\mathcal{L}_{\Phi} &= \frac{1}{2} \partial_\mu h \partial^\mu h  +\frac{g^2}{4} (v + h)^2 \left[W_\mu^+ W^{- \mu} +\frac{1}{2 \cos^2{\theta_W}} Z_\mu Z^\mu \right]  \nonumber \\  
&- \mu^2 ~ \frac{(v + h)^2}{2} ~- \lambda~  \frac{(v + h)^4}{4},
\end{align}
where the Weinberg angle is defined by the relation $\tan{\theta_W} \equiv g'/g$. We also have defined the mediator 
fields of the charged weak interactions $W^\pm_\mu$,
\begin{equation}\label{W-W}
W^\pm_\mu = (W_\mu^1\mp i
W_\mu^2)/\sqrt{2}.
\end{equation}

\noindent These physical fields $W^\pm_\mu$ acquire mass equivalent to $M_W=gv/2=ev/2\sin\theta_W$ 
and the field that mediates the neutral weak interactions $Z_{\mu}$,
\begin{equation}\label{Z-W}
Z_\mu=W_\mu^3 \cos\theta_W- B_\mu
\sin \theta_W,
\end{equation}

\noindent acquires a mass $M_Z=\sqrt{g^2+g'^{2}}v/2=M_W/\cos\theta_W$. The photon field is 
written as the orthogonal neutral combination
\begin{equation}\label{A-electromagnetic}
A_\mu=B_\mu \cos\theta_W + W_\mu^3 \sin \theta_W, 
\end{equation}
\noindent and its mass is $M_A=0$, leaving the $U(1)_{EM}$ unbroken. There are also Yukawa interactions 
between the Higgs doublet $\Phi$ with the quarks and leptons, 
that can be written as 

\begin{equation}\label{eq50}
\mathcal{L}_{\Phi F}= \mathcal{L}_{\mathrm{Y(quarks)}}+\mathcal{L}_{\mathrm{Y (leptons)}},
\end{equation}
where 
\begin{equation}\label{f:Lquarksm}
\mathcal{L}_{\mathrm{Y(quarks)}}= - \sum_{i,j=1}^3\left[ G_{ij}^u ~\overline{\textsf{R}}_{u}^i \left(\tilde{\Phi}^\dag \textsf{L}_q^j \right) +  G_{ij}^d~ \overline{\textsf{R}}_{d}^i \left(\Phi^\dag \textsf{L}_q^j \right)  + \mathrm{h.c.} \right],
\end{equation}
and
\begin{equation}\label{f:Lleptonsm}
 \mathcal{L}_{\mathrm{Y (leptons)}} = -\sum_{i,j=1}^3 \left[G^\ell_{i,j}(\overline{\textsf{L}}_\ell^i\Phi)\textsf{R}_\ell^j + \mathrm{h.c.} \right],
\end{equation}

\noindent where $G_{ij}^u$, $G_{ij}^d$ and $G_{i,j}^\ell$ are the Yukawa couplings, these are not diagonal in the generation 
indexes nor real, and 
$\tilde{\Phi} \equiv i \sigma_2 \Phi^*$. 
We note that all these terms are also gauge invariant under $SU(2)_L$ and have zero net hypercharge $Y.$

In the following we will consider $\mathcal{L}_{\mathrm{Y (quarks)}}$ in (\ref{f:Lquarksm}), and after replacing $\phi$ 
(\ref{f:phigaugeunitario}) in (\ref{f:Lquarksm}), we can identify

\begin{equation}\label{umass-SM}
-\bar{u}_R^{~i}~ M^u_{ij}~ u_L^j ~+~ \mathrm{h.c.},
\end{equation}
\noindent where  $u_L^j$ is the left-handed up-type quark of the doublet 
$\textsf{L}_q^j$ (\ref{f:qL}) and $\bar{u}_R^i = \{\bar{u}_R, \bar{c}_R, \bar{t}_R\}$.\newline 
\newline To the down-type quarks we have also
\begin{equation}\label{downmass-SM}
-\bar{d}_R^{~i}~ M^d_{ij} ~d_L^j ~+ ~\mathrm{h.c.},
\end{equation}

\noindent where $d_L^j$ is the left-handed down-type quark of the doublet 
$\textsf{L}_q^j$ (\ref{f:qL}) and $\bar{d}_R^i = \{\bar{d}_R, \bar{s}_R, \bar{b}_R\}$.\newline
\newline\noindent The matrices $M^u$ and $M^d$, which are generally not diagonal, are given by 
$M^{u(d)}_{ij}= G_{ij}^{u(d)}v/\sqrt{2}$. To diagonalize these mass matrices, 
the unitary matrices $U(D)_{L,R}$ are defined such as:
\begin{equation}\label{matrixL}
u_L^i= U_L^{ij} ~ u_L'^{~j}, ~~~~~~~~d_L^i= D_L^{ij} ~ d_L'^{~j},
\end{equation}
\begin{equation}\label{matrixR}
u_R^i= U_R^{ij} ~ u_R'^{~j}, ~~~~~~~~d_R^i= D_R^{ij} ~ d_R'^{~j},
\end{equation}

Note that the gauge eigenstates ($q$) are linear combinations of the mass eigenstates ($q'$), 
so we have to do the basis changes given by (\ref{matrixL}) and (\ref{matrixR}) to diagonalize $M^{U}$ and $M^{D}$, 
as follows

\begin{equation} \label{f:MUdiag}
M^U_{\mathrm{diag}} ~\equiv  ~ U_R^{\dag}~M^U~U_L~= ~\left( \begin{array}{ccc}
m_u  & 0    & 0   \\
0  & m_c    & 0   \\
0  & 0   & m_t   \\
\end{array} \right),
\end{equation}
\begin{equation} \label{f:MDdiag}
M^D_{\mathrm{diag}} ~\equiv  ~ D_R^{\dag}~M^D~D_L~= ~\left( \begin{array}{ccc}
m_d  & 0    & 0   \\
0  & m_s    & 0   \\
0  & 0   & m_b   \\
\end{array} \right).
\end{equation}

Considering the quark sector in (\ref{eq50}) after the EW symmetry breaking, 
we need to go to the mass basis to diagonalize the Yukawa terms in (\ref{f:Lquarksm}). This is done with the $U_{L,R}$, $D_{L,R}$ 
unitary transformations.

For the case of charged current $J^{+}_{\mu~W}$, we have that it must be proportional to
\begin{equation}\label{JWmu+SM}
\left( \begin{array}{ccc}\bar{u}' & \bar{c}' & \bar{t}' \end{array} \right)_L  U_L^{\dag}~   \gamma_\mu ~D_L  \left(\begin{array}{c} d' \\ s' \\ b' \end{array} \right )_L ~.
\end{equation}

That is, the charged current coupling will not be diagonal anymore, since $U_L^{\dag} D_L \neq \mathbf{1}_3$. 
This unitary matrix that express the mixing between the quarks is known as Cabibbo-Kobayashi-Maskawa (CKM) matrix
\begin{equation}\label{CKM-SM}
\textsf{V} _{\mathrm{CKM}}~ \equiv~  U_L^{\dag} D_L ~= ~ \left( \begin{array}{ccc}
V_{ud}  & V_{us}    & V_{ub}   \\
V_{cd}  & V_{cs}    & V_{cb}   \\
V_{td}  & V_{ts}    & V_{tb}   \\
\end{array} \right).
\end{equation}
In the case of neutral current $J_{\mu~Z}$, considering up-type SM quarks, it must be proportional to
\begin{eqnarray}\label{JZmu0SM}
~\left( \begin{array}{ccc} \bar{u}' & \bar{c}' & \bar{t}' \end{array} \right )_{L,R}  U_{L,R}^{\dag}  ~ \gamma_\mu  ~ U_{L,R}\left(\begin{array}{c} u' \\ c' \\ t' \end{array}\right )_{L,R}=~\left( \begin{array}{ccc} \bar{u}' & \bar{c}' & \bar{t}' \end{array} \right )_{L,R}  ~ \gamma_\mu  ~\left(\begin{array}{c} u' \\ c' \\ t' \end{array}\right )_{L,R} .
\end{eqnarray}
So there is no mixing of the quarks in the sector of neutral currents, due to the fact that $ U_L$ and $U_R$ are 
unitary, namely, 
the couplings are diagonal. Thus, in the Standard Model there are no interactions that change flavor in neutral currents, 
at least at tree level.

Now we will consider $\mathcal{L}_{\mathrm{Y(leptons)}}$, jointly with $\phi$ in (\ref{f:phigaugeunitario}), we can 
obtain mass terms 

\begin{equation}\label{leptonmass-SM}
 -\bar{e}^{~i}_L~M^\ell_{i,j}~ e^{~j}_R+~ \mathrm{h.c.},
\end{equation}

\noindent where $e^i_{L,R}=\{e_{L,R},\mu_{L,R},\tau_{L,R}\}$ and we identify from (\ref{f:Lleptonsm}) 
that the charged lepton mass matrix is given by 
$M^\ell_{i,j} = G^\ell_{i,j} v/ \sqrt{2}$. 

To diagonalize $M^\ell_{i,j}$, the matrices $U_\ell$ and $W_\ell$ are defined as 

\begin{equation}\label{matrixleptons}
e_L^i= U_\ell^{ij} ~ e_L'^{~j}, ~~~~~~~~e_R^i= W_\ell^{ij} ~ e_R'^{~j}.
\end{equation}

Then these basis changes yield the diagonal lepton mass matrix

\begin{equation} \label{LEPTON:MDdiag}
M^{\ell}_{\mathrm{diag}} ~\equiv  ~ D_R^{\dag}~M^D~D_L~= ~\left( \begin{array}{ccc}
m_e & 0    & 0   \\
0  & m_\mu    & 0   \\
0  & 0   & m_\tau   \\
\end{array} \right).
\end{equation}
With this procedure it is possible to obtain the masses terms to the leptons. However, 
the value of $M_\ell$ is not predicted by the theory since the Yukawa couplings $G^\ell_{i,j}$ 
were introduced arbitrarily to reproduce the masses of the observed leptons. Just as in the quark case, 
the theory does not provide the values of $M^{u(d)}_{ij}$. That is, the Yukawa couplings $G_{ij}^{u(d)}$ 
were arbitrarily introduced to correctly give the masses of the fermions observed, once they are diagonalized.\newline

\section{Motivation of Physics Beyond the SM}
In the previous sections we saw that the SM of elementary particles offers a theoretical 
explanation for the interactions of all elementary particles we know of today. It is extremely successful when compared 
with the experimental data we have nowadays. As examples, we have the detection of neutral currents in the decade of 70s, 
the predictions made to the bosons mass $W_\mu^\pm$ and $Z_\mu$, experiments made in LEPI, LEPII and the 
Tevatron~\citep{Yao:2006px,ALTARELLI:2005zv} with high precision ($\lesssim1\%$). 

\noindent The model needs a scalar particle remaining form the process of spontaneous symmetry breaking 
in the electroweak sector, the Higgs boson that couples with the other particles of the SM and that 
recently has been discovered at the LHC~\citep{Aad:2012tfa,Chatrchyan:2012ufa}.\newline
\newline Despite the success of the SM, it has deficiencies in how to explain the hierarchy problem and the 
hierarchy of the SM fermion masses among many others. The hierarchy problem is related to the quantum instability of the vacuum 
that sets 
the electroweak scale around $m_{EW}\sim250GeV$. In addition, the SM does not include the interaction with gravity, which is non-renormalizable 
and can be ignored up to scales of order of Planck mass.

\subsection{The Gauge Hierarchy Problem}

\begin{figure}[H]
   \centering
   \epsfig{file=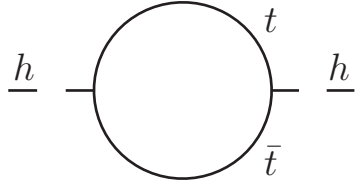}
   \captionsetup{font=small, labelfont=bf, labelsep=period}
   \caption{\label{hierarchy_problem}Quantum contribution to the Higgs boson propagator from the top quark loop in the SM.} 
  \end{figure}

Despite the success of the SM when compared with experiment~\citep{Beringer:1900zz,Montanet:1994xu} we have quite 
important reasons to believe that the SM is not complete, besides the fact that gravity is not included. 
Among all these, the one that requires new physics not too far above the TeV scale is the hierarchy problem. The SM does not provide 
a satisfactory explanation to the hierarchy problem~\citep{Burdman:2007ck,Tavares:2013dga}. 
This problem can be understood when we calculate the quantum corrections to the Higgs mass. For instance, 
the largest of these corrections comes from 
the virtual top pair contribution up to $1$ loop to the Higgs propagator as in Figure (\ref{hierarchy_problem}). This 
contribution is given by

\begin{align}\label{renormalized-mass-higgs}
-i\delta m_h^{2}&=-N_c\left(\frac{\lambda_t}{\sqrt{2}}\right)^2\int \frac{d^4 p}{(2\pi)^4} \mathrm{Tr}\Bigg [\left(\frac{\slashed p+m_t}{p^2-m_t^2}\right)\left(\frac{\slashed p+m_t}{p^2-m_t^2}\right)\Bigg ]
\nonumber\\
&=-2N_c\lambda_t^2\int \frac{d^4 p}{(2\pi)^4} \frac{ p^2+m_t^2}{(p^2-m_t^2)^2}.
\end{align}
Using a Wick rotation, i.e., $p_0\rightarrow ip_4$, $p^2\rightarrow-p_E^2$, jointly with the angular integration, 
we obtain

\begin{align}\label{renormalized-mass-higgs-next}
\delta m_h^{2}&=-\frac{N_c\lambda_t^2}{8\pi^2}\int^{\varLambda_{UV}^2}_0 d p_E^2 \frac{p_E^2(p_E^2-m_t^2)}{(p_E^2+m_t^2)^2}\nonumber\\
&=-\frac{N_c\lambda_t^2}{8\pi^2}\bigg[\varLambda_{UV}^2+...\bigg],
\end{align}

\noindent where $N_c=3$, and $\varLambda_{UV}$ is the highest energy where we believe the SM can be used for this computation. 
For instance, if we consider $\Lambda_{UV}=M_{Planck}$, the scale where gravity becomes strong and needs to be included,


\begin{equation}\label{M_plank}
M_{Planck} = \left(\frac{\hbar c}{G_{\mathrm{Newton}}}\right)^{1/2}\simeq~ 1.2\times 10^{19} ~\mathrm{GeV},
\end{equation}
where in this scale the quantum gravitational effects will become important. On the other hand, 
the recent discovery of a new particle at the Large Hadron Collider (LHC)~\citep{Aad:2012tfa,Chatrchyan:2012ufa} 
confirms that this particle is like the Higgs boson of the SM with mass $m_h\approx125$ GeV, when the 
electroweak symmetry is broken by the VEV of scalar Higgs.

So, as shown in (\ref{renormalized-mass-higgs-next}) this correction to the Higgs mass squared has 
quadratic sensitivity to the cutoff of the theory, and then, if the SM is valid until the Planck scale, we 
will need a large fine tuning to obtain the 
observed Higgs mass in the electroweak scale. This fine tunning is not natural and needs new 
physics beyond the SM to restore naturalness~\citep{Burdman:2007ck}. That is not the case for the masses associated to other 
SM elementary particles, for instance, considering a light fermion or a gauge boson as can be found in Ref.~\citep{Csaki:2016kln}, as we see below.\newline

\begin{figure}[H]
	\centering
        \epsfig{file=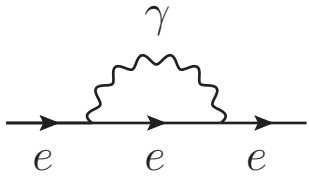}
        \captionsetup{font=small, labelfont=bf, labelsep=period}
        \caption{\label{hierarchy_problem_e}Electron self energy.}
\end{figure}

\noindent Let us consider the quantum corrections to the self-energy of the electron as an example, as 
shown in Figure (\ref{hierarchy_problem_e}). We will compute this correction at zero external momentum, considering the one-loop 
diagram of the electron propagator that comes from the $\gamma$ boson contribution, this result in a contribution given by



\begin{align}\label{renormalized-mass-electron-fig_c_foton_next}
\delta m_{e\gamma}=&m_ee^2\frac{1}{2 \pi^2}\int^{\varLambda}_0 d p_E\frac{p_E^3}{\left(p^2_E\right)\left(p^2_E+m_e^2\right)}\nonumber\\
=&m_ee^2\frac{1}{4 \pi^2}~\mathrm{ln}\left(\frac{\varLambda^2+m^2_e}{m^2_e}\right).
\end{align}

Consequently, from the loop considered to determine $\delta m_e$, we can see that this is divergent but only logarithmically sensitive to the 
cutoff of the theory. If, for instance, we use $\Lambda_{UV}=M_{Planck}$ we have  

\begin{equation}\label{electron_mass_correction}
\delta m_e=0.3m_e, 
\end{equation}

which is not that large. It is also a multiplicative shift, a reflection of chiral symmetry.

\begin{figure}[H]
	\centering
	\captionsetup[subfigure]{font=normalsize, labelfont=bf, labelsep=period}
	\subfloat[]{\includegraphics[width=1.7in]{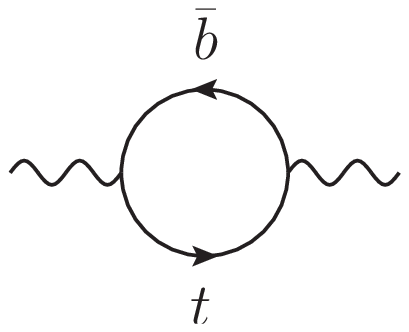}}\quad
	\subfloat[]{\includegraphics[width=1.7in]{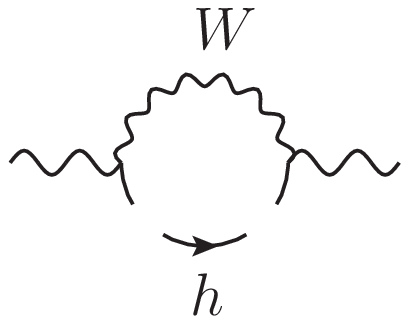}}\quad
        \captionsetup{font=small, labelfont=bf, labelsep=period}
        \caption{\label{hierarchy_problem_W}EW quantum corrections at one-loop for the W boson self energy involving: (a) t and b, 
        (b) W and h.}
\end{figure}

Similarly to the previous calculations, we illustrate the quantum correction at one-loop of the W self energy, at external zero momentum. 
For instance, considering the contributions as indicated in Figure (\ref{hierarchy_problem_W}).

\noindent The corresponding contribution to the $\delta m_{W}^2$ after a Wick rotation and the angular integration associated with 
the first diagram in Figure (\ref{hierarchy_problem_W}) is given by

\begin{align}\label{renormalized-mass-w-a-next}
\delta m_{Wa}^{2}&=N_cg^2m_tm_b\frac{1}{8\pi^2}\int^{\varLambda}_0 d p_E \frac{p^3_E}{(p^2_E+m_t^2)(p^2_E+m_b^2)}\nonumber\\
&=\frac{N_cg^2m_tm_b}{16\pi^2(m^2_t-m^2_b)}\left[m^2_t~\mathrm{ln}\left(\frac{\varLambda^2+m^2_t}{m^2_t}\right)-m^2_b~\mathrm{ln}\left(\frac{\varLambda^2+m^2_b}{m^2_b}\right)\right],
\end{align}

where $N_c=3$ due that we deal with three colors of quarks.\newline\newline 

\noindent Now, from the diagram in Figure (\ref{hierarchy_problem_e}b) gives a contribution to the $\delta m_{W}^2$ given by 



\begin{align}\label{renormalized-mass-w-b-next}
\delta m_{Wb}^{2}&=g^2m^2_W\frac{1}{8\pi^2}\int^{\varLambda}_0 d p_E \frac{p^3_E}{(p^2_E+m_W^2)(p^2_E+m_h^2)}\nonumber\\
&=\frac{g^2m^2_W}{16\pi^2(m^2_W-m^2_h)}\left[m^2_W~\mathrm{ln}\left(\frac{\varLambda^2+m^2_W}{m^2_W}\right)-m^2_h~\mathrm{ln}\left(\frac{\varLambda^2+m^2_h}{m^2_h}\right)\right]
\end{align}

Notice that according to (\ref{renormalized-mass-w-a-next}) and (\ref{renormalized-mass-w-b-next}) for the contributions considered to the 
$\delta m_{W}^2$, these contributions are also divergent but have logarithmic divergence to the cutoff. And it is possible to write these contributions 
as proportional to $m_W^2$. Consequently, in the massless limit of fermions as well as gauge bosons we will have the mass parameters quantum 
corrections go to zero and recover a chiral and EW gauge symmetry, 
respectively.\newline 

We conclude that the quadratically divergence takes place in the Higgs mass squared as indicated in (\ref{renormalized-mass-higgs-next}), and its quantum correction is unnatural, it is 
due that in the SM there is no symmetry that protects the Higgs mass.

\subsection{Problem of Fermion Mass Hierarchy}
When we review in Section \ref{E-W} the electroweak sector of the SM, especially when through 
Higgs mechanism the $\Phi$ acquired an VEV, we obtained mass terms. That is as a consequence of the Yukawa 
interactions given by 
(\ref{f:Lleptonsm}) and (\ref{f:Lquarksm}), 
these mass terms can be identified as
\begin{equation}\label{mass-generic-SM}
 M_f=\frac{G_{f}v}{\sqrt{2}},
\end{equation}

\noindent where $G_f$ are Yukawa couplings of the fermions, and are extremely varied. For instance, we have that the coupling 
for the electron and the top are $G_e \sim 10^{-6}$ and $G_t \sim 1$, respectively. This is a problem, since the SM 
does not explain why the fermions can have masses so different,
after the electroweak symmetry breaking through the Higgs mechanism.\newline 

\noindent Thus, to explain the fact that these couplings are so different it is necessary to introduce physics beyond the Standard Model. 
There are many other problems with the SM. The strong CP problem, origin of baryon asymmetry, dark matter, among others. We will focus on theories 
that address the hierarchy problem, and in particular in which the Higgs mass is protected by a symmetry similar to that protecting the pion mass 
in QCD, i.e, the Higgs will be a pNGB.\newline

\noindent The hierarchy problem (and maybe the problem of the fermion mass hierarchy) seem to be a good guides to construct 
theories Beyond the SM. Let us consider theories that solve these hierarchy problems generating large hierarchy of scales.


\chapter{\bf New Fermions in BSM Theories}


As was pointed out in the previous chapter, the SM does not provide a satisfactory explanation for the hierarchy problem. 
Extensions of the SM that solve the hierarchy problem without supersymmetry~\citep{Martin:1997ns} require the presence of new states 
partners of the SM fields under some symmetry. In particular there will be partners of fermions, especially of the top quark, of gauge bosons, etc. 
We will focus here on the Vector-like quarks that are partners of the SM fermions. 
Several theories have been suggested in the literature, for instance 
Little Higgs models~\citep{Schmaltz:2005ky,Perelstein:2005ka}, composite Higgs models~\citep{Contino:2010rs,Panico:2015jxa,Carena:2014ria} and 
quiver theories~\citep{Burdman:2012sb,Burdman:2014ixa}.\newline

\noindent As a first example, we will study how the hierarchy problem is addressed in a Little Higgs model~\citep{Schmaltz:2005ky,Perelstein:2003wd}. Firstly, 
we consider two independent $SU(3)$ global symmetries, with two nonlinear sigma fields that parametrize the spontaneous symmetry breaking associated with 
the coset $[SU(3)/SU(2)]^2$ and are given by

\begin{align}\label{scalar1-parametrization-convenient}
\phi_1&=\exp\left \{\frac{i}{f}\left(
\renewcommand{\arraystretch}{1.6}
\begin{array}{ccc}
0 & 0 & k_1\\
0 & 0 & k_2\\
k^*_1 & k^*_2 & 0 \\
\end{array}
\right)\right \}\exp\left \{\frac{i}{f}\left(
\renewcommand{\arraystretch}{1.6}
\begin{array}{ccc}
0 & 0 & H_1\\
0 & 0 & H_2\\
H^*_1 & H^*_2 & 0 \\
\end{array}
\right)\right \}\left(
\renewcommand{\arraystretch}{1.6}
\begin{array}{c}
0 \\
0 \\
f \\
\end{array}
\right),\\
\phi_2&=\exp\left \{\frac{i}{f}\left(
\renewcommand{\arraystretch}{1.6}
\begin{array}{ccc}
0 & 0 & k_1\\
0 & 0 & k_2\\
k^*_1 & k^*_2 & 0 \\
\end{array}
\right)\right \}\exp\left \{-\frac{i}{f}\left(
\renewcommand{\arraystretch}{1.6}
\begin{array}{ccc}
0 & 0 & H_1\\
0 & 0 & H_2\\
H^*_1 & H^*_2 & 0 \\
\end{array}
\right)\right \}\left(
\renewcommand{\arraystretch}{1.6}
\begin{array}{c}
0 \\
0 \\
f \\
\end{array}
\right),\label{scalar2-parametrization-convenient}
\end{align}

\noindent where $H_1$, $H_2$, $k_1$ and $K_2$ are complex fields, with the same symmetry breaking scale given by $f$. The symmetry breaking pattern is 
the coset $[SU(3)/SU(2)]^2$, therefore, after the breaking symmetry of the two $SU(3)$ we will identify $10$ spontaneously broken generators, resulting in $10$ 
NGBs. Notice that two singlet fields of $SU(2)$ were ignored for simplicity in the parametrization of both scalar fields $\phi_1$ and $\phi_2$. 
Then, we add the interaction between the scalar fields and gauge bosons associated with the $SU(3)$ through the covariant derivatives acting on 
$\phi_1$ and $\phi_2$. We write the Lagrangian $\mathcal{L}$ associated with these fields as follows

\begin{equation}\label{LescalarLittleH1}
 \mathcal{L} = (D^\mu\phi_1)^\dag(D_\mu\phi_1)+(D^\mu\phi_2)^\dag(D_\mu\phi_2),
\end{equation}
where 
\begin{equation}\label{covariantderivative-LittleH1}
 D_\mu=\partial_\mu - ig A^{a}_\mu T^a,
\end{equation}

\noindent such that $T^a$ are the $SU(3)$ generators with $a= 1,2,\ldots,8$. We can identify the interactions between the scalar fields and the gauge bosons 
by expanding the kinetic terms in (\ref{LescalarLittleH1}),   

\begin{align}
(D_\mu \phi_i)^\dag(D^\mu \phi_i) &= (\partial_\mu \phi_j)^\dag(\partial^\mu \phi_j) + g\left[i\phi_i^\dag A^{a}_\mu T^a \partial^\mu \phi_i+\mathrm{h.c.}\right]\nonumber \\
&+g^{2} \left[\phi_i ^\dag A^a_{\mu} A^{\mu b}T^bT^a\phi_i\right]. \label{scalar-kinetic-term1}
\end{align}

\begin{figure}[H]
	\centering
	\captionsetup[subfigure]{font=normalsize, labelfont=bf, labelsep=period}
	\subfloat[]{\includegraphics[width=1.2in]{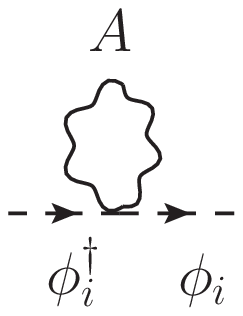}}\quad
	\subfloat[]{\includegraphics[width=1.8in]{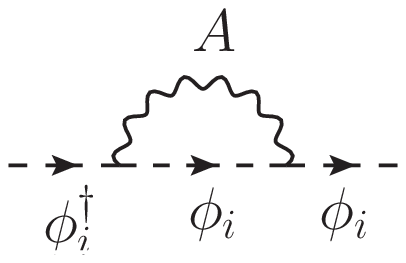}}\quad
        \captionsetup{font=small, labelfont=bf, labelsep=period}
        \caption{\label{LittleH-quadratic}Quantum corrections at one-loop for two-point function of the scalar fields $\phi_i$ due to the $SU(3)$ 
        gauge interaction.}
\end{figure}

In the calculation of the quantum contribution from Figure (\ref{LittleH-quadratic}{\bf a}) the Feynman gauge will be used. This gives



\begin{align}\label{LittleH-fig-a2}
i\Sigma_{(\bf a)}&=\int \frac{d^4 k}{(2\pi)^4}\left[ig^2g^{\mu\nu}\left(T^r_{ij}T^s_{jm}+T^s_{ij}T^r_{jm}\right)\right]\left(\frac{-ig_{\mu\nu}\delta^{rs}}{k^2}\right),\nonumber\\
&=-8ig^2\left(T^r_{ij}T^r_{jm}\right)\frac{1}{16\pi^2}\varLambda^2.
\end{align}

from Figure (\ref{LittleH-quadratic}{\bf b}), we also have

\begin{align}\label{LittleH-fig-b1}
i\Sigma_{(\bf b)}&=\int \frac{d^4 k}{(2\pi)^4}\left(igk^\mu T^r_{ij}\right)\left(\frac{i\delta^{jl}}{k^2}\right)\left(igk^\nu T^s_{lm}\right)\left(\frac{-ig_{\mu\nu}\delta^{rs}}{k^2}\right),\nonumber\\
&=ig^2\left(T^r_{ij}T^r_{jm}\right)\frac{1}{16\pi^2}\varLambda^2.
\end{align}



So the contributions potential  
\begin{equation}\label{quadratic-term-LittleH1}
\varpropto\frac{g^4}{16\pi^2}~\varLambda^2\left(\phi_1^\dag\phi_1+\phi_2^\dag\phi_2\right),
\end{equation}

and then substituting (\ref{scalar1-parametrization-convenient}), (\ref{scalar2-parametrization-convenient}) in (\ref{quadratic-term-LittleH1}) 
we will obtain a constant term that that does not contribute to the potential for $H$, since (\ref{quadratic-term-LittleH1}) does not depend on the 
Higgs.

\begin{figure}[H]
	\centering
	\includegraphics[width=1.2in]{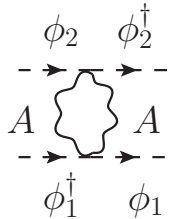}
        \captionsetup{font=small, labelfont=bf, labelsep=period}
        \caption{\label{LittleH-logarithmic}Quantum correction at one-loop from the interaction terms of two gauge boson and two scalars included in (\ref{LescalarLittleH1}).}
\end{figure}

Now, we consider the quantum correction from Figure (\ref{LittleH-logarithmic}) we can write the Feynman integral by considering a power counting
\begin{equation}\label{LittleH-logarithmic-1}
i\Sigma_4\varpropto\int \frac{d^4 k}{(2\pi)^4}\times\left(\frac{g^2}{k^2}\right)\left(\frac{g^2}{k^2}\right),
\end{equation}


which results in

\begin{align}\label{LittleH-logarithmic-2}
&=\frac{g^4}{16\pi^2}~\mathrm{ln}\left(\frac{\varLambda^2}{\mu^2}\right),
\end{align}

the following operator that does not have a quadratic divergence 

\begin{equation}\label{LOGARITHMIC-term-LittleH1}
\frac{g^4}{16\pi^2}~\mathrm{ln}\left(\frac{\varLambda^2}{\mu^2}\right)\arrowvert\phi_1^\dag\phi_2\arrowvert^2.
\end{equation}

To identify the coefficient of $H^\dag H$, i.e., its mass squared we will substitute the parameterizations (\ref{scalar1-parametrization-convenient}), 
(\ref{scalar2-parametrization-convenient}) in (\ref{LOGARITHMIC-term-LittleH1}), for this purpose we compute 

\begin{equation}\label{ph1dag-ph2}
\phi_1^\dag\phi_2=\left(
\renewcommand{\arraystretch}{1.6}
\begin{array}{cc}
0 & f \\
\end{array}
\right)\exp\left \{-\frac{2i}{f}\left(
\renewcommand{\arraystretch}{1.6}
\begin{array}{cc}
0 & H\\
H^\dag & 0 \\
\end{array}
\right)\right \}\left(
\renewcommand{\arraystretch}{1.6}
\begin{array}{c}
0 \\
f \\
\end{array}
\right)
\end{equation}

after expanding the exponential matrix up to order of $\left(\dfrac{1}{f^2}\right)$ we have

\begin{align}\label{ph1dag-ph2-next}
\phi_1^\dag\phi_2&=\left(
\renewcommand{\arraystretch}{1.6}
\begin{array}{cc}
0 & f \\
\end{array}
\right)\left[\left(
\renewcommand{\arraystretch}{1.6}
\begin{array}{cc}
1 & 0\\
0 & 1 \\
\end{array}
\right)-\frac{2i}{f}\left(
\renewcommand{\arraystretch}{1.6}
\begin{array}{cc}
0 & H\\
H^\dag & 0 \\
\end{array}
\right)\right.\nonumber\\
&\left.-\frac{2}{f^2}\left(
\renewcommand{\arraystretch}{1.6}
\begin{array}{cc}
HH^\dag & 0\\
0 & H^\dag H \\
\end{array}
\right)+\mathcal{O}\left(\frac{1}{f^3}\right)\right]\left(
\renewcommand{\arraystretch}{1.6}
\begin{array}{c}
0 \\
f \\
\end{array}
\right)\nonumber\\
&=f^2-2H^\dag H+\cdots
\end{align}

we can substitute this result in (\ref{LOGARITHMIC-term-LittleH1}), obtaining a mass term

\begin{equation}\label{quartic-term-LittleH2}
-\frac{g^4f^2}{4\pi^2}~\mathrm{ln}\left(\frac{\varLambda^2}{\mu^2}\right)H^\dag H.
\end{equation}

\noindent Thus we have a theory that does not have quadratic divergence at one-loop for the mass of $H$, and it is the pseudo-Nambu Goldstone Boson 
associated with two scalar fields that break separately each $SU(3)$ to $SU(2)$. The fact that to generate a genuine contribution to the Higgs 
potential we need the contributions of both $\phi_1$ and $\phi_2$ 
is an example of the so-called collective breaking. Notice that the term given in (\ref{LOGARITHMIC-term-LittleH1}) explicitly 
breaks both $SU(3)$ to the diagonal.\newline

There is a similar mechanism for the Yukawa contributions to the potential. The $\phi_1$ and $\phi_2$ Yukawas are

\begin{equation}\label{Lyuk1}
 \mathcal{L} = -\lambda_1\bar{t}_{1R}\phi^\dag_1\Psi_L-\lambda_2\bar{t}_{2R}\phi^\dag_2\Psi_L+\mathrm{h.c.},
\end{equation}

where $\Psi_L$ is a $SU(3)$ triplet given by

\begin{equation}\label{triplet-Little1}
\Psi_L=\left(
\renewcommand{\arraystretch}{1.6}
\begin{array}{c}
t \\
b \\
T \\
\end{array}
\right)_L,
\end{equation}

and taking into account the $SU(2)$ doublet

\begin{equation}\label{doublet-Little1}
Q_L=\left(
\renewcommand{\arraystretch}{1.6}
\begin{array}{c}
t \\
b \\
\end{array}
\right)_L,
\end{equation}

jointly with

\begin{equation}\label{Higgs-doublet-Little1}
H=\left(
\renewcommand{\arraystretch}{1.6}
\begin{array}{c}
H_1 \\
H_2 \\
\end{array}
\right).
\end{equation}

Now, we can substitute (\ref{scalar1-parametrization-convenient}), (\ref{scalar2-parametrization-convenient}) in (\ref{Lyuk1}) and using the rotation where $k_1$ and $k_2$
are removed, i.e., in unitary gauge for $SU(3)$ jointly with (\ref{doublet-Little1}), we have

\begin{align}\label{Lyuk2}
&-\lambda_1\bar{t}_{1R}\left[\exp\left \{\frac{i}{f}\left(
\renewcommand{\arraystretch}{1.6}
\begin{array}{cc}
0 & H\\
H^\dag & 0 \\
\end{array}
\right)\right \}\left(
\renewcommand{\arraystretch}{1.6}
\begin{array}{c}
0 \\
f \\
\end{array}
\right)\right]^\dag\left(
\renewcommand{\arraystretch}{1.6}
\begin{array}{c}
Q \\
T \\
\end{array}
\right)_L
\nonumber\\
&-\lambda_2\bar{t}_{2R}\left[\exp\left \{\frac{-i}{f}\left(
\renewcommand{\arraystretch}{1.6}
\begin{array}{cc}
0 & H\\
H^\dag & 0 \\
\end{array}
\right)\right \}\left(
\renewcommand{\arraystretch}{1.6}
\begin{array}{c}
0 \\
f \\
\end{array}
\right)\right]^\dag\left(
\renewcommand{\arraystretch}{1.6}
\begin{array}{c}
Q \\
T \\
\end{array}
\right)_L,
\end{align}

expanding the exponential matrix functions up to order of $\left(\dfrac{1}{f^2}\right)$

\begin{align}\label{Lyuk2}
&-\lambda_1\bar{t}_{1R}\left(
\renewcommand{\arraystretch}{1.6}
\begin{array}{cc}
0 & f \\
\end{array}
\right)\left[1-\frac{i}{f}\left(
\renewcommand{\arraystretch}{1.6}
\begin{array}{cc}
0 & H\\
H^\dag & 0 \\
\end{array}
\right)-\frac{1}{2f^2}\left(
\renewcommand{\arraystretch}{1.6}
\begin{array}{cc}
HH^\dag & 0\\
0 & H^\dag H \\
\end{array}
\right)\right]\left(
\renewcommand{\arraystretch}{1.6}
\begin{array}{c}
Q \\
T \\
\end{array}
\right)_L
\nonumber\\
&-\lambda_2\bar{t}_{2R}\left(
\renewcommand{\arraystretch}{1.6}
\begin{array}{cc}
0 & f \\
\end{array}
\right)\left[1+\frac{i}{f}\left(
\renewcommand{\arraystretch}{1.6}
\begin{array}{cc}
0 & H\\
H^\dag & 0 \\
\end{array}
\right)-\frac{1}{2f^2}\left(
\renewcommand{\arraystretch}{1.6}
\begin{array}{cc}
HH^\dag & 0\\
0 & H^\dag H \\
\end{array}
\right)\right]\left(
\renewcommand{\arraystretch}{1.6}
\begin{array}{c}
Q \\
T \\
\end{array}
\right)_L,
\end{align}

and considering

\begin{equation}\label{extra-relation-LittleH}
 \lambda_1=\lambda_2=\frac{\lambda}{\sqrt{2}},
\end{equation}

the terms given in (\ref{Lyuk2}) are simplified to 

\begin{equation}\label{Lyuk3}
 -\frac{\lambda}{\sqrt{2}}\left[f\left(\bar{t}_{1R}+\bar{t}_{2R}\right)T_L+i\left(\bar{t}_{2R}-\bar{t}_{1R}\right)H^\dag Q_L-\frac{1}{2f}\left(\bar{t}_{2R}+\bar{t}_{1R}\right)H^\dag HT_L+...+\mathrm{h.c.}\right],
\end{equation}

now, writing the following combinations

\begin{align}
T_R&=\bar{t}_{1R}+\bar{t}_{2R}\label{combination1LH}\\
\bar{t}_R&=\frac{i}{\sqrt{2}}\left(\bar{t}_{2R}-\bar{t}_{1R}\right),\label{combination2LH}
\end{align}

substituting these combinations, we rewrite (\ref{Lyuk3}) as

\begin{equation}\label{Lyuk4}
 -\lambda\left[f\overline{T}_RT_L+\bar{t}_{R}H^\dag Q_L-\frac{1}{2f}\overline{T}_RH^\dag HT_L+...+\mathrm{h.c.}\right],
\end{equation}

where as in Ref.~\citep{Perelstein:2003wd}, here $H$ can be identified as

\begin{equation}\label{H-identified-LittleH}
H= \frac{1}{\sqrt{2}}\left (
\begin{array}{c}
h-i\varphi_3\\
-\sqrt{2}\varphi^-
\end{array} \right ),
\end{equation}

expanding about the symmetric point, $ \langle h\rangle=0$, we identify from (\ref{Lyuk4}) the following terms 

\begin{equation}\label{Lyuk5}
 -\lambda f\overline{T}_RT_L-\frac{\lambda}{\sqrt{2}}\bar{t}_{R}t_Lh+\frac{\lambda}{4f}\overline{T}_RT_Lh^2+...+\mathrm{h.c.}.
\end{equation}

\begin{figure}[H]
	\centering
	\includegraphics[width=3.5in]{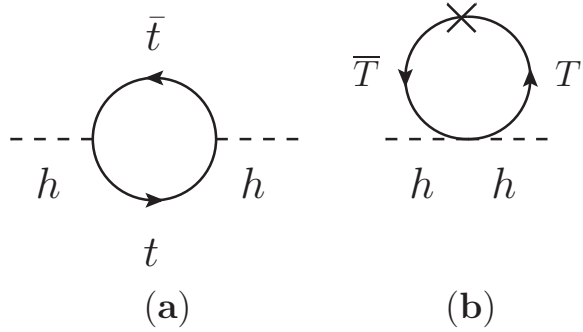}
        \captionsetup{font=small, labelfont=bf, labelsep=period}
        \caption{\label{LittleH-hSelf}Quantum contributions to the Higgs propagator from the top and $T$ at one loop according to the interactions 
        given in (\ref{Lyuk5}).}
\end{figure}

From these, we can compute the corrections to the Higgs mass squared up to 1 loop, these quantum corrections to the Higgs propagator are shown in 
Figure (\ref{LittleH-hSelf}).

The contribution to the Higgs mass squared from Figure (\ref{LittleH-hSelf}a) is given by


\begin{align}\label{q-correction-@1loopfrom-t-next}
\delta m_{h{\bf (a)}}^{2}&=-N_c\lambda^2\frac{1}{4\pi^2}\int^{\varLambda}_0 d p_E \frac{p^3_E}{p^2_E}\nonumber\\
&=-\frac{N_c}{8\pi^2}\lambda^2\varLambda^2,
\end{align}

and from Figure (\ref{LittleH-hSelf}b), we also have



\begin{align}\label{q-correction-@1loopfrom-T-next}
\delta m_{h{\bf (b)}}^{2}&=N_c\lambda^2\frac{1}{4\pi^2}\int^{\varLambda}_0 d p_E \frac{p^3_E}{p^2_E}\nonumber\\
&=\frac{N_c}{8\pi^2}\lambda^2\varLambda^2.
\end{align}

\noindent Thus the quadratic divergent contributions to the Higgs mass (\ref{q-correction-@1loopfrom-t-next}) and (\ref{q-correction-@1loopfrom-T-next}) 
are canceled by the addition of the quantum corrections at one-loop from the top and $T$. And also we saw that the model involves 
three scales $\varLambda$, $f$ 
and $M_{weak}\sim$ such that its squared is given by the mass term in (\ref{quartic-term-LittleH2}), where $\varLambda\lesssim 4\pi f$ as in 
Ref.~\citep{Schmaltz:2005ky}, this could be obtained by 
doing a naive dimensional analysis after expanding the kinetic term of the scalar fields (\ref{scalar1-parametrization-convenient}), 
(\ref{scalar2-parametrization-convenient}). The size of $f$ is of order to the TeV scale if $g$ is equal to the $SU(2)$ gauge coupling. 
Thus, it is important to study theories where we will have fermion masses that are in the TeV scale to solve the hierarchy problem through to the
cancellation of the quadratic divergence as above. This was an example of theories where a Vector-like quark must be present to solve the hierarchy 
problem. In the next sections we will introduce another example where the fermion masses can be in the TeV scale.




\section{Quiver Theories}\label{Quiver-Theories}

This section has attempted to provide a brief summary of the literature relating to quiver theories~\citep{Bai:2009ij,Burdman:2012sb,Burdman:2014ixa}, 
because these are another possibility to address the hierarchy problem and the fermion mass hierarchies, 
in similar way to $\mathrm{AdS_5}$ theories~\cite{Burdman:2007ck,Gherghetta:2000qt,Lillie:2007yh}. To study the more important 
phenomenology associated with the excited SM particles as studied in Ref.~\citep{Burdman:2014ixa}, 
we need to compute the couplings involved by obtaining their wave functions as we 
will see below.  

In the quiver theories approach we consider a four dimensional (4D) gauge theory associated with a product gauge group 

\begin{equation} \label{xgroup}
G=G_0\times G_1\times \ldots \times G_{N-1}\times G_N.
\end{equation}

\noindent Here, we will consider that $G_j=SU(n)_j$ is a gauge symmetry, where $j=0,1,\ldots, N.$ In addition to this framework, 
we include a set of scalar link fields $\Phi_j$, with $j=1,\ldots, N$ , such that $\Phi_j$ transforms under the 
bi-fundamental representation of groups $G_{j-1}\times G_j$, as follows: 

\begin{equation} \label{bifunamental_tranform}
\Phi_j\rightarrow U_{j-1}\Phi_j U^{\dag}_j.
\end{equation}

\begin{figure}[!h]
\center
\epsfig{file=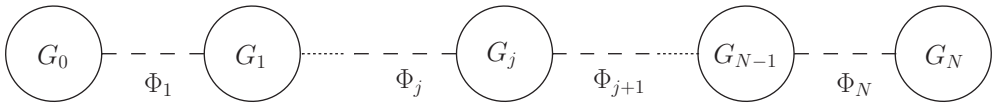,width=14cm}
\caption{\label{moose}Quiver diagram associated with the theory in 
(\ref{action-gauge-fields-QT}).}
\end{figure}

\noindent The action with the considerations mentioned above is given by

\begin{equation} \label{action-gauge-fields-QT}
S = \int d^4 x \left\{\sum_{j=0}^N  - \frac{1}{2} \mathrm{Tr}\left[
F^{j}_{\mu\nu}~F^{j \mu\nu}\right] +
\sum_{j=1}^N\mathrm{Tr}\left[(D_\mu\Phi_j)^\dag
(D^\mu\Phi_j)\right]
 - V(\Phi_j)\right\},
\end{equation}

\noindent where $F^{j}_{\mu\nu}=F^{j\hspace{1mm}a}_{\mu\nu}T^{a}_j$ is the gauge field strength tensor. Here $T^{a}_j$ are the 
generators of the symmetry group $SU(n)_j$ and $F^{j\hspace{1mm}a}_{\mu\nu}$ is more explicitly given by

\begin{equation}\label{gauge_field_strength_tensor}
F^{j\hspace{1mm}a}_{\mu\nu}= \partial_\mu A^{j\hspace{1mm}a}_{\nu} -\partial_\nu A^{j\hspace{1mm}a}_{\mu}
+ g_j~f^{j\hspace{1mm}abc}~A^{j\hspace{1mm}b}_{\mu}~A^{j\hspace{1mm}c}_{\nu},
\end{equation}

\noindent where $f^{j\hspace{1mm}abc}$ are related through the commutation relations $[T_j^a,T_j^b] = if^{j\hspace{1mm}abc}T_j^c.$ 
The covariant derivative $D_\mu\Phi_j$ is given by

\begin{equation}\label{escalar_covariant_derivative}
D_\mu\Phi_j = \partial_\mu~ \Phi_j + ig_{j-1}~A_{\mu, j-1}^{a}~
T_{j-1}^a ~\Phi_j - ig_{j}~\Phi_j~A_{\mu, j}^{a}~ T_{j}^a.
\end{equation}
In addition, we assume that the $\Phi_j$'s 
develop a diagonal VEV, such that $SU(n)_{j-1}\times SU(n)_j$ is broken down to the diagonal group. This means that for 
each VEV we have to have $n^2-1$ Nambu-Goldstone Bosons (NGBs) and we can use the parameterization of the non-linear
sigma models for the $\Phi_j$'s given by

\begin{equation}\label{escalar_parametrization}
\Phi_j = v_j e^{i\pi^{a}_{j}\hat{T}_j^a/ v_j},
\end{equation}

\noindent where the broken generators are $\hat{T}_j^a$'s, the NGBs are $\pi^{a}_{j}$ and the $\Phi_j$'s VEVs $v_j$ are 
related with the breaking of $SU(n)_{j-1}\times SU(n)_j$. The model can be schematically represented by Figure (\ref{moose}). 
We choose to parametrize the $v_j$'s as
\begin{equation}\label{VEVs/Sqrt[2]}
v_j=v q^j,
\end{equation}

\noindent with $0<q<1$, such that $v$ is the UV mass scale and then we have that the $v_{j}$'s decrease as follow $v_1...>v_j...> v_N$. 
These choices have been studied in ~\citep{Bai:2009ij,Burdman:2012sb,Burdman:2013qfa,Burdman:2014ixa}. In this example 
we can consider that each gauge coupling satisfies 

\begin{equation}\label{universal_couoplings}
 g_0=g_1=...g_N=g,
\end{equation}

\noindent and as we have indicated that all gauge groups are identical $T_j^a = T_{j-1}^a=T^a$, with $j=1,\ldots, N.$
The action (\ref{action-gauge-fields-QT}) can be represented by the bosonic quiver diagram of Figure (\ref{moose}), where the 
circles represent the gauge group's $G_j$, $j=0,1,\ldots, N$. Here $j$ is identified with the index of the 
site in the quiver 
diagram, henceforth we will call $j$ as site index. To expand the kinetic term of the $\Phi_j$'s in 
(\ref{action-gauge-fields-QT}), we need to use (\ref{escalar_parametrization}). we obtain

\begin{align}
\mathrm{Tr}\left[(D_\mu \Phi_j)^\dag(D^\mu \Phi_j)\right] &= \mathrm{Tr}\left[(\partial_\mu \Phi_j)^\dag(\partial^\mu \Phi_j)\right] + g\mathrm{Tr}\left[i(\partial_\mu \Phi_j)^\dag(A_{j-1}^{\mu} \Phi_j -\Phi_j A_{j}^{\mu})+\mathrm{h.c.}\right]\nonumber \\
&+v_j^2 g^{2} \mathrm{Tr}\left[A_{\mu,j-1}A^{\mu}_{j-1}\right]-2g^{2} \mathrm{Tr}\left[\Phi_j ^\dag A_{\mu,j-1}\Phi_j A^{\mu}_{j}\right] \nonumber \\
&+v_j^2 g^{2} \mathrm{Tr}\left[A_{\mu,j}A^{\mu}_{j}\right]. \label{scalar-kinetic-term1}
\end{align}

After that, we replace (\ref{escalar_parametrization}) in (\ref{scalar-kinetic-term1}), using the normalization 
$\mathrm{Tr}\left[T^{a}T^{b} \right]=\frac{1}{2}\delta^{ab} $, such that we consider only quadratic terms in the 
fields $\pi^{a}_{j}$ and $A_{\mu,j}^{a}$. Thus we have  

\begin{align}
\sum_{j=1}^N \mathrm{Tr}\left[(D_\mu \Phi_j)^\dag(D^\mu \Phi_j)\right] &=\sum_{j=1}^N\left[\frac{1}{2}(\partial_\mu \pi^{a}_{j})^2-v_j g~\partial_\mu \pi^{a}_{j}(A_{j}^{\mu ~a} - A_{j-1}^{\mu ~a}) +\frac{1}{2}v_j^2 g^{2} (A_{\mu,j-1}^{a})^2\right.\nonumber \\
&\left.-v_j^2 g^{2} A_{\mu,j-1}^{a}A_{j}^{\mu~a}+\frac{1}{2}v_j^2 g^{2} (A_{\mu,j-1}^{a})^2\right]\nonumber \\
&=\sum_{j=1}^N\frac{1}{2} [\partial_\mu \pi^{a}_{j} - v_j g (A_{j}^{\mu ~a} - A_{j-1}^{\mu ~a})]^2 \label{scalar-kinetic-trace}.
\end{align}

As we can see, (\ref{scalar-kinetic-trace}) includes the cross term mixing the NGBs with the gauge bosons in 
(\ref{action-gauge-fields-QT}). To cancel these quadratic terms of the form $\partial_\mu \pi^{a}_{j}(A_{j}^{\mu ~a} 
- A_{j-1}^{\mu ~a})$, we chose to introduce the gauge-fixing term 

\begin{equation}\label{L_GF1}
\mathcal{L}_{GF} =-\frac{1}{2\xi}\sum_{j=0}^N [\partial_\mu A_{j}^{\mu ~a}+\xi g( v_j\pi^{a}_{j} - v_{j+1}\pi^{a}_{j+1})]^2,
\end{equation}

\noindent where $\xi$ is the gauge parameter, and we considered that $\xi$ is the same site for all sites. Then (\ref{L_GF1}) 
can be written as

\begin{align}\label{L_GF2}
\mathcal{L}_{GF} &=-\frac{1}{2\xi}\sum_{j=0}^N(\partial_\mu A_{j}^{\mu ~a})^2+\sum_{j=0}^N g A_{j}^{\mu ~a}( v_j\partial_\mu \pi^{a}_{j}- v_{j+1}\partial_\mu\pi^{a}_{j+1})\nonumber\\
&-\frac{1}{2}\sum_{j=0}^N \xi[g( v_j\pi^{a}_{j} - v_{j+1}\pi^{a}_{j+1})]^2,
\end{align}

\noindent where the second term was integrated by parts. This can be rewritten as follows 

\begin{align}\label{L_GF3}
\mathcal{L}_{GF} &=-\frac{1}{2\xi}\sum_{j=0}^N(\partial_\mu A_{j}^{\mu ~a})^2+\sum_{j=1}^N g v_j\partial_\mu \pi^{a}_{j}(A_{j}^{\mu ~a} - A_{j-1}^{\mu ~a})\nonumber\\
&-\frac{1}{2}\sum_{j=0}^N \xi[g( v_j\pi^{a}_{j} - v_{j+1}\pi^{a}_{j+1})]^2.
\end{align}

\noindent Thus the action (\ref{action-gauge-fields-QT}), using (\ref{scalar-kinetic-trace}), with the inclusion of the gauge-fixing term in the form of (\ref{L_GF3}) is given by
\begin{align} \label{action-gauge-fields-QT-GF}
S &= \int d^4 x \Bigg\{-\sum_{j=0}^N \left[  \frac{1}{4} (F^{j~a}_{\mu\nu})^2 +\frac{1}{2\xi}(\partial_\mu A_{j}^{\mu ~a})^2\right]+\frac{1}{2}\sum_{j=1}^N(\partial_\mu \pi^{a}_{j})^2\nonumber \\
&-\frac{1}{2}\sum_{j=1}^{N-1} \xi[g( v_j\pi^{a}_{j} - v_{j+1}\pi^{a}_{j+1})]^2-\frac{1}{2}\xi g^2v_{1}^2(\pi^{a}_{1})^2-\frac{1}{2}\xi g^2v_{N}^2(\pi^{a}_{N})^2\nonumber \\
&+\frac{1}{2}\sum_{j=1}^N v_j^2 g^2 (A_{j}^{\mu ~a} - A_{j-1}^{\mu ~a})^2\Bigg\}.
\end{align}

\noindent Notice that this action includes the mass terms of NGBs and gauge bosons. The mass term for the NGBs in 
the Lagrangian associated with the action (\ref{action-gauge-fields-QT-GF}) is given by 

\begin{equation}\label{L-NGB's-mass}
\mathcal{L}_{M_{\pi}} = -\frac{1}{2}\sum_{j=1}^{N-1} \xi[g( v_j\pi^{a}_{j} - v_{j+1}\pi^{a}_{j+1})]^2-\frac{1}{2}\xi g^2v_{1}^2(\pi^{a}_{1})^2-\frac{1}{2}\xi g^2v_{N}^2(\pi^{a}_{N})^2\equiv -\frac{1}{2}\pi^{aT}M_{\pi}^{2}\pi^{a},
\end{equation}

\noindent where $\pi^{aT}$ was written in the basis $\pi^{aT}=(\pi^{a}_{1}, \pi^{a}_{2}, \ldots , \pi^{a}_{N}).$ To write the 
mass matrix $M_{\pi}^{2}$ for the NGBs, we used the parametrization (\ref{VEVs/Sqrt[2]}), obtaining

\singlespacing \begin{equation} \label{M_pi^2}
M_{\pi}^2 = g^2 v^2\xi \left( \begin{array}{cccccc}
~~2q^2    & -q^3      & ~~0         &\ldots   &~~0                 & ~~0 \\
-q^3      & 2q^4      & -q^5        &\ldots   &~~0                 & ~~0 \\
~~0       & -q^5      & 2q^6        &\ldots   &~~0                 & ~~0 \\
\vdots    & \vdots    & \vdots      &\ddots   &\vdots              & \vdots \\
~~0       & ~~0       & ~~0         &\ldots   &2q^{2(N-1)}         & -q^{2N}   \\
~~0       & ~~0       & ~~0         &\ldots   &-q^{2N}             &2q^{2N}
\end{array} \right).
\end{equation}

\noindent The determinant of $M_{\pi}^2$ is given by

\begin{equation}
 \mathrm{Det}[M_{\pi}^2]=(g^2 v^2\xi)^N(N+1)q^{N(N+1)}.
\end{equation}

\begin{figure}[!htb]
   \center
   \epsfig{file=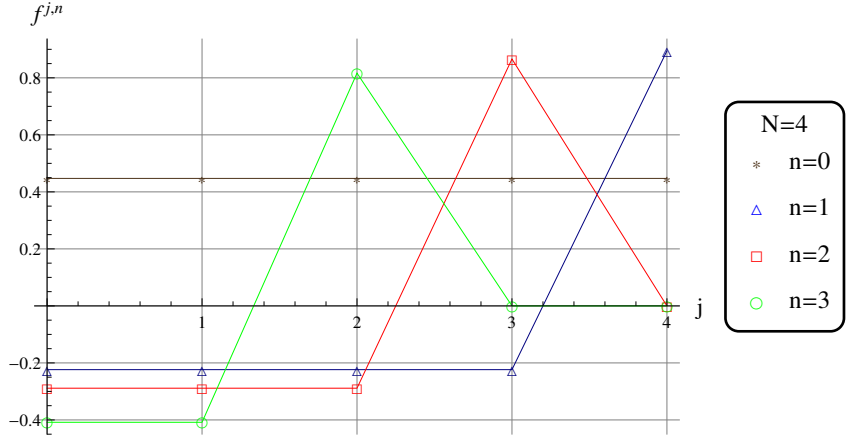}
   \captionsetup{font=small, labelfont=bf, labelsep=period}
   \caption{\label{wfGB_N4}Gauge boson wave functions, 
   where j is a index of site and n is a index of Kaluza-Klein mode. 
   In this case $N=4$ and some allowed $n$ are shown. For the visualization, 
   we choose by the opposite signs of $f^{j,0}$ and $f^{j,2}$.} 
  \end{figure}

\begin{figure}[H]
   \center
   \epsfig{file=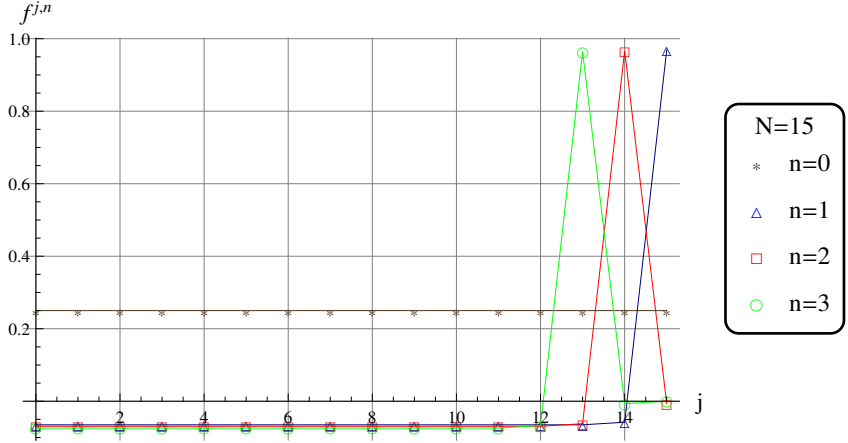}
   \captionsetup{font=small, labelfont=bf, labelsep=period}
   \caption{\label{wfGB_N15}Gauge boson wave functions, 
   where j is a index of site and n is a index of Kaluza-Klein mode. 
   In this case $N=15$ and some allowed $n$ are shown. For the visualization, 
   we choose by the opposite signs of $f^{j,2}$ and $f^{j,3}$.} 
  \end{figure}

\noindent We can see that it is different from zero. In other
words, the NGBs have no zero mode in their mass eigenstate basis, 
such that these NGB masses are proportional to $\sqrt{\xi}$, so by taking 
the limit $\xi \rightarrow \infty$, it corresponds to the unitary gauge. In this limit the NGBs 
disappear from the theory. We will use this gauge, and we say that the NGBs are eaten by the gauge 
bosons which become massive. We will see later that it is possible to extend one of the NGBs to be the Higgs by choosing differently the 
boundary conditions. 
\newline\newline
To determine the spectrum of the massive gauge bosons, we consider the mass term for the gauge bosons in 
the Lagrangian associated with the action (\ref{action-gauge-fields-QT-GF}) that is given by 
\begin{equation}\label{L-gauge-bosons-mass}
\mathcal{L}_{M_{A}} = \frac{1}{2}\sum_{j=1}^N v_j^2 g^2 (A_{j}^{\mu ~a} - A_{j-1}^{\mu ~a})^2\equiv -\frac{1}{2}(A^{a}_\mu)^T M_{A}^{2}A^{\mu~a}.
\end{equation}
Analogously to the case of the NGBs, $(A^{a}_\mu)^T$ was written in the basis 
$(A^{a}_\mu)^T=(A_{\mu,0}, A_{\mu,1}, \ldots, A_{\mu,N})$, and using the parametrization 
(\ref{VEVs/Sqrt[2]}), $M_{A}^{2}$ can be written as follows
\newline\newline\begin{equation} \label{M_A^2}
M_{A}^2 = g^2 v^2 \left( \begin{array}{ccccccc}
~~q^2     & -q^2      & ~~0     &~~0    &\ldots   &~~0                 & ~~0 \\
-q^2      & q^2+q^4   & -q^4    &~~0    &\ldots   &~~0                 & ~~0 \\
~~0       & -q^4      & q^4+q^6 & -q^6  &\ldots   &~~0                 & ~~0 \\
\vdots    & \vdots    & \vdots  &\vdots &\ddots   &\vdots              & \vdots \\
~~0       & ~~0       & ~~0     &~~0    &\ldots   &q^{2(N-1)}+ q^{2N}  & -q^{2N}   \\
~~0       & ~~0       & ~~0     &~~0    &\ldots   &-q^{2N}     &q^{2N}
\end{array} \right).
\end{equation}
\newline\newline \noindent The spectrum of masses can be obtained by diagonalizing this matrix. For this purpose, we will define the orthonormal 
rotation 
\begin{equation}\label{rotation-gauge}
A_{\mu,j}^{a} = \sum_{n=0}^N f^{j,n} A_{\mu}^{a~(n)},
\end{equation}
\noindent where the $A_{\mu}^{a~(n)}$ are the mass eigenstates, such that in this basis $\mathcal{L}_{M_{A}} $ can be written 
as 
\begin{equation}\label{L-gauge-bosons-mass_eigenstate}
 \mathcal{L}_{M_{A}}= \frac{1}{2}\sum_{n=0}^N m_{n}^{2}(A_{ \mu}^{a~(n)})^2.
\end{equation}
Here we perform a numerical calculation of 
$f^{j,n}$ for hypothetical gauge bosons, following our previous formulation. Considering 
$v\lesssim M_P=10^{19}$ GeV and $v_N\cong\mathcal{O}$(1) TeV, We show their wave-functions
 in Figures (\ref{wfGB_N4}) and (\ref{wfGB_N15}) for $N=4$ and $N=15$, respectively. In both cases, the zero mode of 
gauge bosons will be flat as in $\mathrm{AdS_5}$ theories with fields in the bulk~\cite{Gherghetta:2000qt,victor-mestrado}.  
  

In order to understand better the behavior of the gauge boson wave functions, we now look at the equations 
they satisfy. \newline
\newline \noindent The coefficients $f^{j, n}$ can be obtained from the equations of motion 
for the fields $A_{j}^{\mu}$ by using the Lagrangian given for
\begin{equation}\label{eq91}
\mathcal{L}_A =  \sum_{j=0}^N \left\{ - \frac{1}{4} F_{\mu\nu,j
}~F^{\mu\nu}_{j} + \frac{g^2}{2}[v_j(A_{\mu,j-1}-
A_{\mu,j})]^2  \right\},
 \end{equation}
 
\noindent where for simplicity, the abelian case was supposed. and then we can use the Euler-Lagrange equations
\begin{equation}\label{eq92}
 \frac{\partial \mathcal{L}}{\partial A_{\nu,j}} -\partial_\mu
\left( \frac{\partial \mathcal{L} }{\partial(\partial_\mu
A_{\nu,j})}\right) = 0,
\end{equation}

we obtain the following equation
\begin{equation}\label{eq93}
(\partial^2 A_{j}^\nu -\partial^\nu \partial_\mu A_j^\mu)~~ +
~~g^2v_j^2 ~(A_{j}^\nu -A_{j-1}^\nu) ~~+~~
g^2v_{j+1}^2~(A_{j}^\nu -A_{ j+1}^\nu) =0.
\end{equation}
Now, we will use Lorentz gauge $\partial_\mu A_{j}^\mu =0$ and substituting (\ref{rotation-gauge}) in (\ref{eq93}) 
we obtain
\begin{equation}\label{eq94}
f^{j,n}\partial^2 A^{\nu,(n)}~+ ~g^2v^2q^{2j}[(1+q^2)f^{j,n} ~-~f^{j-1,n}~-~
q^2 f^{j+1,n}]A^{\nu,(n)}~=~0.
\end{equation}

Imposing that $A^{\nu,(n)}$ satisfies the Proca equation, that is
\begin{equation}\label{eq95}
\partial ^2 A_{\nu,n} = -m_n^2 A_{\nu,n},
\end{equation}
thus, we substitute (\ref{eq95}) and using the definition $x_n^2\equiv m_n^2/g^2v^2$ in (\ref{eq94}), 
it happens that we obtain
\begin{equation}\label{eq96}
[q + q^{-1}- q^{-1}(x_nq^{-j})^2]f^{j,n} - q f^{j+1,n} - q^{-1}
f^{j-1,n}=0,
\end{equation}
jointly with the discrete Neumann boundary conditions
\begin{equation}\label{eq97}
f^{0,n}=f^{-1,n},~~~f^{N,n}=f^{N+1,n},
\end{equation}
with the normalization condition
\begin{equation}\label{eq98}
\sum_{j=0}^N (f^{j,n})^2 = 1.
\end{equation}
We will now concentrate in the zero mode, $n=0$ and $m_{0}=0$.\newline

\noindent The equation (\ref{eq96}) for the zero mode is
\begin{equation}\label{eq99}
[q + q^{-1}]f^{j,0} - q f^{j+1,0} - q^{-1}
f^{j-1,0}=0,
\end{equation} 
such that for $j=0$ in (\ref{eq96}), we have 
\begin{equation}\nonumber
[q + q^{-1}]f^{0,0} - q f^{1,0} - q^{-1}
f^{-1,0}=0,
\end{equation} 

\noindent and using (\ref{eq97}) we obtain
\begin{equation}\label{eq100}
f^{0,0}=f^{1,0},
\end{equation}
iterating to the others $j's$ we have
\begin{equation}\label{eq101}
f^{0,0}=f^{1,0}=f^{2,0}=...=f^{N,0},
\end{equation}

and by using the normalization condition (\ref{eq98}), we obtain 

\begin{equation}\label{WF-gauge0}
f_{j,0}= \frac{1}{\sqrt{N+1}}.
\end{equation}

\noindent This means that the components of zero mode are equal in all sites. 
It is analogous to what happens in $\mathrm{AdS_5}$ 
theory, where the zero modes of the gauge bosons are delocalized, as can be seen in Figures 
(\ref{wfGB_N4}) and (\ref{wfGB_N15}) for $n=0$. 
For massive modes the equation (\ref{eq96}) has solution as shown in Ref.~\citep{deBlas:2006fz}. For this purpose 
we define the following variables 
\begin{equation}\label{eq103}
t[j]=x_nq^{-j},
\end{equation}
\begin{equation}\label{eq104}
F(t[j])=q^jf^{j,n},
\end{equation}

\noindent and we substitute (\ref{eq103}) and (\ref{eq104}) in the equation of q-differences (\ref{eq96}) we obtain
\begin{equation} \label{eq105}
(q +  q^{-1} - q^{-1}t^2)F(t) - F(tq^{-1}) - F(tq)=0.
\end{equation}

\noindent The equation (\ref{eq105}) is a special case of Hahn-Exton equation~\citep{deBlas:2006fz,qdifer} with solutions 
that are called q-Bessel functions. More general solutions can be written as
\begin{equation}\label{eq106}
F(t)= A J_1(t;q^2)+BY_1(t;q^2), 
\end{equation}

\noindent where in general $J_\nu(t;q^2)$ and $Y_1(t;q^2)$ are the q-Bessel and q-Neumann functions respectively. 
It is interesting to examine the continuum limit $q\rightarrow1^{-}$ and $N\rightarrow\infty$, 
as we will in the next section, we obtain the continuous ordinary functions of Bessel and 
Neumann~\citep{deBlas:2006fz}, 
 
\begin{equation}\label{eq107}
J_\nu(t;q)= t^\nu \frac{(q^{\nu +1};q)_\infty}{(q;q)_\infty}
\sum_{i=0}^\infty \frac{(-1)^i q^{i(i+1)/2}} {(q^{\nu +1};
q)_i(q;q)_i}t^{2i},
\end{equation}
with the factors $(y;q)_i$ are defined as
\begin{equation}\label{definitionqdifer}
(y;q)_k = \left\{ \begin{array}{ll} 1 & \mbox{se $k=0$} \\
 \prod_{n=0}^{k-1}(1 - y q^n)  & \mbox{se $k\geq 1$} \end{array}
\right. \,
\end{equation}

\noindent for $y \in \mathbb{C},~ $ $i \in \mathbb{Z}_+ =
\{0,1,2,\ldots\}~ $ and $(y;q)_\infty \equiv\lim_{i\rightarrow\infty}
(y;q)_i. ~$ 
Meanwhile
\begin{equation}\label{qdiferneuman}
Y_\nu(t;q)= \frac{\Gamma_q(\nu) \Gamma_q(1-\nu)}{\pi}
q^{-\nu^2/2}[\cos(\pi \nu) q^{\nu/2}
J_\nu(t;q)-J_{-\nu}(tq^{-\nu/2};q)],
\end{equation}
where the function $\Gamma_q(\nu)$ is defined for
\begin{equation}\label{definitionof_gamma}
\Gamma_q(\nu)= \frac{(q;q)_\infty}{(q^\nu;q)_\infty}(1 - q)^{1-\nu}.
\end{equation}

Using (\ref{eq104}), (\ref{eq105}) and the boundary conditions (\ref{eq97}), it is possible to find the coefficients 
$f^{j,n}$ to within a constant $N_n$ that can be obtained from the normalization condition (\ref{eq98}) as follow
\begin{equation}\label{eq108}
f^{j,n}= N_n q^{-j} [Y_0(x_n;q^2) J_1(x_n q^{-j};q^2)- J_0(x_n;q^2)Y_1(x_n q^{-j}; q^2)].
\end{equation}
Additionally, the spectrum of masses is obtained from the equation
\begin{equation}\label{eq109}
J_0(x_n;q^2)Y_0(q^{-(N+1)}x_n;q^2)~-~Y_0(x_n;q^2)J_0(q^{-(N+1)};
q^2)=0.
\end{equation}
In the continuum limit, when $q\rightarrow1^{-}$, these coefficients (\ref{eq108}) coincide 
to the wave functions 
of the excited gauge bosons in $AdS_{5}$ theory. Thus with the deconstruction of a 5-dimensional gauge theory 
it is possible to produce a correspondent quiver theory in four dimensions.


\subsection{Relation to $AdS_{5}$}\label{Relation to AdS_{5}}

We know that the $\mathrm{AdS_5}$ theories solve the gauge hierarchy problem, as well as the hierarchy of fermion masses. 
However, these theories are non renormalizable, so it is interesting to obtain a higher universe 
of theories that solve large hierarchies. We will start considering a continuous 5-dimensional gauge 
action in the Abelian case. The extension to the non-Abelian case is straightforward. Working with $\mathrm{AdS_5}$ theories, 
where the extra dimension is compactified on the orbifold $S_1/Z_2$ with 
$-L\leq y\leq L$ and the metrics is given by
\begin{equation}\label{metric}
 ds^2= e^{-2 k|y|} \eta_{\mu \nu} dx^\mu dx^\nu - dy^2,
\end{equation}
where $k$ is the $\mathrm{AdS_5}$ curvature. The action for gauge bosons is given by

\begin{equation}\label{8}
S_A =\int d^5 x\sqrt{g}\, \left[-\frac{1}{4 g_5^2} F_{MN}F^{MN}\right],
\end{equation}

\noindent where $g_5$ is the gauge coupling in 5 dimensions, and $F_{MN} = \partial_M A_N-\partial_N 
A_M$ with $M= 0, 1, 2, 3$ and $5$ ($y$).\newline
\newline\noindent The action (\ref{8}) can be simplified to

\begin{align}
S_5^A &= \int d^5 x\sqrt{g}\, \left[-\frac{1}{4 g_5^2}g^{MO}g^{NP}F_{MN}F_{OP}\right],\nonumber \\
 &= \int d^4 x \int^L_0 dy \sqrt{g} \left [-\frac{1}{4g_5^2} e^{4ky} F_{\mu \nu}F^{\mu \nu} +\frac{1}{2g_5^2}e^{2ky} \left(F_{5 \mu}\right)^2 \right],\nonumber\\
 &= \int d^4 x \int^L_0 dy \left [-\frac{1}{4g_5^2} F_{\mu \nu}F^{\mu \nu} +\frac{1}{2g_5^2}e^{-2ky} \left(\partial_5 A_\mu - \partial_\mu A_5\right)^2 \right].\label{9}
\end{align}

We will discretize the compact dimension, 
with spacing $\ell$. So the action (\ref{9}) will now be
\begin{align}\label{eq71}
S_5^A &= \ell\int d^4 x \left[ \sum_{j=0}^N -\frac{1}{4 g_{5}^2}(F^{j}_{\mu\nu} F^{j\mu \nu})\nonumber \right.\\
 &+\left.\sum_{j=1}^N\frac{1}{2 g_{5}^2}  e^{-2k \ell j} \left(\frac{A^{j}_{\mu}-A^{j-1}_{\mu}}{\ell}-\partial_\mu A_5^j\right)^2~\right],
\end{align}
\noindent where the derivatives with respect to $y$ taken to be discretized. This action can be 
compared to the action from the quiver theory, To make this clear we will rescale the gauge fields 
as

\begin{align}\label{eq111}
S_4^A &= \int d^4 x \left[ - \sum_{j=0}^N\frac{1}{4 g^2} (F_{\mu\nu,j}~F^{\mu\nu}_{j})\nonumber \right.\\ 
&+\left. \frac{1}{2}\sum_{j=1}^N[\partial_\mu\pi_j +v_j (A_{\mu,j-1}-A_{\mu,j})]^2  \right].
\end{align}

We can see the equivalence of both theories setting the dictionary between discretized five-dimensional gauge 
theory with the purely four-dimensional gauge theory (\ref{eq111}) 
as shown in Ref.~\citep{Bai:2009ij}. The dictionary identifying both theories is shown in Table (\ref{t:tabela1}). 
In this way, we identify the sites zero 
and N as branes UV and IR respectively.

\begin{table}[h]
        \centering\doublespacing
        \begin{tabular}{ccc}
          Theory with 4 dimensions & ~ & Theory with 5 dimensions\\[1pt]\hline
          $\frac{1}{g^2}  $&  $\leftrightarrow$ &  $\frac{\ell}{g_{5}^2}$ \\
          $v $&  $\leftrightarrow$ &  $\frac{1}{\sqrt{\ell}g_{5}}$ \\
          $q $&  $\leftrightarrow$ &  $e^{-k\ell}$ \\
          [8pt]\hline          
      \end{tabular}
        \captionsetup{font=small, labelfont=bf, labelsep=period}
        \caption[]{\label{t:tabela1} Dictionary between a quiver theory and a 
        gauge theory with 5 dimensional 
        discretized with a curve extra dimension.}
      \end{table}
We know that $\mathrm{AdS_5}$ theories solve the hierarchy problem of particle physics for 
$kL\approx37$~\citep{Gherghetta:2000qt,Randall:1999ee,Goldberger:1999uk,Goldberger:1999un}. 
The continuous theory ($\mathrm{AdS_5}$) is obtained from a quiver theory when $N\rightarrow \infty$, such that 
$N\ell = L$, where $L$ is the size of the extra dimension and $\ell$ is the network spacing.
So we have
\begin{equation} \label{gg}
k N \ell \sim 37.
\end{equation}
Now, we will see that happens by using (\ref{VEVs/Sqrt[2]}), the matching in Table (\ref{t:tabela1}), jointly with 
$v\lesssim M_P=10^{19}$ GeV and $v_N\cong\mathcal{O}$(1) TeV, 
\begin{align}\label{calcula-to-simplified}
e^{-kN\ell}&=\frac{v_N}{v}\simeq 10^{-16},  
\end{align}
and then
\begin{equation}\label{q-used}
q\simeq10^{-16/N}. 
\end{equation}

In the continuum limit we have that (\ref{calcula-to-simplified}) corresponds to the expression in $\mathrm{AdS_5}$ 
theories with metric given by (\ref{metric}). So we see that the deconstruction of $\mathrm{AdS_5}$ can be seen as a 
way to obtain four-dimensional theories that solve the hierarchy problem. This will be the case as long as 
(\ref{q-used}) is satisfied. However, for large values of $N$ the four-dimensional theory is very similar to 
$\mathrm{AdS_5}$. In order to obtain a very different theory from deconstruction, $N$ must be small.\newline

From (\ref{q-used}) we can infer the following relation 

\begin{equation}\label{q-used1}
k\ell\sim \frac{37}{N}, 
\end{equation}

identifying two cases, the first one is if $N>37$, this meas that $k\ell<1$ or $k<\Lambda_{UV}$, we still have an $\mathrm{AdS_5}$ theory.
But if $N<37$, then $k\ell>1$ or $k>\Lambda_{UV}$, then $N<37$, we will have a pure four-dimensional theory different from $\mathrm{AdS_5}$, since no 
continuum limit possible since curvature is larger than $M_{Planck}$.
 
\newpage\newpage

\subsection{Higgs in Quiver Theories}\label{higgs-pNGB}

In the SM, the Higgs boson is required in order to trigger EWSB. However we do not know how the Higgs sector was obtained 
in low energies. One possibility to consider is the Higgs as a pNGB as shown in 
Refs.~\citep{Contino:2010rs,Contino:2003ve,Agashe:2004rs,Bellazzini:2014yua}. The 
discovery of the Higgs boson of the SM~\citep{Aad:2012tfa,Chatrchyan:2012ufa}, suggests that we
need to focus not only on gauge bosons and fermions, as we will see in the next subsection, but on the Higgs sector 
of quiver theories as recently considered in 
Refs.~\citep{Burdman:2012sb,Burdman:2013qfa,Burdman:2014ixa,Lima-Burdman,Gabriela-Burdman}. In this section we include 
the Higgs as a pNGB. This is achieved by switching on the gauge fields associated to $G_j$, with $j = 1, . . . , N-1 $, 
and reducing the gauge groups $G_0$ to $H_0$ and $G_N$ to $H_N$ so that 
some of the NGBs are not eaten by gauge bosons. Here $H_0$ and $H_N$ are subgroups of $G_0$ and $G_N$ respectively. 
Consequently, to consider 
the Higgs as a pNGB, the gauge structure given in Figure (\ref{moose}) needs to be modified.
\begin{figure}[H]
\center
\epsfig{file=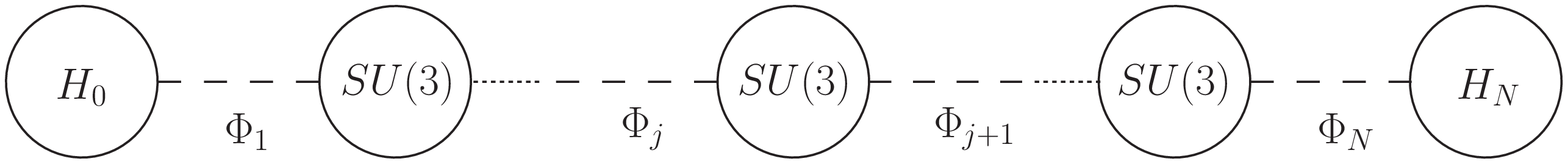,width=14cm}
\caption{\label{moosepNGB}Quiver diagram associated with a quiver theory considering the Higgs as a pNGB, where $G_j=SU(3)$, $0<j<N$ and 
$H_0=H_N=SU(2) \times U(1)$ will be gauged.}
\end{figure}
\noindent Here, we would have four degrees of freedom that will be identified with the degrees of freedom of the Higgs doublet, 
after the breaking of the quiver symmetry. In particular, we will switch on the gauge fields associated 
to $SU(3)_j$ for $1<j<N-1$ and $H_0=H_N=SU(2) \times U(1)$, the gauge structure in this case is given by the quiver diagram 
in Figure (\ref{moosepNGB}). In this way, in the $0$ 
and $N$ sites, the symmetry $SU(2)\times U(1)$ not include the generators associated to $SU(3)/SU(2)\times U(1)$, 
where the matrices in the fundamental representation of $SU(3)$ are given by the eight Gell-Mann matrices
\begin{equation}\label{generators-SU(3)}
 T^a=\frac{\lambda^a}{2}.
\end{equation}\newline\newline
These matrices are given by\newline
\begin{align} \label{Gell-Mall}
&\lambda^1=\left( \begin{array}{ccc}0 & ~~1 & ~~0\\1 & ~~0 & ~~0\\0 & ~~0 & ~~0 \end{array} \right),~\lambda^2=\left( \begin{array}{ccc}0 & -i & ~~0\\i & ~~0 & ~~0\\0 & ~~0 & ~~0 \end{array} \right),~\lambda^3=\left( \begin{array}{ccc}0 & -1 & ~~0\\0 & ~~1 & ~~0\\0 & ~~0 & ~~0 \end{array} \right),\nonumber\\
&\lambda^4=\left( \begin{array}{ccc}0 & ~~0 & ~~1\\0 & ~~0 & ~~0\\1 & ~~0 & ~~0 \end{array} \right),~\lambda^5=\left( \begin{array}{ccc}0 & ~~0 & -i\\0 & ~~0 & ~~0\\i & ~~0 & ~~0 \end{array} \right),~\lambda^6=\left( \begin{array}{ccc}0 & ~~0 & ~~0\\0 & ~~0 & ~~1\\0 & ~~1 & ~~0 \end{array} \right),\nonumber\\
&\lambda^7=\left( \begin{array}{ccc}0 & ~~0 & ~~0\\0 & ~~0 & -i\\0  & ~~i & ~~0 \end{array} \right),~\lambda^8=\frac{1}{\sqrt{3}}\left( \begin{array}{ccc}1 & ~~0 & ~~0\\0 & ~~1 & ~~0\\0 & ~~0 & -2 \end{array} \right).
\end{align}\newline\newline
We define the matrices 
\begin{equation}\label{genratorsEW}
Y^a\equiv \{T^1,T^2,T^3,T^8\}
\end{equation}
and 
\begin{equation}\label{genratorsnotEW}
X^\alpha\equiv \{T^4,T^5,T^6,T^7\},
\end{equation}
are the generators associated to $SU(2)\times U(1)$ and $SU(3)/SU(2)\times U(1)$ respectively. 
We use the convention that Latin and Greek indices take values of $1$, $2$, $3$, $8$ and $4$, $5$, $6$, $7$, respectively. 
Notice that the generators $X^\alpha$ will be associated with the 
degrees of freedom of the Higgs doublet. Now, we will expand the kinetic term of the $\Phi_j$'s, %
such that we will consider only quadratic terms in the NGBs and gauge fields associated to the generators $X^\alpha$
\begin{align}\label{kinetic-termT4567}
     & \frac{1}{2} [\partial_\mu\pi_1^\alpha- v_1 gA^\alpha_{\mu 1} ]^2
     + \sum\limits_{j=2}^{N-1} \frac{1}{2} [\partial_\mu\pi_j^\alpha- v_j g (A^\alpha_{\mu j} - A^\alpha_{\mu, j-1}) ]^2 \nonumber\\
 &\qquad {}   +\frac{1}{2} [\partial_\mu\pi_N^\alpha+ v_NgA^\alpha_{\mu N} ]^2.
\end{align}
\noindent The cross terms mixing the NGB’s with the gauge bosons associated to $X^\alpha$ can be canceled by 
introducing the gauge-fixing term
\begin{equation}\label{fixgaugeT4567}
\mathcal{L}_G = - \sum\limits_{j=1}^{N-1} \frac{1}{2\xi} [\partial_\mu A_j^{\mu \alpha} - \xi g( v_j \pi^\alpha_j - v_{j+1} \pi^\alpha_{j+1})]^2.
\end{equation}
After adding this gauge-fixing term to expansion in (\ref{kinetic-termT4567}), we obtain
\begin{align} \label{kinetic-termT4567+fixgaugeT4567}
&\frac{1}{2}\sum_{j=1}^N(\partial_\mu \pi^{a}_{j})^2 -\sum_{j=1}^{N-1}\frac{1}{2\xi}(\partial_\mu A_{j}^{\mu ~\alpha})^2 -\frac{1}{2}\sum_{j=1}^{N-1} \xi[g( v_j\pi^{a}_{j} - v_{j+1}\pi^{a}_{j+1})]^2\nonumber \\
&+\frac{1}{2}\sum_{j=2}^{N-1} v_j^2 g^2 (A_{j}^{\mu ~\alpha} - A_{j-1}^{\mu ~\alpha})^2-\frac{1}{2}g^2v_{1}^2(A_{1}^{\mu ~\alpha})^2-\frac{1}{2}g^2v_{N}^2(A_{N-1}^{\mu ~\alpha})^2.
\end{align}
\noindent Here, we identify the mass matrix $M_{A^\alpha}^{2}$ for the gauge bosons associated to $X^\alpha$, this is 
given by
\begin{equation}\label{MatMasA-T4567}
M_{A^\alpha}^2 = g^2 v^2 \begin{pmatrix}
q^2 + q^4 & -q^4 & 0 & 0 & \cdots & 0 & 0\\
-q^4  &  q^4 + q^6 & - q^6& 0 & \cdots & 0 & 0 \\
\vdots & \vdots &\vdots &\vdots &\ddots & \vdots & \vdots \\
0 & 0 & 0 & 0 & \cdots & -q^{2N} & q^{2(N-1)} +q^{2N}
\end{pmatrix},
\end{equation}
where we used the parametrization (\ref{VEVs/Sqrt[2]}). We also identify the mass matrix $M^2_\pi$ for the NGBs 
associated to $X^\alpha$ as follow 
\begin{equation}\label{MatMasNGB-T4567}
M^2_{\pi^{\alpha}} = g^2 v^2 \xi \begin{pmatrix}
q^2  & -q^3 & 0 & 0 & \cdots & 0 & 0\\
-q^3  &  2q^4  & - q^5& 0 & \cdots & 0 & 0 \\
0 & -q^5& 2q^6  & -q^7& \cdots & 0 & 0 \\
\vdots & \vdots &\vdots &\vdots &\ddots & \vdots & \vdots \\
0 & 0 & 0 & 0 &\cdots &2q^{2(N-1)}  & - q^{2N-1}\\
0 & 0 & 0 & 0 & \cdots & -q^{2N-1} &  q^{2N}
\end{pmatrix}.
\end{equation}
But now, unlike in the case of (\ref{M_pi^2}), the determinant vanishes: 
\begin{equation}\label{determinant-pi-T4567}
\mathrm{Det}[M_{\pi^\alpha}^2]=0.
\end{equation}
This indicates that the NGBs associated to $X^\alpha$ have a zero mode in
their mass eigenstate basis. This mode is a physical state due the fact that, in the unitary gauge, that is, 
$\xi \rightarrow \infty$, this state will not disappear from the theory.\newline   
Now, we will define the orthonormal rotation that diagonalizes $M_{\pi^\alpha}^2$
\begin{equation}\label{rotation-pi-T4567}
\pi_{j}^{\alpha} = \sum_{n=1}^N b^{j,n} \pi^{\alpha~(n)},
\end{equation}

\noindent where the index $n$ indicates the eigenmode. Since we are interested in the zero mode, we focus on $b^{j,0}$, 
henceforth we will denote it as $b^j$. It is possible to show (using the eigenvalue equation associated to 
$M_{\pi^\alpha}^2$ for the zero eigenvalue) that,

\begin{equation}\label{eigenvector-ZM-pi-T4567}
b^j = \frac{q^{N-j}}{\sqrt{\sum\limits_{j=1}^N q^{2(j-1)}}},
\end{equation}

\noindent where we used the normalization condition 

\begin{equation}\label{normalization-ZM-pi-T4567}
\sum\limits_{j=1}^N |b^j|^2 = 1. 
\end{equation}

\noindent Then, by using (\ref{rotation-pi-T4567}) and (\ref{eigenvector-ZM-pi-T4567}), we obtain

\begin{equation}\label{expansion-ZM-pi-T4567}
\pi^{\alpha~(0)}= \sum_{j=1}^N \frac{q^{N-j}}{\sqrt{\sum\limits_{j=1}^N q^{2(j-1)}}} \pi_j^\alpha.
\end{equation}

\noindent The expression (\ref{eigenvector-ZM-pi-T4567}) indicates that the Higgs is always localized close 
the N-th site. This fact will be used in 
Subsection \ref{Couplings to Higgs}.

Also, we notice that the combination $\pi_j^{\alpha}X^\alpha$ gives 

\begin{equation} \label{combination-pi-T4567}
\pi_j^{\alpha}X^\alpha=\frac{1}{2}\left( \begin{array}{ccc}0 & 0 & \pi_j^4-i\pi_j^5\\0 & 0 & \pi_j^6-i\pi_j^7\\\pi_j^4+i\pi_j^5 & \pi_j^6+i\pi_j^7& 0 \end{array} \right),
\end{equation}

\noindent where the Higgs doublet $H$, as shown in Ref.~\citep{Burdman:2014ixa}, is identified as $(h_1~h_2)^T$, 
where $h_1$ and $h_2$ are given by $\frac{1}{\sqrt{2}}(\pi_j^4-i\pi_j^5)$ 
and $\frac{1}{\sqrt{2}}(\pi_j^6-i\pi_j^7)$ respectively.\newline
\newline \noindent In this way, we have obtained a Higgs out of the breaking of the (partially gauged) global 
symmetry $H_0\times G_1\times \ldots \times G_{N-1}\times H_N$. Is this global symmetry that protects the Higgs mass.

\subsection{Fermions in Quiver Theories}\label{Fermions-in-Quiver-Theories}

We will study the spectrum of Vector-like quarks in quiver theories and their couplings to gauge bosons and the Higgs as can be 
found in Ref.~\citep{Burdman:2014ixa}. 
Then the phenomenology will be done in the following chapter.

The fermions are included in the quiver theories by the following action:
\begin{align}\label{action-fermion-quiver}
S_{\psi}=\int d^4 x \Bigg\{&\sum_{j=0}^N 
\left[\bar{\psi}_{L,j} i \slash\hspace*{-0.22cm} D_{j}\psi_{L,j}+\bar{\psi}_{R,j} i \slash\hspace*{-0.22cm} D_{j}\psi_{R,j}
- (\mu_j \bar{\psi}_{L,j}\psi_{R,j} + \mathrm{h.c.})\right] \nonumber\\
& -\sum_{j=1}^N\lambda ~ (\bar{\psi}_{R,j-1}\Phi_j \psi_{L,j} + \mathrm{h.c.})\Bigg\},
\end{align}

where the fermions $\psi_{j}$ are vector-like, transform in the fundamental representation of $SU(n)_j$, $\lambda$ 
are Yukawa couplings, and $\mu_j$ is the mass term in the interaction eigenstates.

\begin{figure}[!h]
\center
\epsfig{file=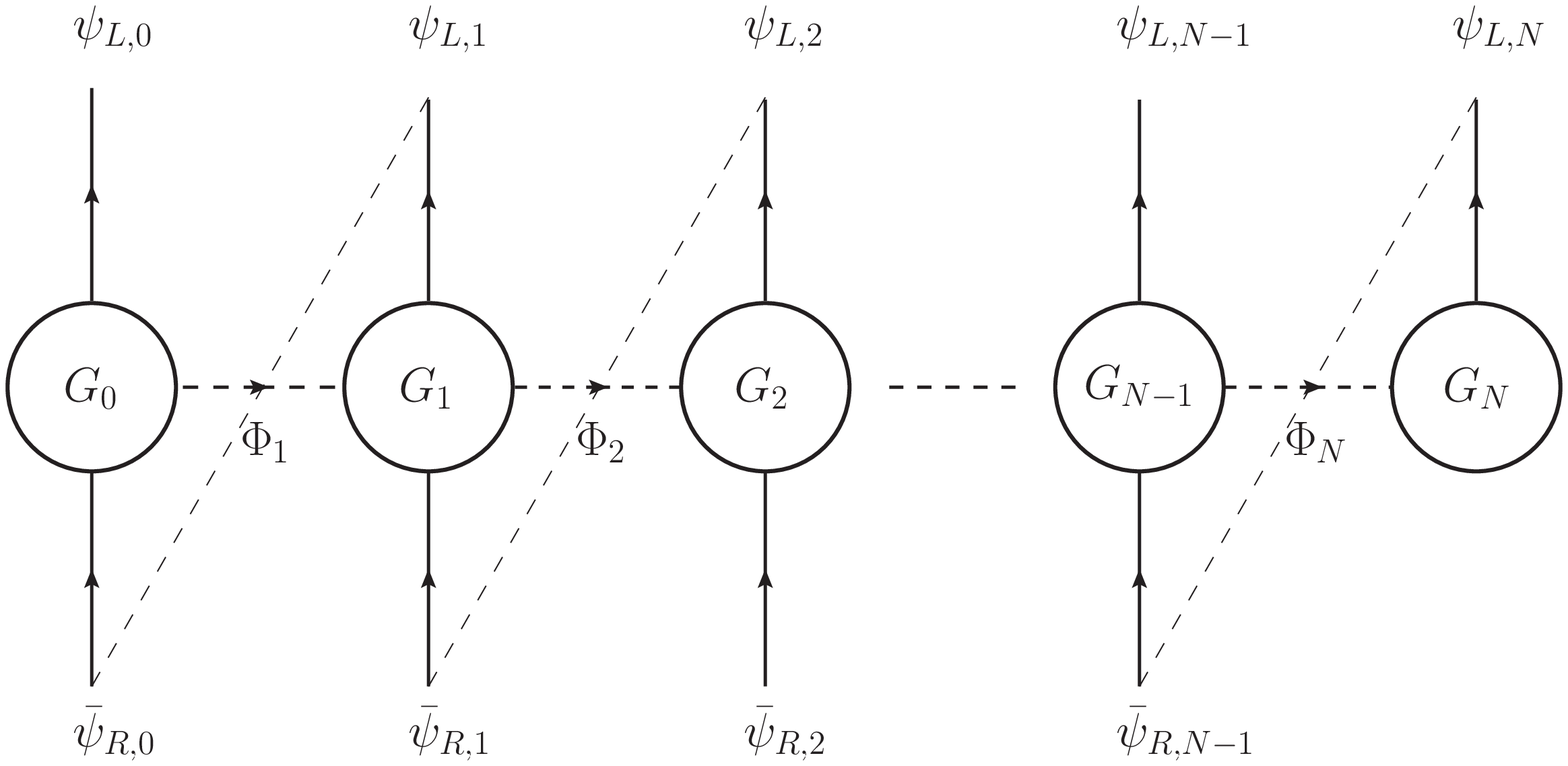,width=14cm}
\caption{\label{moose1}Quiver diagram including the fermions in the action in 
(\ref{action-fermion-quiver}), with the condition that the spectrum includes a fermion with 
left-handed zero mode.}
\end{figure}

\begin{figure}[!h]
\center
\epsfig{file=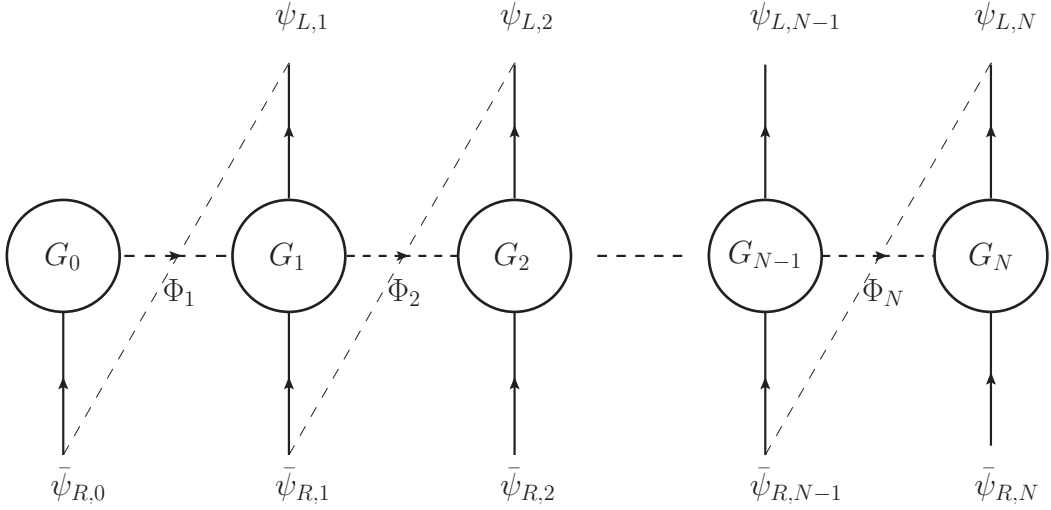,width=14cm}
\caption{\label{moose2}Quiver diagram including the fermions in the action in 
(\ref{action-fermion-quiver}), with the condition that the spectrum includes a fermion with 
right-handed zero mode.}
\end{figure}

Using the link fields in terms of their VEVs (\ref{VEVs/Sqrt[2]}), as shown in 
Refs.~\citep{Bai:2009ij,Burdman:2012sb,Lima-Burdman,Burdman:2014ixa}. In addition, we need to consider 
the next relations that were 
shown in Ref.\citep{Bai:2009ij} 

\begin{equation}\label{paramtrizations-to-fermions}
\mu_j = -gvq^{c+j+1/2}, ~~~~~ \lambda = g, ~~~~~
\end{equation}

\noindent where $c$ is the localization parameter for fermions associated to $\mathrm{AdS_5}$ 
theories for fermions in the Bulk. Now, we can identify 
the mass term for the fermions in the Lagrangian associated with the action (\ref{action-fermion-quiver}) 
as 

\begin{equation}\label{L-fermion-mass}
 \mathcal{L}_{M_{\psi}} =-\sum_{j=0}^N 
(\mu_j \bar{\psi}_{L,j}\psi_{R,j} + \mathrm{h.c.})-\sum_{j=1}^N\lambda v_j ~ (\bar{\psi}_{R,j-1}\psi_{L,j} + \mathrm{h.c.})\equiv -\overline{\Psi}_{L} M_{\psi}^{T}\Psi_{R}+\mathrm{h.c.},
\end{equation}

where $\Psi_{L/R}^T$ was written in the basis $\Psi_{L/R}^{T}=(\psi_{L/R,0}, \psi_{L/R,1}, \ldots,\psi_{L/R,N})$, 
we can change from this basis to the mass eigenstate basis as follow

\begin{equation}\label{WFL-expansion}
\psi_{L/R,j}=  \sum_{n=0}^N
h_{L/R}^{j,n} \chi_{L/R}^{(n)},
\end{equation}

where $\chi_{L/R}^{(n)}$ are the mass eigenstate. Thus, to find $\psi_{L,j}$ as a linear combination of 
$\chi_{L}^{(n)}$, we can obtain it by 
diagonalizing the matrix 

\begin{equation}\nonumber M_{\psi}^T M_{\psi}=\,\left(
\renewcommand{\arraystretch}{1.6}
\begin{array}{ccccccc}
\mu_0^2             & \lambda\mu_0v_1                            &       0                   & \cdots     & 0                                                      &    0     \\
\lambda\mu_0v_1     &  \lambda^2v_1^2+\mu^2_1                    &  \lambda\mu_1v_2          & \cdots     & 0                                                      &0      \\
0                   &   \lambda\mu_1v_2                          &  \lambda^2v_2^2+\mu^2_2   & \cdots     & 0                                                      & 0   \\
\vdots              & \vdots                                     & \vdots                    & \ddots     & \vdots                                                 & \vdots   \\
0                   & 0                                          & 0                         & \cdots     & \lambda^2v^2_{N-1}+\mu^2_{N-1}                         &  \lambda\,\mu_{N-1}\,v_N\,  \\
0                   & 0                                          & 0                         & \cdots     & \lambda\mu_{N-1}\,v_N                                  & \lambda ^2 v _N^2+\mu_N^2
\end{array}
\right). 
\end{equation}

Analogously we diagonalize the matrix $M_{\psi}M_{\psi}^T $ to obtain $\psi_{R,j}$ as a linear combination of 
$\chi_{R}^{(n)}$. Now, we need to indicate if the action (\ref{action-fermion-quiver}) belongs to a fermion with 
left- or right- handed zero mode. Thus, the case where we will have a fermion with left-handed zero mode is 
achieved by taking $\mu_N=0$, it corresponds to the quiver diagram of Figure (\ref{moose1}), in this case the 
linear combination 

\begin{equation}\label{WFL-expansionleft}
\psi_{L,j}=  \sum_{n=0}^N
h_{L}^{j,n}(c_L) \chi_{L}^{(n)}
\end{equation}

is found  by diagonalizing the matrix
\newline\newline \begin{equation} \label{M_psi1^2}
M_{\psi}^T M_{\psi} = g^2 v^2 \left( \begin{array}{ccccccc}
~~q^{2c_L+1}          & -q^{c_L+\frac{3}{2}}                          &\ldots     &0                                   & 0 \\
-q^{c_L+\frac{3}{2}}  & q^{2 c_L+3}+q^2                               &\ldots     &0                                   & 0 \\
~~0                   & -q^{c_L+\frac{7}{2}}                          &\ldots   &0                                   & 0 \\
\vdots                & \vdots                                        &\ddots     &\vdots                                & \vdots \\
~~0                   & ~~0                                           &\ldots     & q^{2 c_L+2N-1}+q^{2(N-1)}            & -q^{c_L+\frac{1}{2}(4N-1)}   \\
~~0                   & ~~0                                           &\ldots     &-q^{c_L+\frac{1}{2}(4N-1)}             &q^{2N}
\end{array} \right).
\end{equation}

To write this matrix explicitly, the parametrizations (\ref{VEVs/Sqrt[2]}) and (\ref{paramtrizations-to-fermions}) 
were used.

On the other hand, if we take $\mu_0=0$, we will have a fermion with right-handed zero mode, for this case 
the quiver diagram corresponds to Figure (\ref{moose2}), such that the linear combination

\begin{equation}\label{WFL-expansionright}
\psi_{R,j}=  \sum_{n=0}^N
h_{R}^{j,n}(c_R) \chi_{R}^{(n)}
\end{equation}

can be found by diagonalizing the matrix
\newline\newline \begin{equation} \label{M_psir^2}
M_{\psi} M_{\psi}^T = g^2 v^2 \left( \begin{array}{ccccccc}
~~q^{2}               & -q^{c_R+\frac{5}{2}}                          &\ldots     &0                                      & 0 \\
-q^{c_R+\frac{5}{2}}  & q^4+q^{2 c_R+3}                               &\ldots     &0                                      & 0 \\
~~0                   & -q^{c_R+\frac{9}{2}}                          &\ldots     &0                                      & 0 \\
\vdots                & \vdots                                        &\ddots     &\vdots                                 & \vdots \\
~~0                   & ~~0                                           &\ldots     & q^{2 c_R+2N-5}+q^{2(N-2)}             & -q^{c_R+\frac{1}{2}(4N-7)}   \\
~~0                   & ~~0                                           &\ldots     &-q^{c_R+\frac{1}{2}(4N-7)}             &q^{2c_R+2N-3}
\end{array} \right),
\end{equation}
\newline\newline where we used the parametrizations (\ref{VEVs/Sqrt[2]}) and (\ref{paramtrizations-to-fermions}). 
Notice that the values of $c_{L,R}$ for fermions zero mode of the SM were found in Ref.~\citep{Burdman:2012sb}
 \subsubsection{Equation of Motion for $\psi_{L,j}$ and $\psi_{R,j}$}
 
We saw that $\chi_{L}^{(n)}$ and $\chi_{R}^{(n)}$ are the mass eigenstates. We will impose that they satisfy 
the Dirac equation
\begin{align}\label{dirac1}
&i \slash\hspace*{-0.22cm} \partial \chi_{L}^{(n)} - m_n  \chi_{R}^{(n)} =0, \\ 
&i \slash\hspace*{-0.22cm} \partial \chi_{R}^{(n)} - m_n\chi_{L}^{(n)}=0.\label{dirac2}
\end{align}
On the other hand the equations of motion can be obtained from (\ref{action-fermion-quiver}). 
Expressing the link fields in terms of their VEVs, we obtain

\begin{align} \label{eq126}
 \text{for}~~\bar{\psi}_{R,j}&:~~~~i~\slash\hspace*{-0.22cm}\partial ~\psi_{R,j} + \lambda v_{j+1}~ \psi_{L,j+1} + \mu_j~ \psi_{L,j} =0, \\ 
 \text{for}~~\bar{\psi}_{L,j}&:~~~~i~\slash\hspace*{-0.22cm}\partial ~\psi_{L,j} + \lambda v_{j}~ \psi_{R,j-1}+ \mu_j~ \psi_{R,j} =0.\label{eq127}
\end{align}
Using (\ref{WFL-expansion}), (\ref{eq126}) and (\ref{eq127}) we obtain
\begin{align} \label{eq128} 
&m_n~ h^{j,n}_R + \mu_j~ h^{j,n}_L +\lambda v_{j+1}~h^{j+1,n}_L =0,\\ 
&m_n~ h^{j,n}_L+ \mu_j~ h^{j,n}_R + \lambda v_{j}~h^{j-1,n}_R =0,\label{eq129}
\end{align}
where the equations (\ref{eq128}) and (\ref{eq129}) are coupled. After decoupling these we obtain
 \begin{align}  &\left (\mu_j^2+ \lambda^2 v_{j}^2- m_n^2 \right)h_L^{j,n} +
\lambda  \mu_j v_{j+1} h_L^{j+1,n} +\lambda\mu_{j-1} v_{j}h_L^{j-1,n} = 0, \label{eq130}\\  
& \nonumber\\ 
&\left(\mu_j^2+ \lambda^2 v_{j+1}^2- m_n^2 \right)h_R^{j,n} +
\lambda \mu_{j+1} v_{j+1} h_R^{j+1,n} +
\lambda \mu_{j}v_{j}h_R^{j-1,n} = 0.\label{eq131}
\end{align}
In the next subsection we will obtain the analytical zero mode wave functions by imposing conditions to have a left-handed zero mode 
or a right-handed zero mode.
 
\subsubsection*{Zero Mode Wave-Function}

To obtain the SM fermion spectrum as the zero modes, we can impose as boundary condition $h_R^{N,n}=0$, that is $\psi_{R,N}=0$ 
to obtain a left-handed zero mode, or we can impose $h_L^{0,n}=0$, that corresponds to $\psi_{L,0}=0$ 
to obtain a right-handed zero mode.

In the case of a left-handed zero mode, we use (\ref{eq128}) to obtain
\begin{equation}\nonumber
\mu_j~h^{j,0}_L + \lambda v_{j+1} h^{j+1,0}_L =0,
\end{equation}
 
\noindent which is equivalent to
\begin{equation}\label{eq137}
\frac{h^{j+1,0}_L}{h^{j,0}_L} = q^{c_L - 1/2}. 
\end{equation}

Since $0 < q < 1$, for $c_L>1/2$ the left-handed zero mode wave function will be “localized” close to the zero 
site corresponding to the left side of quiver diagram in Figure (\ref{moose1}), and for $c_L<1/2$ 
it will be localized close to the N site, that 
corresponds to the right side of the quiver diagram in Figure (\ref{moose2}). 
The $0$ site corresponds to the UV scale because $v_1$ is largest and the $N$ site corresponds to the IR scale 
because $v_N$ is the smallest VEV. Nothing to do with fermion localization.\newline 
\newline Alternatively, if we consider a right-handed zero mode, we use (\ref{eq129}) to obtain
\begin{equation}\nonumber
\mu_j~h^{j,0}_R + \lambda v_{j} h^{j-1,0}_R =0,
\end{equation}
and then
\begin{equation}\label{eq138}
\frac{h^{j,0}_R}{h^{j-1,0}_R} = q^{-(c_R +1/2)}. 
\end{equation}
So for $c_R>-1/2$ the right-handed zero mode wave function is localized in the N site (IR), 
on the other hand for $c_R<-1/2$ it will be localized closer to the zero site (UV).

To obtain an analytical expression for the zero mode wave functions for the fermions, we write

\begin{equation}\label{wf-ZM1}
h^{j,0}_{L,R}=z^{j}_{L,R}h^{0,0}_{L,R}, 
\end{equation}

\noindent where we have considered the definitions $z_L\equiv q^{c_L-1/2}$ and $z_R\equiv q^{-(c_R+1/2)}$, such that the 
normalization conditions for the zero mode wave functions can be written as follows 

\begin{equation}\label{wf-ZM2}
 \sum^{N}_{j=0}|h^{j,0}_{L,R}|^2=|h^{0,0}_{L,R}|^2\sum^{N}_{j=0}z_{L,R}^{2j}=1,
\end{equation}

\noindent and then we obtain 

\begin{equation}\label{wf-ZM3}
h^{0,0}_{L,R}=\sqrt{\frac{1-z_{L,R}^{2}}{1-z_{L,R}^{2(N+1)}}}. 
\end{equation}

So if the zero mode wave function of a fermion is closer to the N site associated to the scale (IR), this has more coupling with the Higgs.
In analogous way, if the zero mode wave function of a fermion is localized closer to the zero site, such that it 
is associated to the scale UV, it has less coupling with the Higgs~\citep{Burdman:2012sb}, given that the Higgs is localized 
close to the $N$-th site.\newline
\noindent The quiver theories will have characteristics similar to $\mathrm{AdS_5}$ theories, but from this point of view we can 
say that they have different phenomenology, in important aspects of the theory. We mention that the $\mathrm{AdS_5}$ 
theories are a particular case of quiver theories in the continuum limit.
\subsubsection*{Spectrum of Excited States}
\begin{figure}[H]
    \begin{center}    
    \resizebox{4.0in}{!}
    {
    \includegraphics{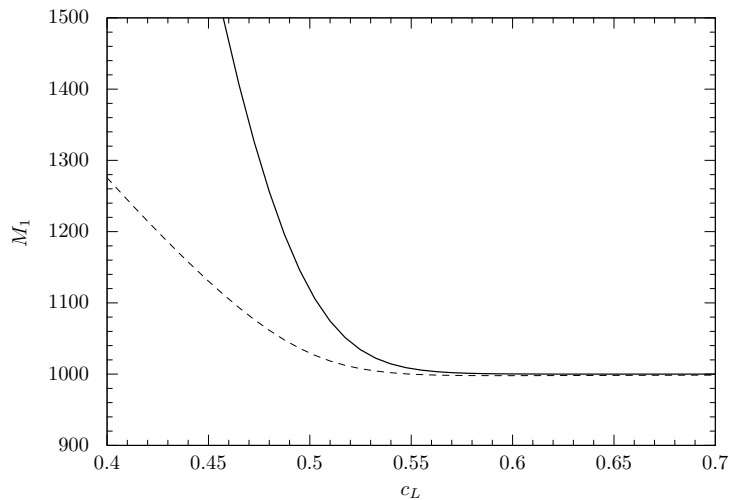}
    }
    \captionsetup{font=small, labelfont=bf, labelsep=period}
    \caption{\label{mass-left}Masses of the first excited states of the fermions that belong to their tower with left-handed zero 
    mode, the solid and dashed lines correspond to N=4, 15, respectively.}
    \label{figure}
    \end{center}
    \end{figure}

\begin{figure}[H]
    \begin{center}
    \resizebox{4in}{!}
    {
    \includegraphics{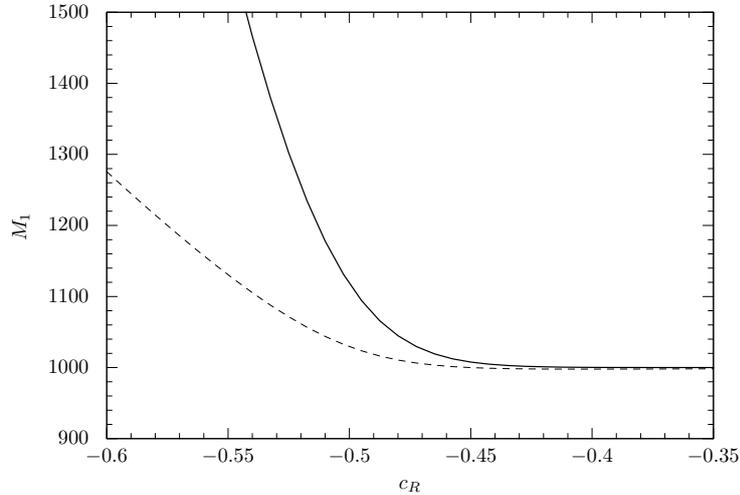}
    }
    \captionsetup{font=small, labelfont=bf, labelsep=period}
    \caption{\label{mass-right}Masses of the first excited states of the fermions that belong to their tower with right-handed zero 
    mode, the solid and dashed lines correspond to N=4, 15, respectively.}
    \label{figure}
    \end{center}
    \end{figure}

To understand the phenomenology, such as the production or decay of the excited fermions in quiver theories, 
we need to compute the spectrum of these excited states. 
Based on Subsection \ref{Fermions-in-Quiver-Theories}, we computed the masses of the excited fermions ($n>0$) 
by diagonalizing the fermion mass matrices 
(\ref{M_psi1^2}) or (\ref{M_psir^2}) to have left- and right-handed zero modes, respectively. Once we have fixed $N$ 
jointly with $v= M_P=10^{19}$ GeV and $v_N=$1 TeV, then these matrices will depend on the localization parameters.\newline 
\newline Thus the masses for the first excited states of fermions with left-handed zero modes are shown 
in Figure (\ref{mass-left}), for $c_L>0.5$, i.e., in which their left-handed zero modes wave functions are localized 
in the UV sites; the masses will be of order $v_N$, differently, for $c_L<0.5$, the masses will be exponentially 
heavier.\newline
\noindent The case when there are right-handed zero modes, the masses for the first excited states of fermions are shown in 
Figure (\ref{mass-right}). We see that for $c_R>-0.5$, i.e., the case 
in which their right-handed zero modes wave functions 
are localized in the IR sites; similarly to the previous case, the masses will be of order $v_N$ and be exponentially 
heavier for $c_R<-0.5$. 

Having computed the wave functions and mass spectrum for the fermion excitations, their couplings will be computed below.  

\subsection{Couplings of Fermionic Excited States}
Based in the approach of ~\citep{Burdman:2014ixa,Nayara-Burdman,Lima-Burdman} 
we can now compute the couplings of the excited fermions to the gauge bosons and the 
Higgs sector in the frame of quiver theories. The rest of this subsection is organized as follows: First, we show how to compute 
the couplings of the excited fermions to gauge boson excitations, for both cases left- and right-handed fermion zero 
modes. We will concentrate in computing the couplings of the zero-mode fermions, for the 
cases of the first and third generations of quarks, to the first excited state of a gluon. 
In addition, the couplings of the excited fermions to their zero mode and the first excited gauge bosons. Afterwards we shall focus on 
obtaining the couplings of the fermions considered in 
the previous subsection to the Higgs sector. The relevant wave functions showing in Appendix \ref{WF-Excited} will be used below.
\subsubsection{Couplings to Gauge Bosons}\label{Couplings to Gauge Bosons}

We will obtain the coupling of the excited fermions to gauge boson excitations in the 
quiver theories. These couplings are included in the kinetic terms in (\ref{action-fermion-quiver}). 
We first consider the case where the spectrum includes a fermion with a left-handed zero mode, as follows

\begin{equation}\label{couplings-gauge-fermion-L}
\mathcal{L}_{\Psi_L A}=\sum_{j=0}^N \widetilde{g}_j\bar{\psi}_{L,j} ~\gamma^{\mu}A_{\mu,j}\psi_{L,j},
\end{equation}

where $\widetilde{g}_j$ is the gauge coupling associated with the $SU(n)_j$ gauge group, $\psi_{L,j}$
and $A_{\mu,j}$ are interaction eigenstates associated to the site $j$, such that $j=0,1,\ldots, N$. 

Now, We can write the fields included in (\ref{couplings-gauge-fermion-L}) by using their mass eigenstates basis 
(\ref{rotation-gauge}) and (\ref{WFL-expansionleft}). Note also in this case that $h_L^{j,n}$ is given in 
(\ref{WFL-expansion}) and obtained by diagonalizing the matrix 
$M_{\psi}^T M_{\psi}$ (\ref{M_psi1^2}) by taking $\mu_N=0$, such that (\ref{couplings-gauge-fermion-L}) can be written as 

\begin{equation}\label{couplings-gauge-fermion-L1}
\mathcal{L}_{\Psi_L A}=\sum_{j,n,m,p=0}^N \left[\widetilde{g}(h_L^{j,n})^{*}f^{j,m}h_L^{j,p}\right]\bar{\chi}_{L}^{(n)} ~\gamma^{\mu}A_{\mu}^{(m)}\chi_{L}^{(p)},
\end{equation}

where we have used the expansion given in (\ref{rotation-gauge}), $A_{\mu}^{(m)}=A_{\mu}^{a~(m)}T^a$ and $\widetilde{g_j}$ 
was assumed to be equal to $\widetilde{g}$. We also define the effective coupling of the $n$ and $p$ fermion excitations to the $m$ 
gauge boson excitation as

\begin{equation}\label{couplings-gauge-fermion-definition}
g_L^{n,m,p}\equiv\widetilde{g}\sum_{j=0}^N \left[(h_L^{j,n})^{*}f^{j,m}h_L^{j,p}\right].
\end{equation}

Notice that in (\ref{couplings-gauge-fermion-definition}), for a given number of sites in this model, this $g_L^{n,m,p}$ 
depends on the value of $c_L$ according to (\ref{WFL-expansionleft}). In addition, it allows the mixing between 
different modes of a left-handed fermion to a gauge boson excitation.

Now, to establish the relation between $\widetilde{g}$ and the SM couplings, $g_L^{n,m,p}$ in 
(\ref{couplings-gauge-fermion-definition}), we impose that for $n=m=p=0$ 
\begin{equation}\label{couplings-gauge-fermion-definition000}
g_L^{0,0,0}=\widetilde{g}\sum_{j=0}^N \left[(h_L^{j,0})^{*}f^{j,0}h_L^{j,0}\right],
\end{equation}
we obtain the corresponding SM gauge coupling of the zero modes. 
We can use (\ref{WF-gauge0}) and the fact the $h_L^{j,0}$ satisfies the normalization condition for a given value of $c_L$. 
The coupling $g_L^{0,0,0}$ in (\ref{couplings-gauge-fermion-definition000}) may be written in the form 
\begin{equation}\label{couplings-gauge-fermion-definition000_first}
g_L^{0,0,0}=\frac{\widetilde{g}}{\sqrt{N+1}}\sum_{j=0}^N \left[(h_L^{j,0})^{*}h_L^{j,0}\right],
\end{equation}
from which we obtain
\begin{equation}\label{couplings-gauge-fermion-definition000_second}
g_L^{0,0,0}=g=\frac{\widetilde{g}}{\sqrt{N+1}},
\end{equation}
where $g$ is the gauge coupling of the SM, so (\ref{couplings-gauge-fermion-definition000_second}) yields
\begin{equation}\label{couplings-gauge-fermion-definition000_second}
\widetilde{g}=g\sqrt{N+1},
\end{equation}
and then substituting (\ref{couplings-gauge-fermion-definition000_second}) into (\ref{couplings-gauge-fermion-definition}). 
We find then that
\begin{equation}\label{couplings-gauge-fermion-definition-withg}
g_L^{n,m,p}=g\sqrt{N+1}\sum_{j=0}^N \left[(h_L^{j,n})^{*}f^{j,m}h_L^{j,p}\right].
\end{equation}
On the other hand, in the case of a fermion with right-handed zero mode, we proceed in a similar way, 
analogously to the previous case. We have 

\begin{equation}\label{couplings-gauge-fermion-R}
\mathcal{L}_{\Psi_L A}=\sum_{j=0}^N \widetilde{g}_j\bar{\psi}_{R,j} ~\gamma^{\mu}A_{\mu,j}\psi_{R,j}.
\end{equation}


Thus, considering this type of spectrum we can write
\begin{equation}\label{couplings-gauge-fermion-definition-withgR}
g_R^{n,m,p}=g\sqrt{N+1}\sum_{j=0}^N \left[(h_R^{j,n})^{*}f^{j,m}h_R^{j,p}\right],
\end{equation}

\noindent where for a fixed number of sites, $g_R^{n,m,p}$ will depend on the value of $c_R$, since in this case $h_R^{j,n}$ 
is obtained diagonalizing the matrix $M_{\psi}M_{\psi}^T$ (\ref{M_psir^2}) by taking $\mu_0=0$.
In subsections (\ref{usp1}) and (\ref{usp2}) we will show some results, for couplings that involve a first excited 
state of a gauge boson to excited fermions. 

\subsubsection*{Zero Mode}\label{usp1}

\begin{figure}[H]
\center
\resizebox{5.7in}{!}
{
\epsfig{file=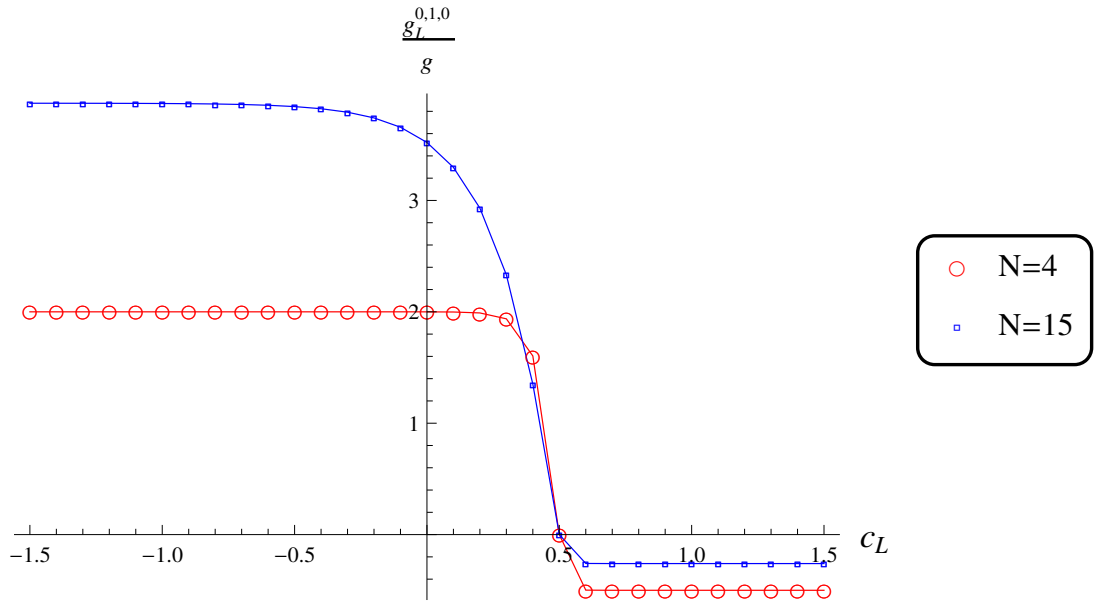}
}
\captionsetup{font=small, labelfont=bf, labelsep=period}
\caption{\label{ZML}Couplings of the left-handed zero mode to a first excited state of a gauge boson,
in units of the SM gauge boson coupling, as a function of the localization parameter. For N=4, N=15.}
\end{figure}

Here we perform a numerical calculation, first to obtain the couplings of the left-handed zero mode 
to a first excited state of a gauge boson, i.e. $g^{0,1,0}_L(c_L)$. Using (\ref{couplings-gauge-fermion-definition-withg}), 
we obtain this coupling as a function of the localization parameter $c_L$, for specific quiver theories 
with five and sixteen sites, i.e. $N=4$ and $N=15$ respectively. 
As shown in Figure (\ref{ZML}), that agrees with the analytical calculation obtained in the previous 
Refs.~\citep{Burdman:2012sb,Burdman:2013qfa,Nayara-Burdman}. In this plot there are two plateaus: 
the upper plateau, for $c_L<0.5$ corresponds to left-handed fermions whose zero modes are localized 
close to the $N$-th site, that is, in the IR region. And the lower plateau, for $c_L>0.5$, where the left-handed fermions 
have their zero modes localized close to the zeroth site, that is, in the UV region.
\begin{figure}[H]
\center
\resizebox{5.7in}{!}
{
\epsfig{file=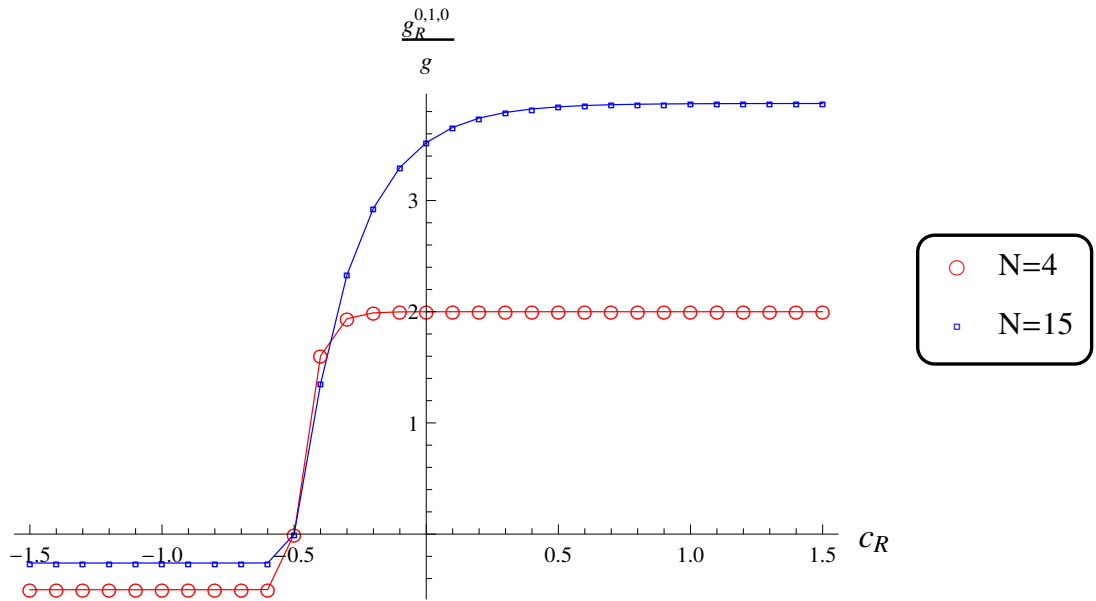}
}
\captionsetup{font=small, labelfont=bf, labelsep=period}
\caption{\label{ZMR}Couplings of the fermions with right-handed zero mode to a first excited state of gauge boson,
in units of the SM gauge boson coupling, as a function of the localization parameter. For N=4, N=15.}
\end{figure}
We also obtain the couplings of the fermions with a right-handed zero-mode 
to a first excited state of gauge boson, i.e. $g^{0,1,0}_R(c_R)$, as a function of the localization parameter $c_R$, for specific quiver theories 
with five and sixteen sites, i.e. $N=4$ and $N=15$ respectively, (\ref{couplings-gauge-fermion-definition-withg}) was used.
\newline\newline\noindent The Figure (\ref{ZMR}) shows this calculation, which is in agreement with the analytical calculation obtained in the previous 
Refs.~\citep{Burdman:2012sb,Burdman:2013qfa,Nayara-Burdman}. In this plot there are also two plateaus: 
the lower plateau, for $c_R<-0.5$, corresponds to right-handed zero-mode fermions localized 
close to the zeroth site, that is, in the UV region. And the upper plateau, for $c_R>-0.5$, corresponds 
to right-handed zero-modes localized close to the $N$-th site, that is, in the IR region.
 
\subsubsection*{Off Diagonal Couplings}\label{usp2}

We will compute the off diagonal couplings involving the first excited fermion, a zero-mode fermion and a first excitation of gauge boson. 
This will be later used to study the phenomenology of the excited fermions. A similar procedure was followed to obtain the values of the 
couplings of the first excited fermions to their zero mode and the first excited of a gauge boson. We use 
(\ref{couplings-gauge-fermion-definition}) and (\ref{couplings-gauge-fermion-definition-withgR}) with n=0, and m=p=1, then 
we can substitute these values to obtain

\begin{equation}\label{couplings-gauge-fermion-definition-withg011}
g_L^{0,1,1}=g\sqrt{N+1}\sum_{j=0}^N \left[(h_L^{j,0})^{*}f^{j,1}h_L^{j,1}\right], 
\end{equation}

\noindent where $h_L^{j,0}$, $h_L^{j,1}$ and $f^{j,1}$ are obtained through 
rotation to the mass eigenstates (\ref{rotation-gauge}) and (\ref{WFL-expansionleft}). Analogously we can 
obtain using (\ref{rotation-gauge}) and (\ref{WFL-expansionright})

\begin{equation}\label{couplings-gauge-fermion-definition-withgR011}
g_R^{0,1,1}=g\sqrt{N+1}\sum_{j=0}^N \left[(h_R^{j,0})^{*}f^{j,1}h_R^{j,1}\right].
\end{equation}

As we have already mentioned, $g^{0,1,1}_L$ and $g^{0,1,1}_R$ depend on the values of $c_L$ and $c_R$. These couplings 
appear in the effective Lagrangian as

\begin{equation}\label{couplings-gauge-fermion-L11}
g_L^{0,1,1}\bar{\chi}_{L}^{(0)} ~\gamma^{\mu}A_{\mu}^{(1)}\chi_{L}^{(1)}+\mathrm{h.c.}
\end{equation}

and 

\begin{equation}\label{couplings-gauge-fermion-R11}
g_R^{0,1,1}\bar{\chi}_{R}^{(0)} ~\gamma^{\mu}A_{\mu}^{(1)}\chi_{R}^{(1)}+\mathrm{h.c.}
\end{equation}

To calculate the couplings in (\ref{couplings-gauge-fermion-definition-withg011}) and 
(\ref{couplings-gauge-fermion-definition-withgR011}), the localization parameters $c_{L,R}$ that are found in 
Refs.~\citep{Burdman:2012sb,Nayara-Burdman} were used. This choice corresponds to a solution that gives the correct quark masses as well as 
the correct CKM matrix. In addition, we considered the values of $c_{L,R}$ for 
the first and third generation of SM quarks. These localization parameters are $c_{u_R}=-0.73$, $c_{u_L}=0.62$, 
$c_{d_R}=-0.96$ and $c_{t_R}=-0.12$, $c_{t_L}=0.51$, $c_{b_R}=-0.61$ for the first and third generation respectively.\newline 
\newline\noindent The results are displayed in Table (\ref{tabela}), as shown in Ref.~\citep{Burdman:2014ixa}. 
The column labeled '$u^{(1)}$' gives the values of the 
couplings between the first excited state of the up quark from the first-generation of the SM to their zero mode, i.e., 
the up quark and the first excited state of a gauge boson $A^{(1)}$ in units of the zero-mode SM gauge coupling 
obtained by using the values of localization parameters $c_{u_L}$ and $c_{u_R}$, for left and right handed chiralities, 
and for quiver theories with $N = 4$ or $N = 15$. 

Analogously, the column labeled '$d^{(1)}$' contains the values of the 
couplings between the first excited state of the down quark from the first-generation of the SM to their zero mode, i.e., 
the down quark and the first excited state of a gauge boson $A^{(1)}$ in units of the zero-mode SM gauge coupling. In 
this case we used the values of localization parameters $c_{u_L}$ and $c_{d_R}$. On the other hand the columns labeled 
respectively '$t^{(1)}$' and '$b^{(1)}$', were obtained by using the values of the localization parameters 
$c_{t_R}$, $c_{t_L}$ and $c_{b_R}$, for left and right handed chiralities, 
and for quiver theories with $N = 4$ or $N = 15$. As can be seen, just the largest 
couplings are found in the third generation, particularly the right-handed top sector for a quiver theory with 
$N=15$. This is in agreement with the overlap of the wave-functions that were shown in Figures (\ref{wfGB_N4}) 
and (\ref{wfGB_N15}) for the gauge boson and top fermion with right-handed zero mode and localization 
parameter $c_{R}$ for a quiver theory with $N=15$.

\begin{table}[H]
 \begin{center}
\begin{tabular}{|c | c |  c c c c|}
\hline\hline
 N & & $u^{(1)}$ & $d^{(1)}$ & $t^{(1)}$  & $b^{(1)}$ \\
\hline\hline
$4$&$L$& $0.028$ & $0.028$ & $0.85$ & $0.85$  \\
&$R$& $5.2\times10^{-4}$ & $1.1\times10^{-7}$ & $0.075$ & $0.04$  \\
\hline
$15$&$L$& $0.033$ & $0.033$ & $0.83$ & $0.83$  \\
&$R$& $7.1\times10^{-4}$ & $1.7\times10^{-7}$ & $1.49$ & $0.046$  \\
\hline
\end{tabular}
\end{center}
\captionsetup{font=small, labelfont=bf, labelsep=period}
\caption{The couplings of the first excited fermions to their zero mode and the first excited gauge boson, in
units of the zero-mode SM gauge coupling, for N=4, N=15 }\label{tabela}
\end{table}
 
\subsection{Couplings to the Higgs Boson}\label{Couplings to Higgs}

Based on Subsection \ref{Fermions-in-Quiver-Theories}, where it was studied how treat the wave-functions associated to fermions 
in quiver theories, we now consider the couplings of the excited fermions to the Higgs sector. We will follow the
approach from Ref.~\citep{Burdman:2014ixa}. Then to obtain these couplings, we need to consider the general form of the
fermion couplings to the link fields that contain the Higgs doublet. As we have seen in (\ref{action-fermion-quiver}), it 
will involve fermions of the same tower $\psi_{L,j}$ and $\psi_{R,j}$ with a common zero mode, that is, 
after the application of the boundary condition to have left- or right- handed zero mode. 

\begin{figure}[!h]
\center
\epsfig{file=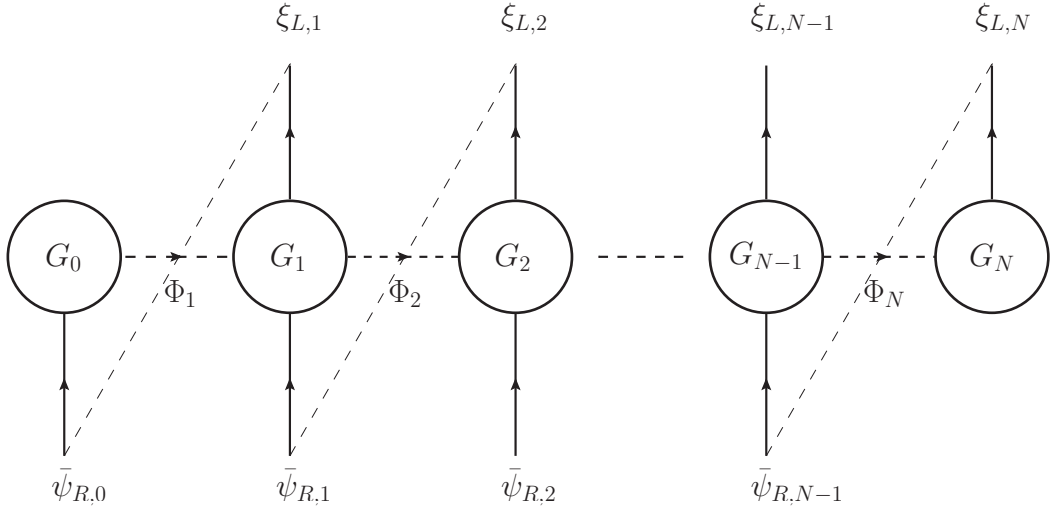,width=14cm}
\captionsetup{font=small, labelfont=bf, labelsep=period}
\caption{\label{moose-chapter3}Quiver diagram including the interaction between fermions of different towers associated to 
(\ref{yukawwa2}).}
\end{figure}

Moreover, we consider another type of Yukawa terms involving fermions of different towers, as follow    

\begin{equation}\label{yukawwa2}
\bar{\psi}_{R,j-1}\Phi_j\xi_{L,j},
\end{equation}

\noindent where $\xi_{L,j}$ is associated to a tower with a zero mode, and $\psi_{R,j-1}$ is associated to a 
tower with a zero mode different from $\xi_{L,j}$. Notice that this term is gauge invariant due also to the transformation 
(\ref{bifunamental_tranform}), so is allowed by the theory.\newline
\newline\noindent Here the term given in (\ref{yukawwa2}) is related to the quiver diagram of Figure (\ref{moose-chapter3}). In 
contrast to the Yukawa terms included in (\ref{action-fermion-quiver}), where these can be obtained from the 
deconstruction of $\mathrm{AdS_5}$ theory with fermions as shown in Ref.~\citep{Bai:2009ij} and then these have 
analog in the continuum limit, the term in (\ref{yukawwa2}) does not have analog in the continuum limit.\newline
\newline \noindent The Higgs is a pNGB extracted from the link fields $\Phi_j$ in the manner explained in Subsection \ref{higgs-pNGB}.\newline
\newline\noindent The coupling in (\ref{yukawwa2}) can be written as

\begin{equation}\label{yukawa3}
-\sum_{j=1}^Ny_j\bar{\psi}_{R,j-1}\Phi_j\xi_{L,j}+\mathrm{h.c.},
\end{equation}

\noindent where the Yukawa couplings $y_j$ in (\ref{yukawa3}) are assumed to be $\mathcal{O}$(1), and the fermion fields $\xi_{L,j}$ and 
$\psi_{R,j-1}$ correspond to different zero modes, characterized by the localization parameters $c_L$ or $c_R$, 
with appropriate quantum numbers. The Higgs doublet is given by 

\begin{equation}\label{doublet-higgs}
H= \frac{1}{\sqrt{2}}\left (
\begin{array}{c}
\sqrt{2}\varphi^+\\
v +H^0+i\varphi_3
\end{array} \right ).
\end{equation}

Then, to  obtain the couplings of the excited fermions to the Higgs sector, the rotations (\ref{WFL-expansionleft}) and 
(\ref{WFL-expansionright}) were substituted into (\ref{yukawa3}), and the Higgs doublet couplings in (\ref{yukawa3}) are 
rewritten as follows

\begin{equation}\label{yukawa3-expansion-nm}
-\sum_{j=1}^N\sum_{n,m=0}^N y_j(h_R^{j-1,n})^{*}h_L^{j,m}b^j\bar{q}_{R}^{(n)}HQ_{L}^{(m)}+\mathrm{h.c.},
\end{equation}

\noindent where $b^j$ is given by (\ref{eigenvector-ZM-pi-T4567}).\newline
\newline Notice that after the quiver symmetry breaking, we still have terms in the theory invariant under $SU(2)\times U(1)$, 
as mentioned above, and then we consider terms that are invariant under $SU(2)$ and have zero net hypercharge $Y$. 
Thus, considering mixing of the excited states of quarks belonging to the same family, we can write

\begin{align}\label{Yukawa4}
\mathcal{L}\supset&-\sum_{n,m=0}^N\sum_{j=1}^N y_{uj}(h_R^{j-1,n}(c_{\textsf{R}_u}))^{*}h_L^{j,m}(c_{\textsf{L}_q})b^j\overline{\textsf{R}}_u^{(n)}\widetilde{H}^\dagger \textsf{L}_q^{(m)}\nonumber\\
&-\sum_{n,m=0}^N\sum_{j=1}^N y_{dj}(h_R^{j-1,n}(c_{\textsf{R}_d}))^{*}h_L^{j,m}(c_{\textsf{L}_q})b^j\overline{\textsf{R}}_d^{(n)}H^\dagger \textsf{L}_q^{(m)}+\mathrm{h.c.},
\end{align}

\noindent where $\widetilde{H}=i\sigma_2 H^{*}$.\newline

\noindent Here $\textsf{R}_u^{(n)}$ and $\textsf{R}_d^{(n)}$ are the excited states of the right-handed up- and 
right-handed down-type quarks fields, respectively with different right-handed zero modes. Also 
in (\ref{Yukawa4}) $\textsf{L}_q^{(m)}$ contains the excited states of the left-handed quarks. After 
that, we select the coupling between the first excited state of the right-handed up-type quark to 
the Higgs doublet and the left-handed doublet of SM quarks from the same family as follow
 
\begin{equation}\label{Yukawa5}
-\sum_{j=1}^N y_{uj}(h_R^{j-1,1}(c_{\textsf{R}_u}))^{*}h_L^{j,0}(c_{\textsf{L}_q})b^j\overline{\textsf{R}}_u^{(1)}\widetilde{H}^\dagger \textsf{L}_q^{(0)}+\mathrm{h.c.},
\end{equation}

\noindent The results are shown in Table (\ref{tabela1}). For instance, considering the third generation of SM quarks, i.e., 
using (\ref{Yukawa5}), with the localization parameters $c_{t_R}$, $c_{t_L}$ and (\ref{eigenvector-ZM-pi-T4567}), 
we obtained the couplings displayed in the column labeled '$t^{(1)}_R\,t^{(0)}_L$' 
for quiver theories with $N=4$ or $N=15$. On the other hand 
the column labeled '$u^{(1)}_R\,u^{(0)}_L$' were obtained by using the values of the localization parameters 
$c_{u_R}$, $c_{u_L}$ (corresponding to the first generation of SM quarks).\newline
\newline We also considered the coupling between the right-handed up-type SM quark to the Higgs doublet and the 
first excited state of the left-handed doublet of quarks associated to the same family as follow
\begin{equation}\label{Yukawa6}
-\sum_{j=1}^N y_{uj}(h_R^{j-1,0}(c_{\textsf{R}_u}))^{*}h_L^{j,1}(c_{\textsf{L}_q})b^j\overline{\textsf{R}}_u^{(0)}\widetilde{H}^\dagger \textsf{L}_q^{(1)}+\mathrm{h.c.},
\end{equation}
\noindent The results are also shown in Table (\ref{tabela1}). For instance, considering the third generation of SM quarks, i.e., 
using (\ref{Yukawa6}), with the localization parameters $c_{t_R}$, $c_{t_L}$ and (\ref{eigenvector-ZM-pi-T4567}), 
for the Higgs wave function, we obtained the couplings displayed in the column labeled '$t^{(1)}_L\,t^{(0)}_R$' 
for quiver theories with $N=4$ or $N=5$. Similarly, the values of the couplings in the column labeled 
'$u^{(1)}_L\,u^{(0)}_R$' was obtained by using the 
values of the localization parameters $c_{u_R}$, $c_{u_L}$ (corresponding to the first generation of SM quarks). The next 
selection of couplings, that we have considered is between the right-handed down-type SM quark to the Higgs doublet and 
the first excited state of the left-handed doublet of quarks associated to the same family

\begin{equation}\label{Yukawa7}
-\sum_{j=1}^N y_{dj}(h_R^{j-1,0}(c_{\textsf{R}_d}))^{*}h_L^{j,1}(c_{\textsf{L}_q})b^j\overline{\textsf{R}}_d^{(0)}H^\dagger \textsf{L}_q^{(1)}+\mathrm{h.c.}.
\end{equation}
The evaluation for this couplings are also shown in Table (\ref{tabela1}). For instance, considering the third generation of SM 
quarks, i.e., using (\ref{Yukawa7}), with the localization parameters $c_{t_R}$, $c_{t_L}$ and 
(\ref{eigenvector-ZM-pi-T4567}), we obtained the couplings displayed in the column labeled '$t^{(1)}_L\,b^{(0)}_R$' 
for quiver theories with $N=4$ or $N=15$. Similarly, by considering the first generation and showed in the column 
labeled '$u^{(1)}_L\,d^{(0)}_R$', where the localization parameters $c_{d_R}$, $c_{u_L}$ were used.\newline
\newline Finally, we also considered the coupling between the first excited state right-handed down-type quark to the 
Higgs doublet and the left-handed doublet of SM quarks associated to the third generation of SM quarks. This coupling was calculated 
by using
\begin{equation}\label{Yukawa8}
-\sum_{j=1}^N y_{dj}(h_R^{j-1,1}(c_{\textsf{R}_d}))^{*}h_L^{j,0}(c_{\textsf{L}_q})b^j\overline{\textsf{R}}_d^{(1)}H^\dagger \textsf{L}_q^{(0)}+\mathrm{h.c.},
\end{equation}

\noindent and the results are shown in the column labeled '$b^{(1)}_R\,t^{(0)}_L$' in Table (\ref{tabela1}).\newline
\newline As can be seen in Table (\ref{tabela1}), the larger couplings are found in the third generation. 
Once again, this agrees with the overlap of the wave-functions associated with the localization 
parameters ($c_{t_L}$, $c_{t_R}$ and $c_{b_R}$, the wave-functions associated with these localization parameters 
were obtained in Appendix (\ref{WF-Excited})) and with the (\ref{eigenvector-ZM-pi-T4567}).


\begin{table}[H]
 \begin{center}
\scalebox{0.9}{
\begin{tabular}{|c |c|c| c| c| c|c|c|}
\hline\hline
$N$& $t^{(1)}_R\,t^{(0)}_L$ & $t^{(1)}_L\,t^{(0)}_R$ & $u^{(1)}_R\,u^{(0)}_L$  & $u^{(1)}_L\,u^{(0)}_R$ & $t^{(1)}_L\,b^{(0)}_R$ & $u^{(1)}_L\,d^{(0)}_R$ & $b^{(1)}_R\,t^{(0)}_L$\\
\hline
$4$& $0.365$ & $0.028$ & $4.1\times10^{-9}$ & $0.00173$ & $0.041$ & $3.02\times10^{-6}$ & $2.64\times10^{-4}$ \\
$15$& $0.179$ & $0.352$ & $3.4\times10^{-5}$ & $3.02\times10^{-4}$ &
$0.0143$ & $1.28\times10^{-7}$ & $0.001$ \\
\hline
\end{tabular}
}
\end{center}
\captionsetup{font=small, labelfont=bf, labelsep=period}
\caption{Couplings of fermions from different towers to the Higgs doublet, for allowed modes 0 and 1.}\label{tabela1}
\end{table}

Now that we have calculated the couplings in this section, we will study in the next chapter the phenomenology 
of the excited fermions.



\section{General Effective Theory for New Fermions}\label{generalEFT-VLQ}

Our aim is to provide alternatives to the study of the phenomenology involving new heavy fermions that, besides quiver theories there are also many 
others Vector-like quarks theories. So it is of interest to study Vector-like quark phenomenology in a general model-independent way. 
Generically, we can consider vector-like quarks as being multiplets of $SU(2)_L$ as shown in Refs.~\citep{AguilarSaavedra:2009es,Cacciapaglia:2010vn,
Okada:2012gy}, such that these new fermions will be coupled to SM fermions and gauge bosons through the Yukawa terms and kinetic terms, respectively. 
Our procedure will be according to Ref.~\citep{Cacciapaglia:2010vn}, where these new fermions couple to the third generations of SM quarks. 
In this section, the cases of a singlet vector-like up-type and SM doublet will be studied. For the first case, we will use the bound associated 
to the $tbW$ coupling~\citep{Chatrchyan:2012ep} and for the second one, the constraint associated to the 
decay $Z\rightarrow b\bar{b}$~\citep{Olive:2016xmw} will be used. It is due to after the EWSB 
the mixing between the vector-like quarks and the third generation 
of the SM quarks induce deviations on the couplings associated with these measurements. In addition to these cases, for each one the interaction 
with a heavy gluon will be included.\newline\newline\newline\newline\newline

\subsection{Vector-like quark SU(2)-singlet up-type, T}\label{subvlqsinglet}

Let us consider a vector-like up-type fermion $T'$, a singlet of $SU(2)_L$ with hypercharge equal to $2/3$. It couples to the SM quarks through the Yukawa 
couplings as follows

\begin{equation}\label{YukawaVLTLquarksm}
\mathcal{L}_{\mathrm{Y(VLT)}}= - Y_{i\beta}^u ~\overline{q}'_{Li} \tilde{\Phi} u'_{R\beta} -  Y_{ij}^d~ \overline{q}'_{Li} \Phi d'_{Rj}   + \mathrm{h.c.},
\end{equation}

\noindent where $i,j=~1,2,3$ correspond to indexes of generations in the SM ($q'_{Li}$ and $u'_{Ri}$ are $SU(2)_L$-doublet and -singlet, respectively) 
and the index $\beta=~1,2,3,4$, such that $u'_{4}=T'$. To indicate that the fermions are not in their 
mass eigenstate primes are used. In addition, there is a vector-like term of the form

\begin{equation}\label{MassVLTL}
\mathcal{L}_{\mathrm{M(VLT)}}= - M ~\overline{T}'_{L} T'_{R}+\mathrm{h.c.}.
\end{equation}

\noindent The Higgs doublet is given by 

\begin{equation}\label{doublet-higgs}
\Phi= \frac{1}{\sqrt{2}}\left (
\begin{array}{c}
0\\
v +H
\end{array} \right ).
\end{equation}

\begin{figure}[H]
    \begin{center}
    \resizebox{4.0in}{!}
    {
    \epsfig{file=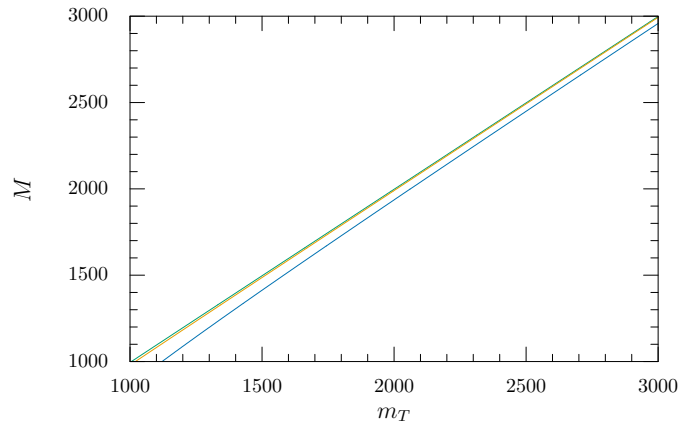}
    }
    \captionsetup{font=small, labelfont=bf, labelsep=period}
    \caption{\label{mTSinglet}Correlation between the mass vector-like and $m_T$, for $\dfrac{v}{\sqrt{2}}|Y^u_{34}|=100$, $200$ and $500$ from left to right.}
    \end{center}
   \end{figure}

\noindent We substitute (\ref{doublet-higgs}) in (\ref{YukawaVLTLquarksm}) and considering the mixing between $T'$ and the third generation, 
we can identify the mass term that is written as follows

\begin{equation} \mathcal{L}_{Mass}=-\,\left(
\renewcommand{\arraystretch}{1.6}
\begin{array}{cc}
\overline{t}'_{L} & \overline{T}'_{L}\\
\end{array}
\right){\bf M_{t'T'}}
\left(
\renewcommand{\arraystretch}{1.6}
\begin{array}{c}
t'_R \\
T'_R \\
\end{array}
\right)+\mathrm{h.c.},
\end{equation}

\noindent where 

\begin{equation}\label{MassMatrixTSinglet1}
{\bf M_{t'T'}}=\left(
\renewcommand{\arraystretch}{1.6}
\begin{array}{cc}
\dfrac{v}{\sqrt{2}}Y^u_{33} & \dfrac{v}{\sqrt{2}}Y^u_{34}\\
0     &  M \\
\end{array}
\right).
\end{equation}
\noindent We have that the masses in the mass eigenstate are given by
\begin{align}
m^2_t=&\dfrac{M^2+\dfrac{v^2}{2}\left(|Y^u_{33}|^2+|Y^u_{34}|^2\right)-\sqrt{\left(M^2+\dfrac{v^2}{2}\left(|Y^u_{33}|^2+|Y^u_{34}|^2\right)\right)^2-2v^2M^2|Y^u_{33}|^2}}{2}\label{mtopSinglet}\\
m^2_T=&\dfrac{M^2+\dfrac{v^2}{2}\left(|Y^u_{33}|^2+|Y^u_{34}|^2\right)+\sqrt{\left(M^2+\dfrac{v^2}{2}\left(|Y^u_{33}|^2+|Y^u_{34}|^2\right)\right)^2-2v^2M^2|Y^u_{33}|^2}}{2},\label{MTSinglet}
\end{align}
by using (\ref{mtopSinglet}), (\ref{MTSinglet}), jointly with 
$v^2\ll M^2$, we can expand the square roots in powers of $\left(\dfrac{v}{M}\right)^2$ as follow
\begin{align}
m^2_t=&\dfrac{M^2+\dfrac{v^2}{2}\left(|Y^u_{33}|^2+|Y^u_{34}|^2\right)-M^2\left[1+\dfrac{1}{2}\left(\dfrac{v}{M}\right)^2\left(|Y^u_{34}|^2-|Y^u_{33}|^2\right)+\mathcal{O}\left(\dfrac{v}{M}\right)^4\right]}{2}\label{expansionmtopSinglet}\\
m^2_T=&\dfrac{M^2+\dfrac{v^2}{2}\left(|Y^u_{33}|^2+|Y^u_{34}|^2\right)+M^2\left[1+\dfrac{1}{2}\left(\dfrac{v}{M}\right)^2\left(|Y^u_{34}|^2-|Y^u_{33}|^2\right)+\mathcal{O}\left(\dfrac{v}{M}\right)^4\right]}{2},\label{expansionMTSinglet}
\end{align}
and then $m^2_t$ and $m^2_T$ go as $\left(\dfrac{v}{\sqrt{2}}Y^u_{33}\right)^2$ and $M^2+\left(\dfrac{v}{\sqrt{2}}Y^u_{34}\right)^2$, respectively. 
\noindent Therefore, this means that $m_T$ is greater than the unphysical vector-like mass. Also, using (\ref{mtopSinglet}) and (\ref{MTSinglet}), 
we can deduce the relation 
\begin{equation}\label{MassRelation}
m^2_T= M^2\left(1+\dfrac{\dfrac{v^2}{2}|Y^u_{34}|^2}{M^2-m^2_t}\right).
\end{equation}
This relation is shown in Figure (\ref{mTSinglet}). We see that the differences between $m_T$ and $M$ decrease for large $M$, whereas for small $M$, 
theses differences increase.

Now, we write ${\bf M_{t'T'}}$ as factorized as follows
\begin{equation}\label{MassMatrixTSinglet2} {\bf M_{t'T'}}=V^\dagger_L\,\left(
\renewcommand{\arraystretch}{1.6}
\begin{array}{cc}
m_t & 0\\
0     &  m_T \\
\end{array}
\right)V_R,
\end{equation}
where $V_{L,R}$ are given by
\begin{equation}\label{rotationLR-Tsinlet}
V_{L,R}=\left(
\renewcommand{\arraystretch}{1.6}
\begin{array}{cc}
\cos\theta^{L,R}_u & \sin\theta^{L,R}_u\\
-\sin\theta^{L,R}_u    &  \cos\theta^{L,R}_u \\
\end{array}
\right),
\end{equation} 
in such a way that the next relations are obtained by using (\ref{rotationLR-Tsinlet}), (\ref{MassMatrixTSinglet2}) and (\ref{MassMatrixTSinglet1})
\begin{align}
\dfrac{v}{\sqrt{2}}Y^u_{33}\cos\theta^{L}_u\cos\theta^{R}_u-\dfrac{v}{\sqrt{2}}Y^u_{34}\cos\theta^{L}_u\sin\theta^{R}_u+M\sin\theta^{L}_u\sin\theta^{R}_u&=m_t,   \nonumber\\
\dfrac{v}{\sqrt{2}}Y^u_{33}\sin\theta^{L}_u\cos\theta^{R}_u-\dfrac{v}{\sqrt{2}}Y^u_{34}\sin\theta^{L}_u\sin\theta^{R}_u-M\cos\theta^{L}_u\sin\theta^{R}_u&=0,   \nonumber\\
\dfrac{v}{\sqrt{2}}Y^u_{33}\cos\theta^{L}_u\sin\theta^{R}_u-\dfrac{v}{\sqrt{2}}Y^u_{34}\cos\theta^{L}_u\cos\theta^{R}_u-M\sin\theta^{L}_u\cos\theta^{R}_u&=0,   \nonumber\\
\dfrac{v}{\sqrt{2}}Y^u_{33}\sin\theta^{L}_u\sin\theta^{R}_u-\dfrac{v}{\sqrt{2}}Y^u_{34}\sin\theta^{L}_u\cos\theta^{R}_u+M\cos\theta^{L}_u\cos\theta^{R}_u&=m_T,\label{RelationsVLTSinglet}
\end{align}
where, assuming real Yukawa couplings, we have
\begin{equation}\label{sLuTsinglet}
\sin\theta^{L}_u=\dfrac{vM}{\sqrt{2}}\dfrac{Y^u_{34}}{\sqrt{\left(M^2-m_t^2\right)^2+\dfrac{v^2M^2|Y^u_{34}|^2}{2}}}
\end{equation} 
and 
\begin{equation}\label{RelationsRusLu}
\sin\theta^{R}_u=\dfrac{m_t}{M}\sin\theta^{L}_u.
\end{equation}
\noindent To obtain the couplings between $T$ to the SM third generations and the EW gauge bosons we first write the fermion-gauge boson interaction 
in the EW basis 
\begin{align}\label{TsingletEW}
& \left(
\renewcommand{\arraystretch}{1.6}
\begin{array}{cc}
\overline{t}'_{L} & \overline{b}'_{L}\\
\end{array}
\right) ~i \gamma^\mu \left(- ig A^a_\mu\tau^a -ig' B_\mu Y\right)\left(
\renewcommand{\arraystretch}{1.6}
\begin{array}{c}
t'_{L}\\
b'_{L}\\
\end{array}
\right)\nonumber \\
 &+\overline{t}'_{R} ~ i \gamma^\mu \left( -i g' B_\mu Y \right) t'_{R}+
\overline{b}'_{R} ~ i \gamma^\mu \left( -i g' B_\mu Y \right) b'_{R}\nonumber \\
 &+\overline{T}'_{L,R} ~ i \gamma^\mu \left( -i g' B_\mu Y \right) T'_{L,R},
\end{align}
 
\noindent where $A^a_\mu$ and $B_\mu$ are the gauge bosons of $SU(2)_L$ and $U(1)_Y$ respectively. In the following we use the vector mass eigenstates 
jointly with the rotations defined by (\ref{rotationLR-Tsinlet}) and, furthermore we will 
neglect the prime in $b'_L$, because we have considered the mixing between the up sector of the SM third generation and $T'$ according to 
(\ref{YukawaVLTLquarksm}) where the down-type quarks are not affected by the mixing. After these substitutions in (\ref{TsingletEW}), we obtain in the mass eigenstates

\begin{align}
&\left(\cos\theta^{L}_u\frac{g}{\sqrt{2}}\bar{t}_L\slashed{W}^+ b_L+\mathrm{h.c.}\right)\nonumber\\
&+\frac{g}{2c_W}\bar{t}\slashed{Z}\left(\cos^2\theta^{L}_u P_L-\frac{4}{3}s_W^2\right)t\nonumber\\
&+\frac{g}{2c_W}\bar{b}\slashed{Z}\left(-P_L+\frac{2}{3}s_W^2\right)b\nonumber\\
&+\frac{g}{2c_W}\bar{T}\slashed{Z}\left(\sin^2\theta^{L}_u P_L-\frac{4}{3}s_W^2\right)T\nonumber\\
&+\left(\sin\theta^{L}_u\frac{g}{\sqrt{2}}\overline{T}_L\slashed{W}^+ b_L+\mathrm{h.c.}\right)\nonumber\\
&+\left(\cos\theta^{L}_u\sin\theta^{L}_u\frac{g}{2c_W}\bar{t}\slashed{Z} P_LT+\mathrm{h.c.}\right)\nonumber\\
&+\frac{2gs_W}{3}\bigg(\bar{t}\slashed{A}t+\bar{T}\slashed{A}T\bigg)-\frac{gs_W}{3}\bigg(\bar{b}\slashed{A}b\bigg),\label{VLTsingletEW1}
\end{align}

\noindent while the Higgs interaction involving $T$ and the SM third generation comes from the Yukawa couplings (\ref{YukawaVLTLquarksm}), 
by using (\ref{rotationLR-Tsinlet}), as follows

\begin{align}
&-\frac{gm_T}{2m_W}\sin^2\theta^{L}_u\overline{T}TH\nonumber\\
&-\frac{gm_T}{2m_W}\overline{T}\left(\frac{m_t}{m_T}\cos\theta^{L}_u\sin\theta^{L}_uP_R+\cos\theta^{L}_u\sin\theta^{L}_uP_L\right)tH+\mathrm{h.c.}\nonumber\\
&-\frac{gm_t}{2m_W}\cos^2\theta^{L}_u\bar{t}tH-\frac{gm_b}{2m_W}\cos^2\theta^{L}_u\bar{b}bH.\label{VLTsingletEW2}
\end{align}
\noindent We can use a CMS measurement of single top cross sections at 7 TeV~\citep{Chatrchyan:2012ep}, that is the $V_{tb}$ ($tbW$ coupling), such that 
$\lvert V_{tb}\arrowvert>0.92$, which results in the constraint

\begin{equation}\label{bound1sLu}
\lvert\sin\theta^{L}_u\arrowvert<0.4. 
\end{equation}

\noindent We point out that $\sin^2\theta^{L}_u$ is always further suppressed by $\dfrac{v}{M}$. The QCD interaction of the Vector-like quark is just as in 
the SM:
$T$, the interaction following is considered

\begin{equation}
\mathcal{L}_{gTT}= g_s\bar{T}\gamma_\mu\textsf{T}^ag^{\mu,a}T.\\
\end{equation}

\noindent Finally, we also allow for the interaction of the Vector-like quark with a heavy gluon $G$, given by
\begin{equation}
\mathcal{L}_{GTt}= g_s\left(f_L\bar{t}_L\gamma_\mu\textsf{T}^aG^{\mu,a}T_L+f_R\bar{t}_R\gamma_\mu\textsf{T}^aG^{\mu,a}T_R+\mathrm{h.c.}\right),\\
\end{equation}

\noindent where $g_sf_L$ and $g_sf_R$ are left- and right-handed couplings between $t$, $T$ and $G$.\newline\newline\newline\newline\newline\newline\newline
\subsection{Vector-like quark SM SU(2)-doublet}

Considering a vector-like SM $\left(
\renewcommand{\arraystretch}{1.6}
\begin{array}{cc}
T' & B'\\
\end{array}
\right)^{{\bf T}}$, where it is a doublet of $SU(2)_L$ and has hypercharge $1/6$, this couples to the SM quarks in the weak eigenstate through the Yukawa 
couplings as follow

\begin{equation}\label{YukawaVLTBquarksm}
\mathcal{L}_{\mathrm{Y(VLT)}}= - Y_{\alpha i}^u ~\overline{q}'_{L\alpha} \tilde{\Phi} u'_{Ri} -  Y_{\alpha j}^d~ \overline{q}'_{L\alpha} \Phi d'_{Rj}   + \mathrm{h.c.},
\end{equation}

where $i,j=~1,2,3$ correspond to indexes of generations in the SM ($q'_{Li}$ is a doublet of $SU(2)_L$ and $u'_{Ri}$ is a 
singlet of $SU(2)_L$) and the index $\alpha=~1,2,3,4$, such that $q'_{L,R4}=\left(
\renewcommand{\arraystretch}{1.6}
\begin{array}{cc}
T_{L,R}' & B_{L,R}'\\
\end{array}
\right)^{{\bf T}}$, to indicate that the fermions are not in their 
mass eigenstate primes are used, jointly with the mass vector-like term 

\begin{equation}\label{MassVLTB}
\mathcal{L}_{\mathrm{M(VLT)}}= - M \overline{q}'_{L4} q_{R4}+\mathrm{h.c.}.
\end{equation}

\noindent Thus we substitute (\ref{doublet-higgs}) in (\ref{YukawaVLTBquarksm}) and considering the mixing between $T'$, $B'$ 
and the SM third generation, we can identify the mass term that is written as follows

\begin{equation}\label{MassMatrixTBDoublet1} \mathcal{L}_{Mass}=\,\left(
\renewcommand{\arraystretch}{1.6}
\begin{array}{cc}
\overline{t}'_{L} & \overline{T}'_{L}\\
\end{array}
\right){\bf M^u}
\left(
\renewcommand{\arraystretch}{1.6}
\begin{array}{c}
t'_R \\
T'_R \\
\end{array}
\right)+\left(
\renewcommand{\arraystretch}{1.6}
\begin{array}{cc}
\overline{b}'_{L} & \overline{B}'_{L}\\
\end{array}
\right){\bf M^d}
\left(
\renewcommand{\arraystretch}{1.6}
\begin{array}{c}
b'_R \\
B'_R \\
\end{array}
\right)+\mathrm{h.c.},
\end{equation}

Now, we write ${\bf M^{u,d}}$ as factorized as follows
\begin{equation}\label{MassMatrixTBDoublet2} {\bf M^{u,d}}=V^{u,d\dagger}_L\,\left(
\renewcommand{\arraystretch}{1.6}
\begin{array}{cc}
m_{t,b} & 0\\
0     &  m_{T,B} \\
\end{array}
\right)V^{u,d}_R+\mathrm{h.c.},
\end{equation}
where $V^{u,d}_{L,R}$ are given by
\begin{equation}\label{rotationLR-TBdoublet}
V^{u,d}_{L,R}=\left(
\renewcommand{\arraystretch}{1.6}
\begin{array}{cc}
\cos\theta^{L,R}_{u,d} & \sin\theta^{L,R}_{u,d}\\
-\sin\theta^{L,R}_{u,d}    &  \cos\theta^{L,R}_{u,d} \\
\end{array}
\right).
\end{equation}
we also have the relations 
\begin{equation}\label{MassRelationTB}
M^2_{T,B}= M^2\left(1+\dfrac{\dfrac{v^2}{2}|Y^{u,d}_{34}|^2}{M^2-m^2_{t,b}}\right).
\end{equation}
Here the next relations are obtained by using (\ref{rotationLR-TBdoublet}), (\ref{MassMatrixTBDoublet2}) and (\ref{MassMatrixTBDoublet1})
\begin{align}
\dfrac{v}{\sqrt{2}}Y^{u,d}_{33}\cos\theta^{L}_{u,d}\cos\theta^{R}_{u,d}-\dfrac{v}{\sqrt{2}}Y^{u,d}_{34}\cos\theta^{L}_u\sin\theta^{R}_{u,d}+M\sin\theta^{L}_{u,d}\sin\theta^{R}_{u,d}&=m_{t,b},   \nonumber\\
\dfrac{v}{\sqrt{2}}Y^{u,d}_{33}\sin\theta^{L}_{u,d}\cos\theta^{R}_{u,d}-\dfrac{v}{\sqrt{2}}Y^{u,d}_{34}\sin\theta^{L}_{u,d}\sin\theta^{R}_{u,d}-M\cos\theta^{L}_{u,d}\sin\theta^{R}_{u,d}&=0,   \nonumber\\
\dfrac{v}{\sqrt{2}}Y^{u,d}_{33}\cos\theta^{L}_{u,d}\sin\theta^{R}_{u,d}-\dfrac{v}{\sqrt{2}}Y^{u,d}_{34}\cos\theta^{L}_{u,d}\cos\theta^{R}_{u,d}-M\sin\theta^{L}_{u,d}\cos\theta^{R}_{u,d}&=0,   \nonumber\\
\dfrac{v}{\sqrt{2}}Y^{u,d}_{33}\sin\theta^{L}_{u,d}\sin\theta^{R}_{u,d}-\dfrac{v}{\sqrt{2}}Y^{u,d}_{34}\sin\theta^{L}_{u,d}\cos\theta^{R}_{u,d}+M\cos\theta^{L}_{u,d}\cos\theta^{R}_{u,d}&=m_{T,B},\label{RelationsVLTDoublet}
\end{align}
where
\begin{equation}
\sin\theta^{R}_{u,d}=\dfrac{vM}{\sqrt{2}}\dfrac{Y^{u,d}_{43}}{\sqrt{\left(M^2-m_{t,b}^2\right)^2+\dfrac{v^2M^2|Y^{u,d}_{43}|^2}{2}}}
\end{equation} 
and 
\begin{equation}\label{RelationAnglesTB-doublet}
\tan\theta^{L}_{u,d}=\dfrac{m_{t,b}}{M_{T,B}}\tan\theta^{R}_{u,d}.
\end{equation}
To obtain the couplings between $T$, $B$ to the SM third generations and the EW gauge bosons, we first write the fermion-gauge boson interaction in the EW basis 
\begin{align}\label{TBdoubletEW}
&~~\left(
\renewcommand{\arraystretch}{1.6}
\begin{array}{cc}
\overline{t}'_{L} & \overline{b}'_{L}\\
\end{array}
\right) ~i \gamma^\mu \left(- ig A^a_\mu\tau^a -ig' B_\mu Y\right)\left(
\renewcommand{\arraystretch}{1.6}
\begin{array}{c}
t'_{L}\\
b'_{L}\\
\end{array}
\right)\nonumber \\
 &+\overline{t}'_{R} ~ i \gamma^\mu \left( -i g' B_\mu Y \right) t'_{R}+
\overline{b}'_{R} ~ i \gamma^\mu \left( -i g' B_\mu Y \right) b'_{R}\nonumber \\
 &+\left(
\renewcommand{\arraystretch}{1.6}
\begin{array}{cc}
\overline{T}'_{L} & \overline{B}'_{L}\\
\end{array}
\right) ~i \gamma^\mu \left(- ig A^a_\mu\tau^a -ig' B_\mu Y\right)\left(
\renewcommand{\arraystretch}{1.6}
\begin{array}{c}
T'_{L}\\
B'_{L}\\
\end{array}
\right)\nonumber \\
 &+\left(
\renewcommand{\arraystretch}{1.6}
\begin{array}{cc}
\overline{T}'_{R} & \overline{B}'_{R}\\
\end{array}
\right) ~i \gamma^\mu \left(- ig A^a_\mu\tau^a -ig' B_\mu Y\right)\left(
\renewcommand{\arraystretch}{1.6}
\begin{array}{c}
T'_{R}\\
B'_{R}\\
\end{array}
\right),
\end{align}
in the following we rewrite all interaction terms\footnote{The passing from the EW to the mass eigenstate can be seen in Appendix (\ref{EW-MassDoubletCase}) by using 
the definitions given in (\ref{mixingdoublet1}) and (\ref{mixingdoublet2}).} in the vector mass eigenstates, to do this the following definitions will be considered

\begin{align}
&c^{u,d}_{L,R}\equiv\cos\theta^{L,R}_{u,d},\label{mixingdoublet1}\\
&s^{u,d}_{L,R}\equiv\sin\theta^{L,R}_{u,d},\label{mixingdoublet2}
\end{align}

Now, we write in the mass eigenstate the Higgs interaction involving $T$, $B$ and the SM third generations,
which comes from the Yukawa couplings (\ref{YukawaVLTBquarksm}), 
by using (\ref{rotationLR-TBdoublet}), as follows

\begin{align}
&-\frac{gm_T}{2m_W}(s^{u}_{R})^2\overline{T}TH\nonumber\\
&-\frac{gm_B}{2m_W}(s^{d}_{R})^2\overline{B}BH\nonumber\\
&-\frac{gm_T}{2m_W}\overline{T}\left(\frac{m_t}{m_T}c^{u}_{R}s^{u}_{R}P_L+s^{u}_{R}c^{u}_{R}P_R\right)tH+\mathrm{h.c.}\nonumber\\
&-\frac{gm_B}{2m_W}\overline{B}\left(\frac{m_b}{m_B}c^{d}_{R}s^{d}_{R}P_L+s^{d}_{R}c^{d}_{R}P_R\right)bH+\mathrm{h.c.}\nonumber\\
&-\frac{gm_t}{2m_W}(c^{u}_{R})^2\bar{t}tH-\frac{gm_b}{2m_W}(c^{d}_{R})^2\bar{b}bH,\label{VLTBdoubletEW3}
\end{align}

\noindent Thus, we can identify from (\ref{VLTBdoubletEW1}) the coupling to study the single production of $T$ via $W$, that is given by 

\begin{equation}
 \frac{g}{\sqrt{2}}\sin\left(\theta^{u}_{L}-\theta^{d}_{L}\right)\overline{T}_L\slashed{W}^+ b_L-\frac{g}{\sqrt{2}}\cos\theta^{u}_{R}\sin\theta^{d}_{R}\overline{T}_R\slashed{W}^+ b_R+\mathrm{h.c.}
\end{equation}

Notice that the angle $\theta^{d}_{L}$ is negligible, because is suppressed by the $m_b$ according to (\ref{RelationAnglesTB-doublet}).


In the mass eigenstate basis we have also found terms as

\begin{equation}\label{constraintDoublet1}
-\frac{g}{2c_W}\bar{b}_L\slashed{Z}b_L+\frac{gs_W^2}{3c_W}\bar{b}\slashed{Z}b-\frac{g}{2c_W}(s^{d}_{R})^2\bar{b}_R\slashed{Z}b_R,
\end{equation}

and 

\begin{equation}\label{constraintDoublet2}
\frac{g}{\sqrt{2}}\left(c^{u}_{L}c^{d}_{L}+s^{u}_{L}s^{d}_{L} \right)\bar{t}_L\slashed{W}^+ b_L+\mathrm{h.c.}
\end{equation}
respectively.

First, we can compute the $s^{d}_{R}$, by using the constraint associated to $Z\rightarrow b\bar{b}$, $R_b=0.21629 \pm 0.00066$ that is found 
in Ref.~\citep{Olive:2016xmw}, the width is as follows

\begin{align}\label{Zbbar1}
\Gamma(Z\rightarrow b\bar{b})=&\frac{e^2 \left(m_Z^2 \left(m_Z^2-4 m_b^2\right)\right)^{1/2}}{288 \pi  c_W^2 s_W^2 m_Z^3}\times\left(-9 c_W^4 m_b^2-42 c_W^2 s_W^2 m_b^2\right.\nonumber\\
&\left.-54 c_W^2 (s^{d}_{R})^2 m_b^2-17 s_W^4 m_b^2-30 s_W^2 (s^{d}_{R})^2 m_b^2-9 (s^{d}_{R})^4 m_b^2+9 c_W^4 m_Z^2\right.\nonumber\\
&+\left.6 c_W^2 s_W^2 m_Z^2+5 s_W^4 m_Z^2+12 s_W^2 (s^{d}_{R})^2 m_Z^2+9 (s^{d}_{R})^4 m_Z^2\right)
\end{align}

then we obtain, 
\begin{equation}\label{sRd}
0.09<s^{d}_{R}<0.16~\text{or}~-0.16<s^{d}_{R}<-0.09.
\end{equation}

Now, we used it jointly with (\ref{RelationAnglesTB-doublet}) and $M_B=2250GeV$, obtaining

\begin{equation}\label{sLd}
0.0002<s^{d}_{L}<0.0003~\text{or}~-0.0003<s^{d}_{R}<-0.0002.
\end{equation}

After that, we use (\ref{sLd}) in (\ref{constraintDoublet2}), and applying the measurement of the $V_{tb}$ ($tbW$ coupling) given in Ref.~\citep{Chatrchyan:2012ep},

and then we have

\begin{equation}\label{sLu}
c^{u}_{L}>0.92,
\end{equation}

which results in the constraint

\begin{equation}\label{bound2sLu}
s^{u}_L<0.4. 
\end{equation}

and for the QCD pair production of $T$, the following interaction is considered

\begin{equation}
\mathcal{L}_{gTT}\supset g_s\bar{T}\gamma_\mu\textsf{T}^ag^{\mu,a}T,\\
\end{equation}

and also we will consider the next interaction
\begin{equation}
\mathcal{L}_{GTt}\supset g_s\left(f_L\bar{t}_L\gamma_\mu\textsf{T}^aG^{\mu,a}T_L+f_R\bar{t}_R\gamma_\mu\textsf{T}^aG^{\mu,a}T_R+\mathrm{h.c.}\right),\\
\end{equation}

where $g_sf_{L,R}$ are the left- and right-handed couplings between $T$, $t$ and $G$. Similarly for the QCD pair production of $B$, the following interaction is considered

\begin{equation}
\mathcal{L}_{gBB}\supset g_s\bar{B}\gamma_\mu\textsf{T}^ag^{\mu,a}B,\\
\end{equation}

and also we can consider the next interactions
\begin{equation}
\mathcal{L}_{GBb}\supset g_s\left(g_L\bar{b}_L\gamma_\mu\textsf{T}^aG^{\mu,a}B_L+g_R\bar{b}_R\gamma_\mu\textsf{T}^aG^{\mu,a}B_R+\mathrm{h.c.}\right),\\
\end{equation}

where $g_sg_{L},R$ are the left- and right-handed couplings between $B$, $b$ and $G$. Now that we have know the couplings associated for the single production of T, we will study the production and decay of this heavy state.

\chapter{\bf Vector-Like Heavy Quarks at the LHC}

In this chapter we will study the production and decay of vector-like quarks, motivated by extensions of the SM where the Higgs is a pNGB. 
We will focus on quiver theories as well as a model-independent effective Lagrangian approach. In particular, inspired by quiver theories we will 
add a possible decay mode of the heavy quark into a heavy gluon which has not been considered in the literature. It because the possibility of a 
first exited state of fermion as in quiver theories can be heavier than the first excited gluon as in Ref.~\citep{Burdman:2014ixa}. 
We will first consider the 
phenomenology of the quiver theories and comment on the model-independent case, which only requires a rescaling of our results. We will focus on the phenomenology associated 
to single production of the first excited state of the top quark with left-handed zero 
mode (we denoted it by $T_L$) through the EW interactions because this mode of production becomes dominant over the pair production at higher masses~\citep{Backovic:2015bca}, 
as saw in Table (\ref{tabela1}) the more relevant couplings were found 
in the third generation. Considering the following interactions involving $T_L$
\begin{align}
\mathcal{L}_{gT_LT_L}\supset&g_s\bar{T}_L\gamma_\mu\textsf{T}^ag^{\mu,a}T_L,\label{equ-0}\\
\mathcal{L}_{GT_Lt}\supset&g_s\bar{T}_L\gamma_\mu\textsf{T}^aG^{\mu,a}(g^L_{T_LGt}P_L)t,\label{equ-1}\\
\mathcal{L}_{T_LEW}\supset&-c_{T_Lth}\bar{T}_LHP_Rt+\mathrm{h.c.}\nonumber\\
&+c_{T_LbW}\frac{e}{\sqrt{2}s_W}\bar{T}_L\gamma_\mu W^{\mu,+}P_Lb+\mathrm{h.c.}\nonumber\\
&+c_{T_LZt}\frac{e}{2s_Wc_W}\bar{T}_L\gamma_\mu Z^{\mu}P_Lt+\mathrm{h.c.},\label{equ-2}
\end{align}
where $g^L_{T_LGt}$ and $c_{T_Lth}$ are inferred from Tables (\ref{tabela}) e (\ref{tabela1}), respectively, $G^{\mu,a}$ is the first excited state of 
the gluon, $H$ is the Higgs boson, $ W^{\mu,+}$ and $Z^{\mu}$ 
are the massive EW gauge bosons. Now, to find a relation between $c_{T_Lth}$, $c_{T_LbW}$ and $c_{T_LZt}$ we will use the equivalence theorem. 
Firstly, to obtain a the relation between $c_{T_Lth}$ and $c_{T_LZt}$, the widths of $T_L$ to Higgs-top and Z-top will be assumed to be equal at high 
energies, for $m_T\gg m_t,~m_h,~m_Z$. The $T_L\rightarrow Zt$ partial width is given by

\begin{align}\label{equ-3}
\Gamma_{Zt}=&\frac{1}{96\pi m_{T_L}^3}(m^4_t-2m^2_tm^2_{T_L}-2m^2_tm^2_{Z}+m^4_{T_L}-2m^2_{T_L}m^2_{Z}+m^4_{Z})^{1/2}\times\nonumber\\
&\left[\frac{3e^2}{4m^2_Zc^2_Ws^2_W}m^4_t-\frac{3e^2}{2m^2_Zc^2_Ws^2_W}m^2_tm^2_{T_L}+\frac{3e^2}{4c^2_Ws^2_W}m^2_t+\frac{3e^2}{4m^2_Zc^2_Ws^2_W}m^4_{T_L}\right.\nonumber\\
&+\left.\frac{3e^2}{4c^2_Ws^2_W}m^2_{T_L}-\frac{3e^2}{2c^2_Ws^2_W}m^2_Z\right](c_{T_LZt})^2
\end{align}

expanding in powers of $\frac{m_t}{m_{T_L}}$ or $\frac{m_Z}{m_{T_L}}$, we obtain

\begin{align}\label{equ-4}
\Gamma_{Zt}=&\frac{m_{T_L}}{96\pi}\Bigg\{1+\mathcal{O}\left[\left(\frac{m_t}{m_{T_L}}\right)^2\right]+\mathcal{O}\left[\left(\frac{m_Z}{m_{T_L}}\right)^2\right]\Bigg\}\left[\frac{3e^2}{4c^2_Ws^2_W}\left(\frac{m^2_t}{m_Zm_{T_L}}\right)^2\right.\nonumber\\
&-\frac{3e^2}{2c^2_Ws^2_W}\left(\frac{m_t}{m_Z}\right)^2+\frac{3e^2}{4c^2_Ws^2_W}\left(\frac{m_t}{m_{T_L}}\right)^2+\frac{3e^2}{4c^2_Ws^2_W}\left(\frac{m_{T_L}}{m_Z}\right)^2\nonumber\\
&+\left.\frac{3e^2}{4c^2_Ws^2_W}-\frac{3e^2}{2c^2_Ws^2_W}\left(\frac{m_Z}{m_{T_L}}\right)^2\right](c_{T_LZt})^2.
\end{align}

The partial width corresponding to the decay mode of $T_L$ into $h$ $t$ is as follows

\begin{align}\label{equ-5}
\Gamma_{ht}=&\frac{1}{96\pi m_{T_L}^3}(m^4_h-2m^2_hm^2_{t}-2m^2_hm^2_{T_L}+m^4_{t}-2m^2_{t}m^2_{T_L}+m^4_{T_L})^{1/2}\times\nonumber\\
&\left[-3m^2_{h}+3m^2_{t}+3m^2_{T_L}\right](c_{T_Lth})^2,
\end{align}

this can be expanded in powers of $\frac{m_h}{m_{T_L}}$ or $\frac{m_h}{m_{T_L}}$, we obtain

\begin{align}\label{equ-6}
\Gamma_{ht}=&\frac{m_{T_L}}{96\pi}\Bigg\{1+\mathcal{O}\left[\left(\frac{m_h}{m_{T_L}}\right)^2\right]+\mathcal{O}\left[\left(\frac{m_t}{m_{T_L}}\right)^2\right]\Bigg\}\times\nonumber\\
&\left[-3\left(\frac{m_h}{m_{T_L}}\right)^2_{h}+3\left(\frac{m_t}{m_T}\right)^2+3\right](c_{T_Lth})^2.
\end{align}

The Goldstone equivalence theorem as in Ref~\citep{Backovic:2015bca} implies that the ratios of the branching fractions of $T_L\rightarrow$ $th$, $tZ$ 
and $W^+b$ are approximately 1:1:2. Then, for $m_{T_L}\gg m_h,m_t,m_W$ we have $\Gamma_{Zt}\simeq\Gamma_{ht}$, which results in a 
relation between $c_{T_LZt}$ and $c_{T_Lth}$ as follows


\begin{equation}\label{equ-7}
 \frac{e}{2c_Ws_W}\left(\frac{m_{T_L}}{m_Z}\right)c_{T_LZt}=c_{T_Lth}.
\end{equation}

In addition, to obtain the relation between $c_{T_LtZ}$ and $c_{T_LbW}$, we write the $T_L\rightarrow W^+b$ partial width,

\begin{align}\label{equ-8}
\Gamma_{W^+b}=&\frac{1}{96\pi m_{T_L}^3}(m^4_b-2m^2_bm^2_{T_L}-2m^2_bm^2_{W}+m^4_{T_L}-2m^2_{T_L}m^2_{W}+m^4_{W})^{1/2}\times\nonumber\\
&\left[\frac{3e^2}{2s^2_W}m^4_b+\frac{3e^2}{2s^2_W}m^2_{T_L}+\frac{3e^2}{2s^2_Wm^2_W}m^4_b-\frac{3e^2}{s^2_Wm^2_W}m^2_bm^2_{T_L}\right.\nonumber\\
&+\left.\frac{3e^2}{2s^2_Wm^2_W}m^4_{T_L}-\frac{3e^2}{s^2_W}m^2_W\right](c_{T_LbW})^2.
\end{align}

In an analogous way to the previous expansion (\ref{equ-5}), but now expanding in powers of $\frac{m_W}{m_{T_L}}$ or $\frac{m_b}{m_{T_L}}$, 
we obtain

\begin{align}\label{equ-9}
\Gamma_{W^+b}=&\frac{m_{T_L}}{96\pi}\Bigg\{1+\mathcal{O}\left[\left(\frac{m_b}{m_{T_L}}\right)^2\right]+\mathcal{O}\left[\left(\frac{m_W}{m_{T_L}}\right)^2\right]\Bigg\}\left[\frac{3e^2}{2s^2_W}\left(\frac{m_b}{m_{T_L}}\right)^2\right.\nonumber\\
&+\frac{3e^2}{2s^2_W}+\frac{3e^2}{2s^2_W}\left(\frac{m^2_b}{m_{W}m_{T_L}}\right)^2-\frac{3e^2}{s^2_W}\left(\frac{m_{b}}{m_W}\right)^2\nonumber\\
&+\left.\frac{3e^2}{2s^2_W}\left(\frac{m_{T_L}}{m_W}\right)^2-\frac{3e^2}{2s^2_W}\left(\frac{m_W}{m_{T_L}}\right)^2\right](c_{T_LbW})^2.
\end{align}

Again we use the equivalence theorem to establish the relation between $c_{T_LZt}$ and $c_{T_LbW}$, imposing that $\Gamma_{Zt}\simeq\frac{1}{2}\Gamma_{W^+b}$, 
which results in

\begin{equation}\label{equ-10}
 \frac{1}{c_W}\left(\frac{m_{W}}{m_Z}\right)c_{T_LtZ}=c_{T_LbW}.
\end{equation}

Notice that the coupling $c_{T_Lth}=Y_{T_Lth}$, where $Y_{T_Lth}/\sqrt{2}$ is inferred from the third column in Table (\ref{tabela1}). 
Using (\ref{equ-6}), (\ref{equ-7}), we can study the EW single $T_L$ production for instance at the LHC. Figure (\ref{singleLHCTL}) shows the 
single production of $T_L$ through fusion of $W$ $b$, i.e., involving $c_{T_LbW}$. We will consider both the model-independent case as well as the 
case of quiver theories. On the one hand we will consider the couplings given in Ref.~\citep{Backovic:2015bca}, here we point out that $c_{T_LbW}$ is 
actually $\sin\theta^{L}_{u}$ mentioned in Subsection (\ref{subvlqsinglet}), for the study of the single production of the heavy fermion. 
On the other hand, from the quiver theory the couplings given in Table (\ref{tabela1}) will be also considered. According to quiver theories, 
once we fix the parameters, e.g., $N$, $v_1$ and $v_N$, we are ready to compute all the couplings 
of the excited fermions, relevant for their production and decay at the LHC. In the next section we consider the pair production and the 
single production of $T_L$.
\section{Heavy $T_L$ Production at LHC}
\begin{figure}[bt]
    \begin{center}
    \resizebox{3.0in}{!}
    {
    \epsfig{file=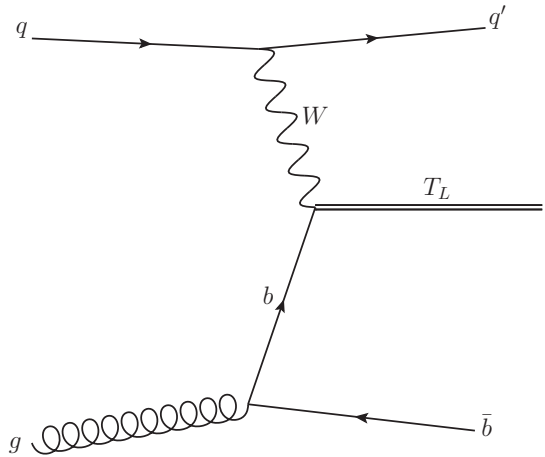}
    }
    \captionsetup{font=normalsize, labelfont=bf, labelsep=period}
    \caption{Production diagram for the process pp$\rightarrow T_L\bar{b}j$ with W, inferred from (\ref{equ-2}).}
    \label{singleLHCTL}
    \end{center}
\end{figure}

In this section, we consider $T_L$ production. On the one hand, there is a heavy quark pair production via QCD, where these pairs come from quark-antiquark and 
gluon-gluon fusion~\citep{Agashe:2014kda,Barger}. On the other hand, we consider the single production channel of $T_L$ via the $W$ $b$ EW fusion. This production mode is shown 
in Figure (\ref{singleLHCTL}).\newline\newline 
We computed the pair and single $T_L$ production cross section at the LHC with $s=\sqrt{13}$ by using 
MadGraph~\citep{Alwall:2011uj}. 
The first production mode is model-independent since is just QCD pair production. Figure (\ref{PairSingle13TeV}) shows these production modes as a 
function of $m_{T_L}$.
We considered the single production channel of $T_L$ via the $W$ exchange process pp$\rightarrow T_L\bar{b}j$ as can be seen in the top panel of 
Figure (\ref{PairSingle13TeV}), the black line shows the pair production of $T_L$ via QCD, whereas the other lines correspond to 
the single production for fixed values of the couplings $c_{T_LbW}\lesssim 0.4$ as considered in Ref.~\citep{Backovic:2015bca}. It shows 
that the single production dominates for $m_{T_L}$ greater than 1 TeV.\newline

Also, the $T_L$ production as a function of the mass of $T_L$ in quiver theories was computed. As shown in the bottom panel of 
Figure (\ref{PairSingle13TeV}), the pair $T_L$ production is the same as in the top panel case. While, the single production cross section 
is given for the cases N=4 and N=15 by the medium turquoise and dark turquoise lines, 
respectively. It is shown that the pair production dominates for $m_{T_L}$ masses below 1 TeV 
for both cases N=4 and N=15; however the single production dominates for $m_{T_L}$ greater 
than 1 TeV and 3.9 Tev for N=15 and N=4 respectively.


\section{Analysis Heavy $T_L$ Decay Modes}

Here, we examine the $T_L$ decay to the Higgs sector and the additional way in which $T_L$ can decay, i.e., 
the decay mode $T_L\rightarrow Gt$, where $G$ the color-octet. To study the contribution of $G$ to the signal, we will first compute the mass difference 
between $T_L$ and $G$, i.e. $\Delta m$, for quiver models couplings. In order to have comparable branching ratios, the following condition is required 

\begin{equation}\label{equ-11}
\Gamma_{W^+b}\lesssim\Gamma_{Gt},
\end{equation}

where $\Gamma_{Gt}$, is the $T_L\rightarrow Gt$ partial width and is given by

\begin{align}\label{equ-10}
\Gamma_{Gt}=&\frac{1}{24 \pi  m_G^2  m_{T_L} ^3}(m_G^4-2 m_G^2 m_{t}^2-2 m_G^2 m_{T_L}^2+m_{t}^4-2 m_t^2 m_{T_L}^2+m_{T_L}^4)^{1/2}\times\nonumber\\
&\left[ m_{T_L}^4+ m_G^2 m_t^2-2  m_G^4+ m_G^2 m_{T_L}^2+ m_t^4-2  m_t^2 m_{T_L}^2\right](g_{T_LGt})^2.
\end{align}

\noindent We consider quiver models with $N=4(15)$, with couplings $g_{T_LGt}=0.85(0.83)$ and $Y_{t_LHt}=0.028(0.352)$, respectively. Notice that the coupling 
$c_{T_Lht}=Y_{t_LHt}/\sqrt{2}$. Now, we can substitute (\ref{equ-8}), (\ref{equ-10}) in (\ref{equ-11}) and considering $G$ mass to be $2$~TeV, and then, 
after solving for $m_{T_L}$, the mass differences are listed in Table (\ref{tabela2}).

\begin{table}[H]
\centering
\begin{tabular}{|c | c | c| c|}
\hline\hline
 N & $m_G$ [GeV]& min $m_{T_L}$ [GeV]& $\Delta m$ \\
\hline\hline
$4$&$2000$& $2172$ & $172$  \\
\hline
$15$&$2000$& $2278$ & $279$   \\
\hline
\end{tabular}
\captionsetup{font=normalsize, labelfont=bf, labelsep=period}
\caption{Masses involved in the decay modes of $T_L$ to the color-octet and EW sector for $N=4,15$.}\label{tabela2}\end{table}

From the information of the bottom panel of Figure (\ref{PairSingle13TeV}) and Table (\ref{tabela2}), we conclude that to produce a heavy $T_L$ 
decaying to $G$ with reasonable branching ratio, we must consider the single production as the dominant channel for $N=15$.

\begin{table}[H]
\vspace{-1.0cm}
\begin{figure}[H]
    \begin{center}
    \resizebox{5.5in}{!}
    {
    \epsfig{file=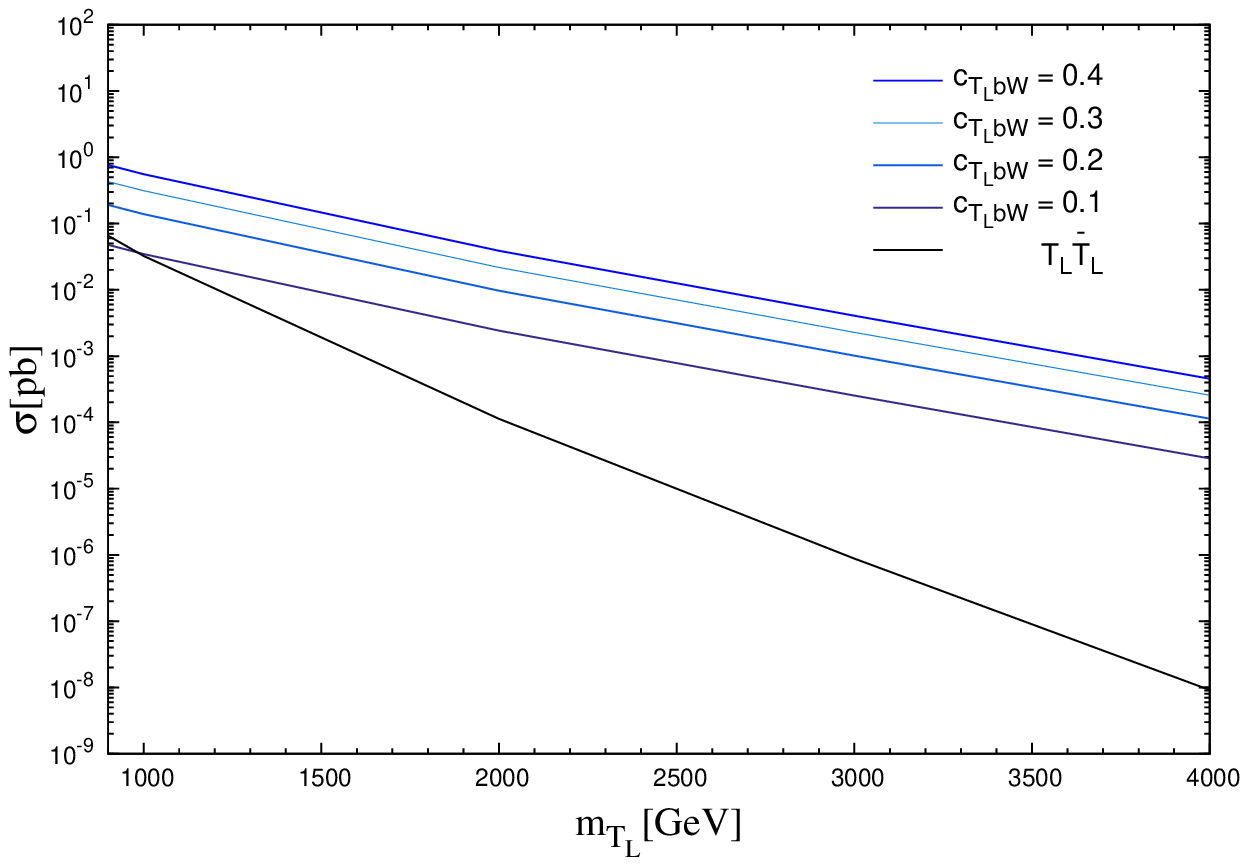}
    }
    \label{figure}
    \end{center}
    \end{figure}
\vspace{-1.0cm}
    \begin{figure}[H]
    \begin{center}
    \resizebox{5.5in}{!}
    {
    \epsfig{file=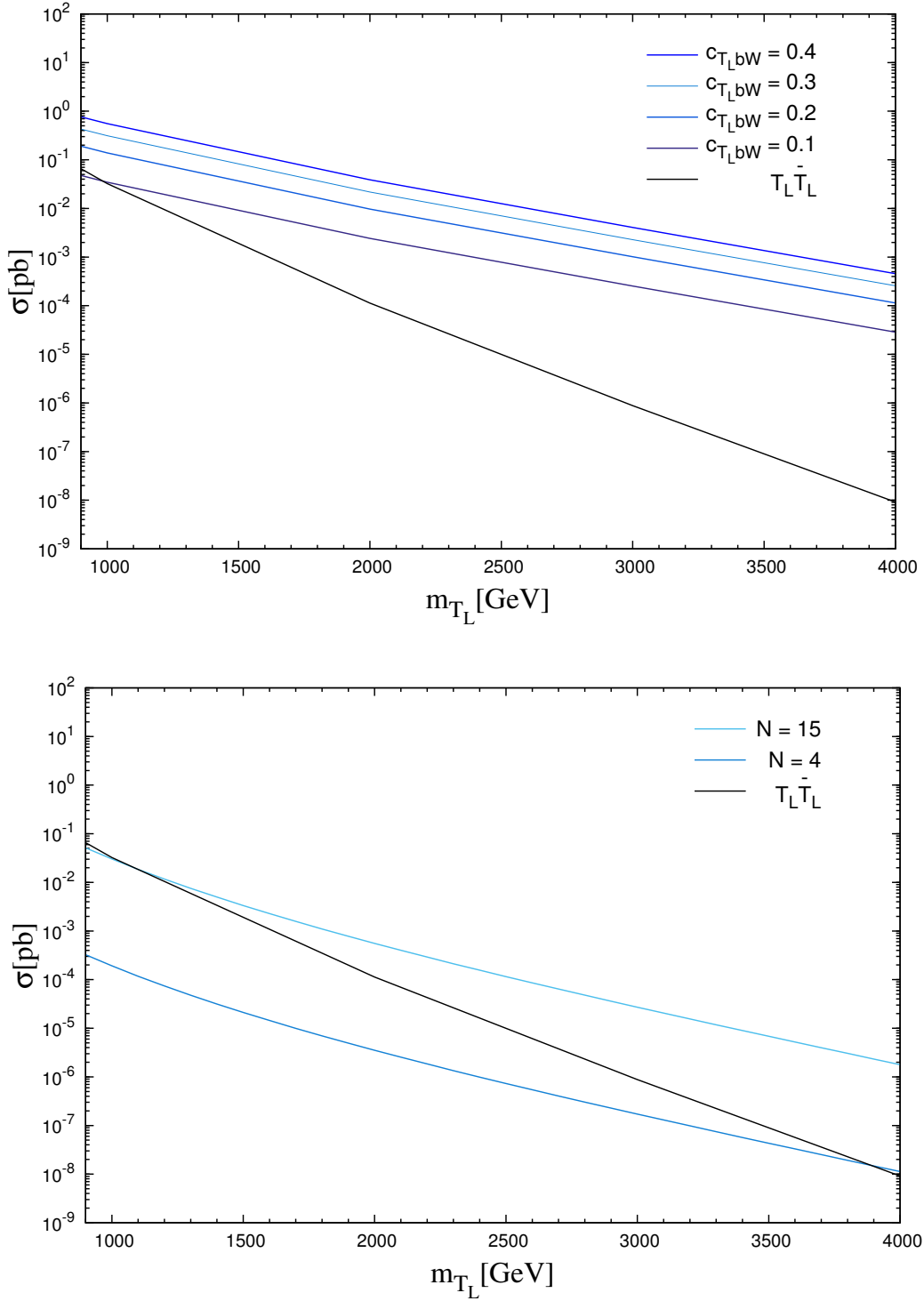}
    }
    \captionsetup{font=normalsize, labelfont=bf, labelsep=period}
    \caption{Cross sections for QCD pair and single production for $T_L$ at LHC with $\sqrt{s}$ = 13 TeV as a function of 
    $m_{T_L}$. For fixed values of $c_{T_{L}bW}$ (top panel) and couplings from quiver theories with $N=4,15$ (bottom panel). In the later 
    $c_{T_Lht}$ was inferred from the third column in Table (\ref{tabela1}) and $c_{T_{L}bW}$ vary according to (\ref{equ-7}) and (\ref{equ-10}) 
    with respect to the $T_L$ mass.}
    \label{PairSingle13TeV}
    \end{center}
    \end{figure}
\end{table}

\section{Prospects for the LHC}\label{Sec3.3}
The previous section has shown that according the condition (\ref{equ-11}), is possible to study the case where both decay modes are comparable. 
The decay modes into the Higgs and EW gauge bosons were studied in the literature
~\citep{AguilarSaavedra:2009es,Cacciapaglia:2010vn,Backovic:2015bca,Aguilar-Saavedra:2013qpa}. Here, we will concentrate on the decay mode 
$T_L\rightarrow Gt$. We will consider both the $N=15$ quiver theory as well as the model-independent case. 
We opted as signal process the production of that heavy top decaying into 
$G$, which afterwards decay into a pair of bottom quarks, i.e. pp$\rightarrow T_Lbj\rightarrow Gtbj\rightarrow b\bar{b}bWbj$. 
To study the feasibility of detecting the signal we consider the example of single 
top production pp$\rightarrow tbj\rightarrow Wbbj$ as a background, this simple background was considered to estimate the required luminosity for the 
discovery of our signal. The couplings considered 
in for our signal were based on quiver theories with $N=15$ and 
the masses were chosen to be $m_T$ and $m_G$ as $2.27$ and $2$ TeV, respectively. 
The cross sections both for signal and background were computed at $\sqrt{s}$ = 13 TeV pp collider jointly with a cut given by 
$H_T>500$ GeV (similar to Ref.~\citep{Backovic:2015bca}), where $H_T$ is the 
total transverse hadronic energy and defined by
\begin{equation}\label{equ-12}
H_T= \sum_{hadronic~particles}\lVert\vec{p}_T\lVert,
\end{equation}
where $\vec{p}_T$ is the transverse momentum.

\begin{figure}[H]
    \begin{center}
    \resizebox{5.9in}{!}
    {
    \input{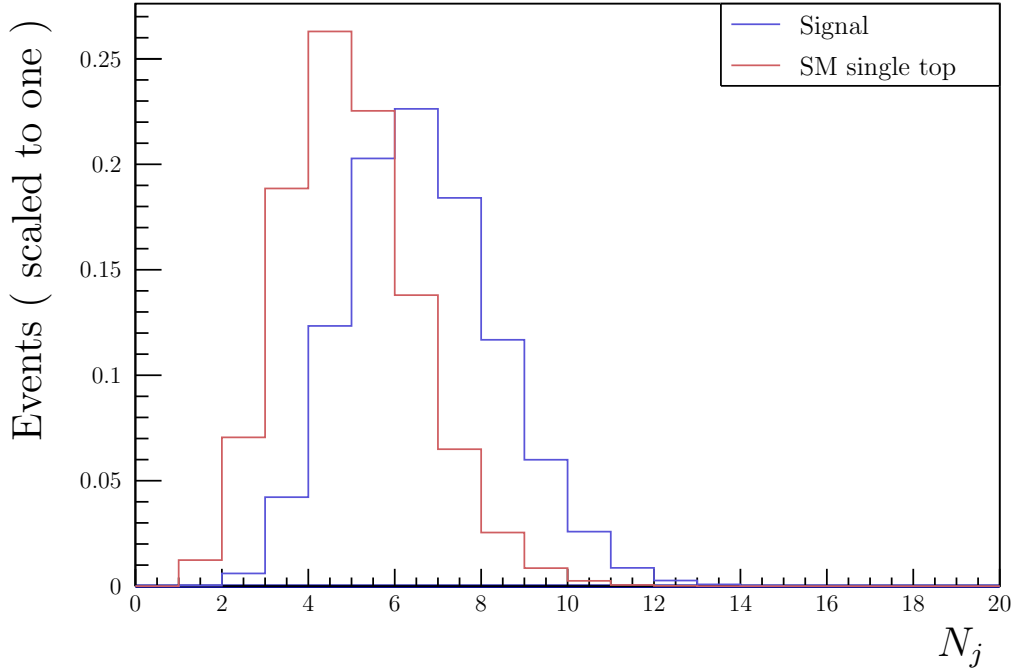}
    }
    \captionsetup{font=normalsize, labelfont=bf, labelsep=period}
    \caption{Number of jets distributions for events normalized to unity before any cuts.
    }
   \label{figureselection0}
    \end{center}
    \end{figure}

\noindent Thus, considering this cut we simulated 50K and 500K events for signal and background, respectively. To generate the events, 
the chain MadGraph~\citep{Alwall:2011uj}-Pythia~\citep{Sjostrand:2014zea}-Delphes~\citep{deFavereau:2013fsa} was used. 
After that, the comparison between the signal and background distributions was obtained by using 
Madanalysis (ma5)~\citep{Conte:2012fm,Conte:2014zja,Dumont:2014tja}. The baseline luminosity was assumed to be $1~ab^{-1}$.  
Then, we implemented other kinematic cuts on the events as can be seen in Table (\ref{tablema5}). It shows a summary of the results of applying 
simple kinematic cuts, the selected cuts were inferred sequentially according to the kinematic distributions showed in 
Figures~(\ref{figureselection0}),~(\ref{figureselection1}), and (\ref{figureselection2}). We considered the Fastjet algorithm anti-kt with 
$\Delta R=0.6$ interfaced in ma5. For details of the recombination algorithm anti-kt see~\citep{Cacciari:2008gp}.

For the first selection after generating events, we took into account the number of jets distributions as shown in Figure~(\ref{figureselection0}). 
To compare the shapes of the curves related to both background and signal datasets, the events were normalized to unity. We select the following cut
\begin{equation}\label{equ-13}
N_j>6, 
\end{equation}
\begin{table}[H]
\centering
\begin{tabular}{| c | c|c|}
\hline
 $ $ & Signal & SM single top  \\
\hline\hline
Cross section & $5.2\times10^{-5}$ pb & $0.2$ pb\\
\hline
Events before cuts& $50\mathrm{K~a.u.}$ & $500\mathrm{K~a.u.}$ \\
\hline
$N_j>6$& $19952$ & $51112$\\
\hline
$H_T>800$ GeV& $ 18272$ & $22290$ \\
\hline
$p_{T,j_1}>400$ GeV& $ 14231$ & $7408$ \\
\hline
Reco. Eff.& $ 0.28$ & $1.48\times10^{-2}$ \\
\hline
\end{tabular}
\captionsetup{font=normalsize, labelfont=bf, labelsep=period}
\caption{Best cut-flow analysis obtained with ma5. Notice that a generator cut of $H_T> 500$ GeV was applied. Fastjet algorithm anti-kt with 
$\Delta R=0.6$.}\label{tablema5}\end{table}
\vspace{-1cm}
where $N_j$, is the number of jets. Figure~(\ref{figureselection1}) shows the total transverse hadronic energy of jets distributions after the selection (\ref{equ-13}). 
To obtain this plot the events also were normalized to unity. So a good discrimination between signal and background can be obtained by using the 
following cut
\begin{equation}\label{equ-14}
H_T>800~\text{GeV}.
\end{equation}

\begin{table}[H]
\vspace{-1.0cm}
\begin{figure}[H]
    \begin{center}
    \resizebox{5.9in}{!}
    {
    \input{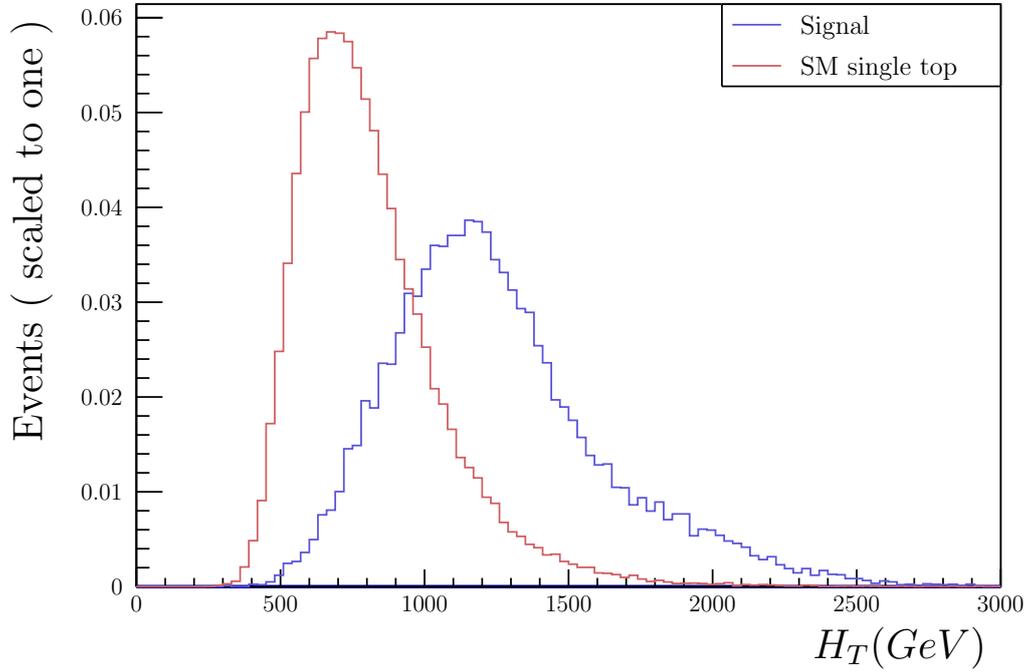}
    }
    \captionsetup{font=normalsize, labelfont=bf, labelsep=period}
    \caption{Total transverse hadronic energy of jets distributions for events normalized to unity after the selection $N_j>6$.
    }
    \label{figureselection1}
    \end{center}
    \end{figure}
\vspace{-1.0cm}
    \begin{figure}[H]
    \begin{center}
    \resizebox{5.9in}{!}
    {
    \input{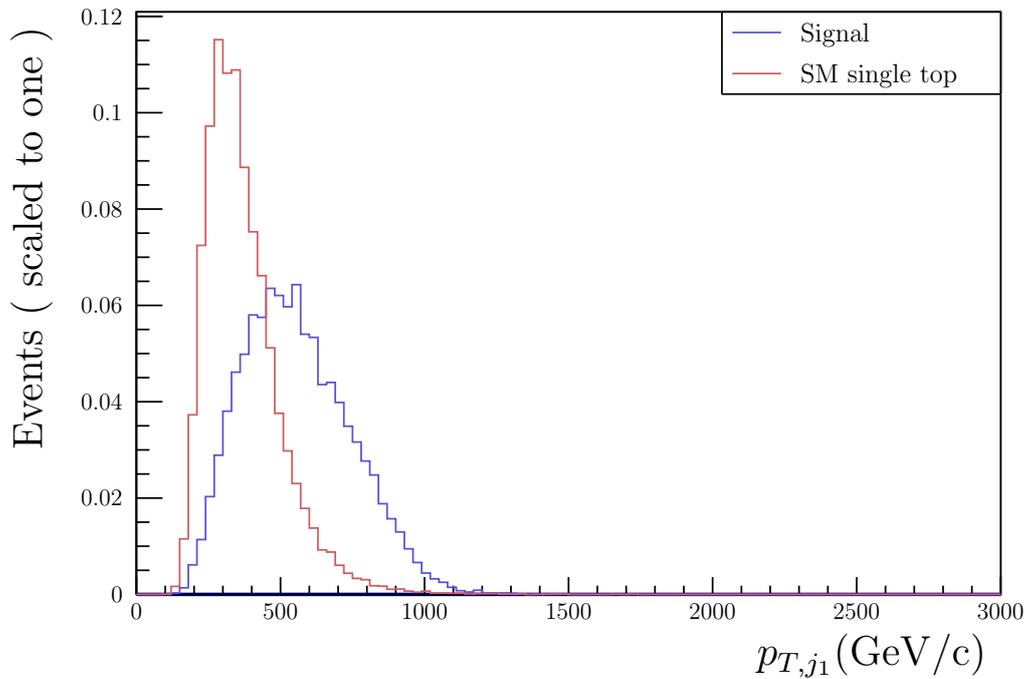}
    }
    \captionsetup{font=normalsize, labelfont=bf, labelsep=period}
    \caption{Transverse-momentum of the hardest jet distributions for events normalized to unity after the selection $H_T>800$ GeV.
    }
    \label{figureselection2}
    \end{center}
    \end{figure} 
\end{table}

After these cuts, we also considered the transverse-momentum of the hardest jet distributions for events as shown in Figure~(\ref{figureselection2}), 
in such a way that to compare both signal and background distributions the events were also normalized to unity in this plot. Then, we select
\begin{equation}\label{equ-15}
p_{T,j_1}>400~\text{GeV},
\end{equation}
where $p_{T,j_1}$ is the transverse-momentum of the hardest jet.




Now, from the values in Table (\ref{tablema5}), we are able to estimate the amount of luminosity that we need in order to have $5\sigma$ discovery. 
To estimate the required luminosity for $5\sigma$ we impose that $S/\sqrt{B}\geqslant 5$, where $S$ and $B$ are the number of events after cuts, which 
scale linearly with the luminosity. Then the previous requirement can be written as,
\begin{align}
\mathcal{L}\geq&~25\frac{\sigma_B\times\epsilon_B}{\left(\sigma_S\times\epsilon_S\right)^2},\label{equ-16}
\end{align}
where $\sigma_B$ and $\sigma_S$ are the background and signal cross sections, respectively. In (\ref{equ-16}) we also have that 
$\epsilon_B$ and $\epsilon_S$ are the efficiencies after cuts (these are given in the last row of Table (\ref{tablema5})) 
for the background and signal, respectively. Then substituting the numerical values in (\ref{equ-16}), we have  

\begin{align}
\geq&~25\frac{0.2\times1.48\times10^{-2}}{\left(5.2\times10^{-5}\times0.28\right)^2}\times\frac{1}{1000}fb^{-1},\nonumber\\
\geq&~\frac{7.4\times10^{-5}}{2.1\times10^{-10}}fb^{-1},\nonumber\\
\geq&~3.52\times10^{5}fb^{-1}~\left(352ab^{-1}\right).\label{equ-16new}
\end{align}
Clearly, this number is too big to be achieved by the LHC, for these parameters the LHC will not be sensitive to quiver theories for high masses of 
the vector-like quarks.

On the other hand, we can estimate the reach of the model-independent scheme presented in Section (\ref{generalEFT-VLQ}). For instance, rescaling the 
calculations above, this is due to the main channel for the signal is proportional to $c^2_{T_LWb}$, so we have that
\begin{align}
\mathcal{L}\geq&~25\frac{\sigma_B\times\epsilon_B}{\left(\sigma_S\times\epsilon_S\right)^2}\times\left(\frac{0.027}{c_{T_LWb}}\right)^4,\label{equ-16follows}
\end{align}

where $0.027$ was the value of $c_{T_LWb}$ for $N=15$. Now, we can see that a $5\sigma$ discovery 
can be achieved with the following values of the parameters: $c_{T_LWb}\geq~0.16$ for 
$300fb^{-1}$, $c_{T_LWb}\geq~0.12$ for $1000fb^{-1}$ and $c_{T_LWb}\geq~0.09$ for a High-Luminosity Large Hadron Collider (HL-LHC) with $3000fb^{-1}$.\newline

\noindent The objects considered are too heavy, such that one possibility to study these kind of particles will be consider pp colliders, 
at high energies, for instance at $\sqrt{s}$ = 100 TeV.



    









\chapter{\bf Vector-Like Heavy Quarks at High Energy Colliders}

As we saw in chapter 4 the LHC will not be sensitive for vector-like quark masses $\gtrsim 2 TeV$ in quiver theories. 
And after estimating the coupling between the heavy top, bottom quark with W gauge boson in a model-independent approach, 
we concluded that more energy was need in cases 
where couplings are smaller than 0.09 and with large vector-like quark masses ($>2$ TeV). Here we look at a hypothetical pp collider with 
$\sqrt{s}$ = 100 TeV. Other center-of-mass energies are being considered, such as 27 TeV at the LHC tunnel.

Just as in the previous chapter we use as benchmark the case of the Vector-like quark $SU(2)$-singlet up-type, $T$, in the case where it is coupled to a heavy gluon $G$, such that $T$ can decay 
into $Gt$ with subsequent decays of $G$ into a pair of bottom or top quarks. The electroweak decays of $T$ obey the approximate relations, 

\begin{equation}\label{EWwidth1}
\Gamma(T\rightarrow Ht)\simeq \Gamma(T\rightarrow Zt)\simeq \frac{1}{2}\Gamma(T\rightarrow W^+b), 
\end{equation}

as seen in~\citep{Backovic:2015bca}. 

Although extensive research has been carried out on the EW channels: $T\rightarrow(H,Z)t$ and $T\rightarrow(b_L,W^+)$, 
no study of the $T\rightarrow Gt$ channel was done, we will focus on it. We will take into account that the interactions
between $T$, $t$ and $G$ as well 
as $b$ and $G$ are 
through the qcd coupling.

We considered as signal process the single $T$ production, such that $T$ decays into $Gt$ 
with subsequent decays of $G$ into a quark pair of $b$ and the quark $t$ decaying semi-leptonically.  We took into account the couplings indicated in 
Section (\ref{generalEFT-VLQ}) and the masses were chosen for $T$ and $G$ as $2.3$ and $2$ TeV, respectively.\newline 

\begin{figure}[H]
    \begin{center}
    \resizebox{5.5in}{!}
    {
    \includegraphics{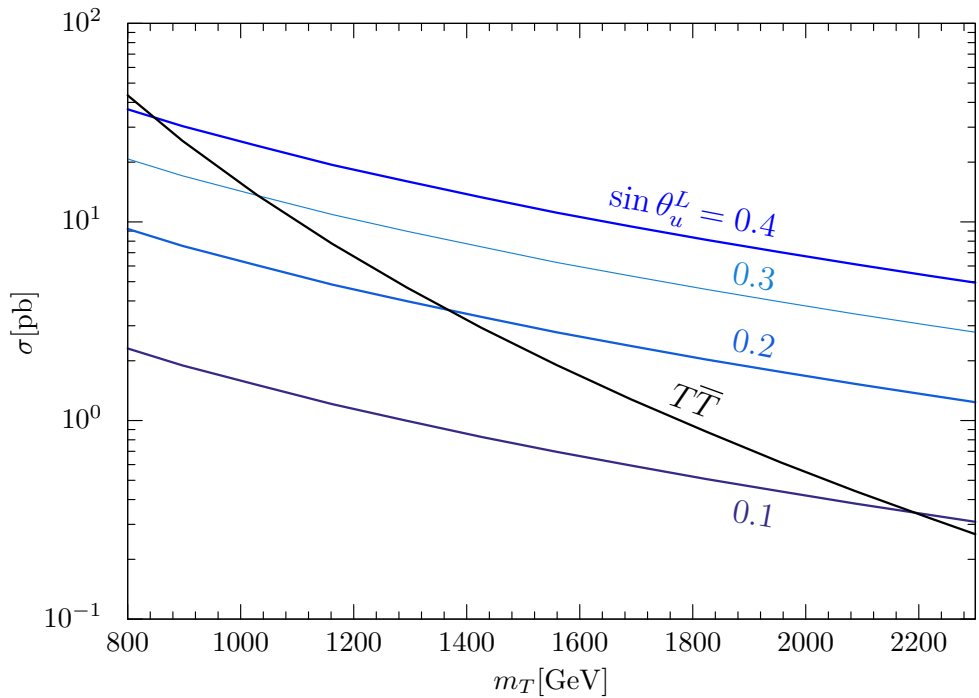}
    }
    \captionsetup{font=small, labelfont=bf, labelsep=period}
    \caption{\label{TTTbar100TeV}Cross section for single and pair $T$ production at 100 TeV \text{pp} collider, as a function of the
 $T$ mass.}
    \end{center}
   \end{figure}

\noindent At this high energies it is still true that the single-$T$ production is dominant over the pair production, as seen in 
Figure (\ref{TTTbar100TeV}). Even for small values of $\sin\theta^{L}_u$ such as $0.1$, single-$T$ production dominates for $m_T>2.2$ TeV, 
at this point of $m_T$ jointly with $\sin\theta^{L}_u = 0.1$, the corresponding 
value of the Yukawa coupling is

\begin{equation}\label{yukawaTsinglet1}
Y^u_{34}=1.315,
\end{equation}

it is obtained by using the relations (\ref{sLuTsinglet}) and (\ref{MassRelation}). The maximum 
value of the coupling $\sin\theta^{L}_u$ comes from the relation $(\ref{bound1sLu})$, and the maximum Yukawa coupling is given in the second 
row of Table (\ref{YukawasSinglet}), this would still be perturbative.\newline
\noindent Table (\ref{YukawasSinglet}) shows the Yukawa couplings that were obtained by using the allowed values of 
$\sin\theta^{L}_u$ according to (\ref{bound1sLu}) as well as the relations given in (\ref{MassRelation}), (\ref{sLuTsinglet}) and considering 
the same mass for $T$ indicated above.

\begin{table}[H]
\centering
\begin{tabular}{| c | c|}
\hline
 $\sin\theta^{L}_u$& $Y^u_{34}$ \\
\hline\hline
$0.4$& $5.25 $ \\
\hline
$0.3$& $3.94$  \\
\hline
$0.2$& $2.63$  \\
\hline
$0.1$& $1.31$  \\
\hline
\end{tabular}
\captionsetup{font=normalsize, labelfont=bf, labelsep=period}
\caption{Values of the Yukawa coupling $Y^u_{34}$ for $m_T=2.3$ TeV as a function of $\sin\theta^{L}_u$.}\label{YukawasSinglet}\end{table}
\vspace{-0.5cm}
\noindent The vector-like heavy quark pair production via QCD was considered, 
where these pairs come from quark-antiquark and gluon-gluon fusion. The pair production dominates for $T$ masses below $0.8$ TeV 
for $T$ as a Vector-like $SU(2)$-singlet; 
however the single production dominates for $m_T$ greater than $1$ TeV, as shown in Figure (\ref{TTTbar100TeV}).\newline

   \noindent The widths and branching ratios of $T$ for the maximum value of $\sin\theta^{L}_u$ are shown in Table (\ref{tableBR04TSinglet}). 
   The branching ratios of $T$ to the EW sector are 
higher, of which the decay of $T$ into a $W^+$ boson and a b quark provides easily the most proportion. Less than $1$ percent of branching ratio consists of the heavy gluon mode.    
   
\begin{table}[H]
\centering
\begin{tabular}{| c | c|c|}
\hline
 $m_{T}=2300$ [GeV] & $\Gamma_T$ [GeV] & Br  \\
\hline\hline
$\Gamma(T\rightarrow W^+b)$& $638.8$ & $0.54$\\
\hline
$\Gamma(T\rightarrow Ht)$& $276.2$ & $0.23$ \\
\hline
$\Gamma(T\rightarrow Zt)$& $265.8$ & $0.22$\\
\hline
$\Gamma(T\rightarrow Gt)$& $4.3$ & $0.005$ \\
\hline
\end{tabular}
\captionsetup{font=normalsize, labelfont=bf, labelsep=period}
\caption{Decays of $T$ to the EW sector and the heavy gluon mode in units of GeV for $m_T=2.3$ TeV, $\sin\theta^{L}_u=0.4$  and $m_G=2$ TeV.}\label{tableBR04TSinglet}\end{table}

\noindent We investigate whether the heavy gluon mode can be observed at this higher energies. 
Branching ratios are similar for other values of $\sin\theta^{L}_u$, as can be seen in Tables (\ref{tableBR03TSinglet}), (\ref{tableBR02TSinglet}) 
and (\ref{tableBR01TSinglet}). Similar results have been found for the branching ratios corresponding to the EW mode. However, for the heavy gluon mode its branching ratio increases 
with the decreases of $\sin\theta^{L}_u$. For instance, for $\sin\theta^{L}_u=0.1$ the EW widths are lower when compared to the other cases, 
this is because the EW widths are increasing with $\sin\theta^{L}_u$.  

\begin{table}[H]
\centering
\begin{tabular}{| c | c|c|}
\hline
$m_{T}=2300$ [GeV] & $\Gamma_T$ [GeV] & Br  \\
\hline\hline
$\Gamma(T\rightarrow W^+b)$& $359.3$ & $0.52$\\
\hline
$\Gamma(T\rightarrow Ht)$& $167.1$ & $0.24$ \\
\hline
$\Gamma(T\rightarrow Zt)$& $160.8$ & $0.23$\\
\hline
$\Gamma(T\rightarrow Gt)$& $6.4$ & $0.009$ \\
\hline
\end{tabular}
\captionsetup{font=normalsize, labelfont=bf, labelsep=period}
\caption{Decays of $T$ to the EW sector and the heavy gluon mode in units of GeV for $m_T=2.3$ TeV, $\sin\theta^{L}_u = 0.3$ and $m_G=2$ TeV.}\label{tableBR03TSinglet}\end{table}
\begin{table}[H]
\centering
\begin{tabular}{| c | c|c|}
\hline
 $m_{T}=2300$ [GeV]& $\Gamma_T$ [GeV] & Br  \\
\hline\hline
$\Gamma(T\rightarrow W^+b)$& $159.7$ & $0.50$\\
\hline
$\Gamma(T\rightarrow Ht)$& $78.3$ & $0.25$ \\
\hline
$\Gamma(T\rightarrow Zt)$& $75.4$ & $0.24$\\
\hline
$\Gamma(T\rightarrow Gt)$& $6.4$ & $0.02$ \\
\hline
\end{tabular}
\captionsetup{font=normalsize, labelfont=bf, labelsep=period}
\caption{Decays of $T$ to the EW sector and the heavy gluon mode in units of GeV for $m_T=2.3$ TeV, $\sin\theta^{L}_u = 0.2$ and $m_G=2$ TeV.}\label{tableBR02TSinglet}\end{table}
\begin{table}[H]
\centering
\begin{tabular}{| c | c|c|}
\hline
$m_{T}=2300$ [GeV] & $\Gamma_T$ [GeV] & Br  \\
\hline\hline
$\Gamma(T\rightarrow W^+b)$& $39.9$ & $0.46$\\
\hline
$\Gamma(T\rightarrow Ht)$& $20.2$ & $0.23$ \\
\hline
$\Gamma(T\rightarrow Zt)$& $19.4$ & $0.23$\\
\hline
$\Gamma(T\rightarrow Gt)$& $6.4$ & $0.07$ \\
\hline
\end{tabular}
\captionsetup{font=normalsize, labelfont=bf, labelsep=period}
\caption{Decays of $T$ to the EW sector and the heavy gluon mode in units of GeV for $m_T=2.3$ TeV, $\sin\theta^{L}_u = 0.1$ and $m_G=2$ TeV.}\label{tableBR01TSinglet}\end{table}

\section{Prospects for Future pp Colliders}   

\begin{figure}[H]
    \begin{center}
    \resizebox{5.9in}{!}
    {
    \input{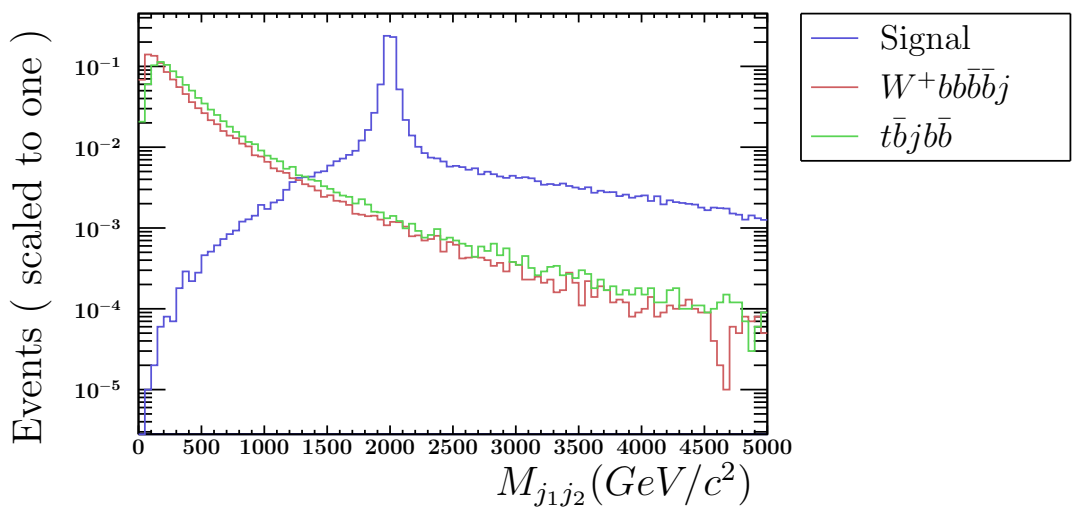}
    }
    \captionsetup{font=normalsize, labelfont=bf, labelsep=period}
    \caption{The invariant mass distribution for the two more energetic jets for the signal with $m_{T,G}=2.3$, $2$ TeV, $\sin\theta^{L}_u=0.4$, and the 
    $W^+bb\bar{b}\bar{b}j$ and $t\bar{b}jb\bar{b}$ backgrounds before cuts.}
    \label{selection04_0}
    \end{center}
\end{figure}

\begin{figure}[H]
    \begin{center}
    \resizebox{5.9in}{!}
    {
    \input{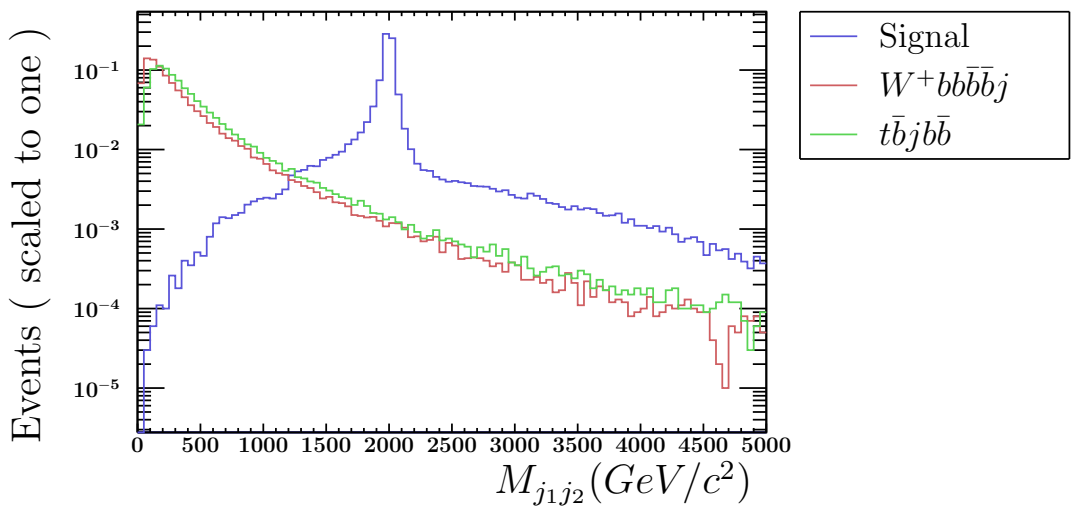}
    }
    \captionsetup{font=normalsize, labelfont=bf, labelsep=period}
    \caption{The mass distribution for the two more energetic jets for the signal with $m_{T,G}=2.3$, $2$ TeV, $\sin\theta^{L}_u=0.2$, and the 
    $W^+bb\bar{b}\bar{b}j$ and $t\bar{b}jb\bar{b}$ backgrounds before cuts.}
    \label{selection02_0}
    \end{center}
\end{figure}

   
In this section a different approach that showed in Section (\ref{Sec3.3}) will be considered, 
this due to the use of $\sqrt{s}$ = 100 TeV pp collider and different 
relations to that considered in (\ref{equ-11}), for instance, as shown in Table (\ref{tableBR04TSinglet}), the widths of $T$ 
to the EW sector are considerably higher than the heavy gluon mode. To study this latter mode of decay at $\sqrt{s}$ = 100 TeV pp collider, 
the signal is pp$\rightarrow T\bar{b}j\rightarrow Gt\bar{b}j\rightarrow b\bar{b}W^+b\bar{b}j\rightarrow b\bar{b}l\nu_lb\bar{b}j$ and the backgrounds considered are 
pp$\rightarrow W^+bb\bar{b}\bar{b}j\rightarrow l\nu_lbb\bar{b}\bar{b}j$ and pp$\rightarrow t\bar{b}jb\bar{b}\rightarrow l\nu_lb\bar{b}jb\bar{b}$. We 
do not consider $\tau$ leptons in the final state because they are harder to identify. 
The cross sections for the signal and backgrounds were computed at $\sqrt{s}$ = 100 TeV pp collider. We simulated 100K events for signal as well as 
backgrounds, these were obtained by using MadGraph. These events were analyzed in Madanalysis, and then the efficiencies were obtained. 
Simple cuts were applied as can be seen in Table (\ref{table1ma5}), which shows a summary of the results. The events before cuts were 
normalized to $3000~fb^{-1}$ of luminosity.\newline\newline 
Firstly, considering $\sin\theta^{L}_u=0.4$, we implemented kinematic cuts on the events as can be seen in Table (\ref{table1ma5}). 
It shows a summary of the results of applying 
simple kinematic cuts, inferred sequentially according to the kinematic distributions showed in 
Figures~(\ref{selection04_0}) and (\ref{selection04_1}).
\begin{table}[H]
\centering
\begin{tabular}{| c | c|c|c|c|}
\hline
 & Signal & {\footnotesize $W^+bb\bar{b}\bar{b}j$}&{\bf\footnotesize$t\bar{b}jb\bar{b}$ }  & Significance\\
\hline\hline
$\sigma$[pb]& $0.0219$ & $13.69$& $5.528$ & $S/\sqrt{B}$\\
\hline
Events before cuts&$65700$&$41070000 $ & $16584000$ & $9$\\
\hline
$M_{j_1j_2} > 1900.0$ GeV& $56607.1$ & $888344$ & $459708$& $49$\\
\hline
$H_T > 2200.0$ GeV& $42641$ & $31213$ & $8623.7$& $214 $ \\
\hline

\end{tabular}
\captionsetup{font=normalsize, labelfont=bf, labelsep=period}
\caption{Sequence of cuts analyzed in ma5 for $\sin\theta^{L}_u$=0.4.}\label{table1ma5}\end{table}
For the first selection in Table (\ref{table1ma5}) after generated events, we took into account the invariant mass distribution for 
the two more energetic jets as shown in 
Figure~(\ref{selection04_0}). 
To compare the shapes of the curves related to both background and signal datasets, the events were normalized to unity. We select the following cut
\begin{equation}\label{equ-17}
M_{j_1j_2} > 1900.0~\text{GeV}, 
\end{equation}
where $M_{j_1j_2}$, is the invariant mass of the two more energetic jets. Figure~(\ref{selection04_1}) shows the distributions of the total 
transverse energy in jets, $H_T$ after the selection (\ref{equ-17}). 
To obtained this plot the events also were normalized to unity. So a good discrimination between signal and background can be obtained by using the 
following cut
\begin{equation}\label{equ-18}
H_T>2200.0~\text{GeV},
\end{equation}
\begin{figure}[H]
    \begin{center}
    \resizebox{5.9in}{!}
    {
    \input{./CHAPTERS/CHAPTER5/Histograms/ANALYSIS04/SelectionCuts/selection_1}
    }
    \captionsetup{font=normalsize, labelfont=bf, labelsep=period}
    \caption{The transverse hadronic energy distribution for the signal with $m_{T,G}=2.3$, $2$ TeV, $\sin\theta^{L}_u=0.4$, and the 
    $W^+bb\bar{b}\bar{b}j$ and $t\bar{b}jb\bar{b}$ backgrounds after the selection $M_{j_1j_2} > 1900.0$ GeV.}
    \label{selection04_1}
    \end{center}
\end{figure}
\noindent Now, assuming $70\%$ $b$-tagging efficiency, 
and then considering $4$ $b$ tags the significance will be 
\begin{equation}\label{significance0.4}
\frac{S}{\sqrt{B}}\simeq50.
\end{equation}  
Secondly, considering $\sin\theta^{L}_u=0.3$, we implemented kinematic cuts on the events as can be seen in Table (\ref{table2ma5}). 
It shows a summary of the results of apply 
simple kinematic cuts, the selected cuts were inferred sequentially according to the kinematic distributions showed in 
Figures~(\ref{selection03_0}) and (\ref{selection03_1}).
\begin{table}[H]
\centering
\begin{tabular}{| c | c|c|c|c|}
\hline
 & Signal & {\small $W^+bb\bar{b}\bar{b}j$}&{\small$t\bar{b}jb\bar{b}$ } & Significance\\
\hline\hline
$\sigma$[pb]& $0.01595$ & $13.69$& $5.528$ & $S/\sqrt{B}$\\
\hline
Events before cuts&$ 47850$&$41070000 $ & $16584000$ & $6 $\\
\hline
$M_{j_1j_2} > 1800$ GeV& $ 42706.6$ & $1004161 $ & $518415 $& $ 35$\\
\hline
$H_T > 2000$ GeV& $34090.7  $ & $50926 $ & $12935 $& $135 $ \\
\hline

\end{tabular}
\captionsetup{font=normalsize, labelfont=bf, labelsep=period}
\caption{Sequence of cuts analyzed in ma5 for $\sin\theta^{L}_u$=0.3.}\label{table2ma5}\end{table}     
For the first selection in Table (\ref{table1ma5}) after generated events, we took into account the invariant mass distribution for 
the two more energetic jets as shown in 
Figure~(\ref{selection03_0}). 
To compare the shapes of the curves related to both background and signal datasets, the events were normalized to unity. We select the following cut
\begin{equation}\label{equ-19}
M_{j_1j_2} > 1800.0~\text{GeV}. 
\end{equation}
\begin{figure}[H]
    \begin{center}
    \resizebox{5.9in}{!}
    {
    \input{./CHAPTERS/CHAPTER5/Histograms/ANALYSIS03/SelectionCuts/selection_0}
    }
    \captionsetup{font=normalsize, labelfont=bf, labelsep=period}
    \caption{The mass distribution for the two more energetic jets for the signal with $m_{T,G}=2.3$, $2$ TeV, $\sin\theta^{L}_u=0.3$, and the 
    $W^+bb\bar{b}\bar{b}j$ and $t\bar{b}jb\bar{b}$ backgrounds before cuts.}
    \label{selection03_0}
    \end{center}
\end{figure}
\begin{figure}[H]
    \begin{center}
    \resizebox{5.9in}{!}
    {
    \input{./CHAPTERS/CHAPTER5/Histograms/ANALYSIS03/SelectionCuts/selection_1}
    }
    \captionsetup{font=normalsize, labelfont=bf, labelsep=period}
    \caption{The transverse hadronic energy distribution for the signal with $m_{T,G}=2.3$, $2$ TeV, $\sin\theta^{L}_u=0.3$, and the 
    $W^+bb\bar{b}\bar{b}j$ and $t\bar{b}jb\bar{b}$ backgrounds after the selection $M_{j_1j_2} > 1800$ GeV.}
    \label{selection03_1}
    \end{center}
\end{figure}
Figure~(\ref{selection03_1}) shows the distributions of the total transverse energy in jets after the selection (\ref{equ-19}). 
To obtained this plot the events also were normalized to unity. So a good discrimination between signal and background can be obtained by using the 
following cut
\begin{equation}\label{equ-20}
H_T>2000.0~\text{GeV},
\end{equation}
\noindent Now, assuming $70\%$ $b$-tagging efficiency, 
and then considering $4$ $b$ tags the significance will be 
\begin{equation}\label{significance0.3}
\frac{S}{\sqrt{B}}\simeq30. 
\end{equation}
Following this, considering $\sin\theta^{L}_u=0.2$, we implemented kinematic cuts on the events as can be seen in Table (\ref{table3ma5}). 
It shows a summary of the results of apply 
simple kinematic cuts, the selected cuts were inferred sequentially according to the kinematic distributions showed in 
Figures~(\ref{selection02_0}) and (\ref{selection02_1}).

\begin{table}[H]
\centering
\begin{tabular}{| c | c|c|c|c|}
\hline
 & Signal & {\small $W^+bb\bar{b}\bar{b}j$}&{\small$t\bar{b}jb\bar{b}$ }  & Significance\\
\hline\hline
$\sigma$[pb]& $0.009938$ & $13.69$& $5.528$ & $S/\sqrt{B}$\\
\hline
Events before cuts&$29814$&$41070000 $ & $16584000$ & $4$\\
\hline
$M_{j_1j_2} > 1800$ GeV& $25922.4$ & $1004161$ & $518415$& $21$\\
\hline
$H_T > 1800$ GeV& $21132.2$ & $70229$ & $17744$& $71$ \\
\hline

\end{tabular}
\captionsetup{font=normalsize, labelfont=bf, labelsep=period}
\caption{Sequence of cuts analyzed in ma5 for $\sin\theta^{L}_u$=0.2.\newline}\label{table3ma5}\end{table}
For the first selection in Table (\ref{table3ma5}) after generated events, we took into account the invariant mass distribution for 
the two more energetic jets as shown in 
Figure~(\ref{selection02_0}). 
To compare the shapes of the curves related to both background and signal datasets, the events were normalized to unity. We select the following cut

\begin{equation}\label{equ-21}
M_{j_1j_2} > 1800.0~\text{GeV}.
\end{equation}
Figure~(\ref{selection02_1}) shows the distributions of the total transverse energy in jets after the selection (\ref{equ-21}). 
To obtained this plot the events also were normalized to unity. So a good discrimination between signal and background can be obtained by using the 
following cut\newline
\begin{equation}\label{equ-22}
H_T>1800.0~\text{GeV},
\end{equation}

\begin{figure}[H]
    \begin{center}
    \resizebox{5.9in}{!}
    {
    \input{./CHAPTERS/CHAPTER5/Histograms/ANALYSIS02/SelectionCuts/selection_1}
    }
    \captionsetup{font=normalsize, labelfont=bf, labelsep=period}
    \caption{The transverse hadronic energy distribution for the signal with $m_{T,G}=2.3$, $2$ TeV, $\sin\theta^{L}_u=0.2$, and the 
    $W^+bb\bar{b}\bar{b}j$ and $t\bar{b}jb\bar{b}$ backgrounds after the selection $M_{j_1j_2} > 1800$ GeV.}
    \label{selection02_1}
    \end{center}
\end{figure}
\noindent Now, assuming $70\%$ $b$-tagging efficiency, 
and then considering $4$ $b$ tags the significance will be    

\begin{equation}\label{significance0.2}
\frac{S}{\sqrt{B}}\simeq17.
\end{equation}



\begin{figure}[H]
    \begin{center}
    \resizebox{5.9in}{!}
    {
    \input{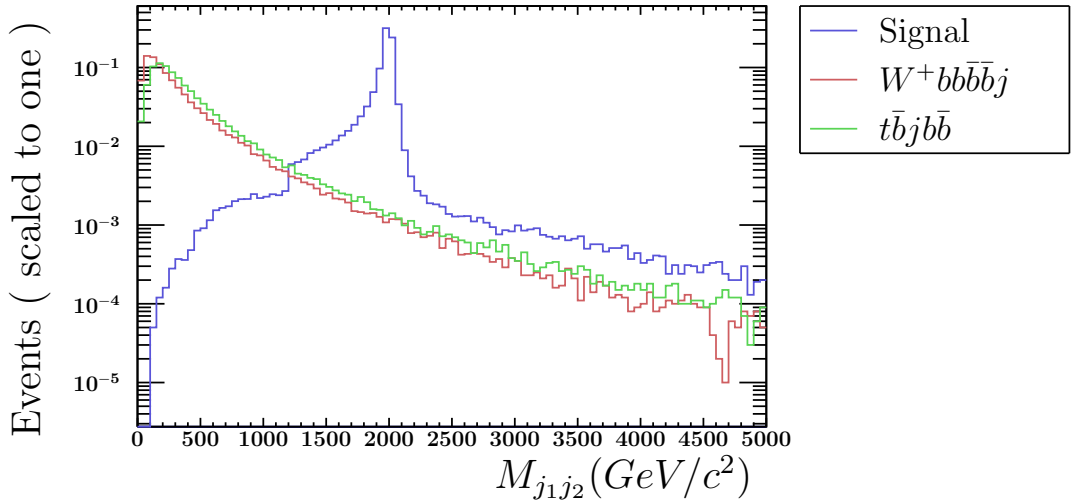}
    }
    \captionsetup{font=normalsize, labelfont=bf, labelsep=period}
    \caption{The mass distribution for the two more energetic jets for the signal with $m_{T,G}=2.3$, $2$ TeV, $\sin\theta^{L}_u=0.1$, and the 
    $W^+bb\bar{b}\bar{b}j$ and $t\bar{b}jb\bar{b}$ backgrounds before cuts.}
    \label{selection01_0}
    \end{center}
\end{figure}

\begin{figure}[H]
    \begin{center}
    \resizebox{5.9in}{!}
    {
    \input{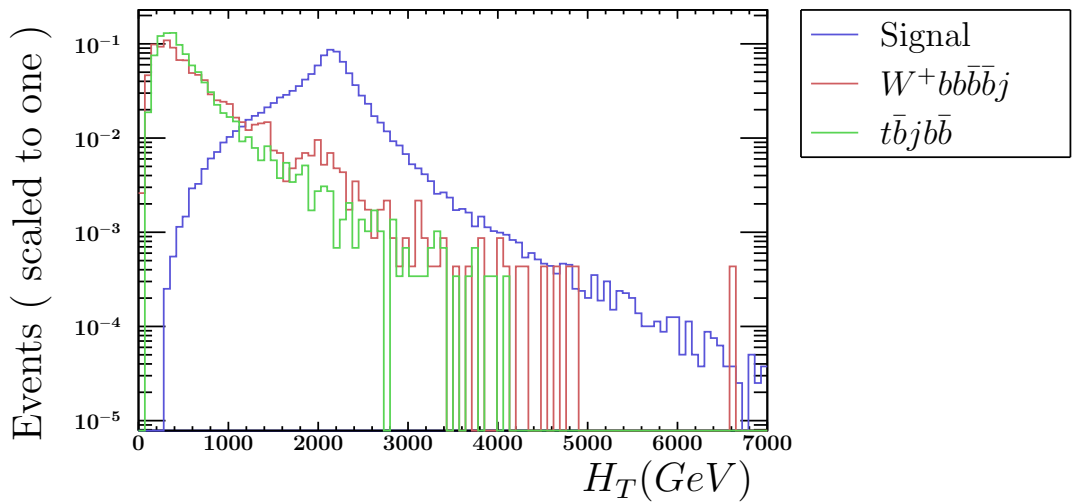}
    }
    \captionsetup{font=normalsize, labelfont=bf, labelsep=period}
    \caption{The transverse hadronic energy distribution for the signal with $m_{T,G}=2.3$, $2$ TeV, $\sin\theta^{L}_u=0.1$, and the 
    $W^+bb\bar{b}\bar{b}j$ and $t\bar{b}jb\bar{b}$ backgrounds after the selection $M_{j_1j_2} > 1850$ GeV.}
    \label{selection01_1}
    \end{center}
\end{figure}

Finally, considering $\sin\theta^{L}_u=0.1$, we implemented kinematic cuts on the events as can be seen in Table (\ref{table4ma5}). 
It shows a summary of the results of apply 
simple kinematic cuts, the selected cuts were inferred sequentially according to the kinematic distributions showed in 
Figures~(\ref{selection01_0}) and (\ref{selection01_1}).
\begin{table}[H]
\centering
\begin{tabular}{| c | c|c|c|c|}
\hline
 & Signal & {\small $W^+bb\bar{b}\bar{b}j$}&{\small$t\bar{b}jb\bar{b}$ }  & Significance\\
\hline\hline
$\sigma$[pb]& $0.005148$ & $13.69$& $5.528$ & $S/\sqrt{B}$\\
\hline
Events before cuts&$15443$&$41070000 $ & $16584000$ & $ 2$\\
\hline
$M_{j_1j_2} > 1850$ GeV& $12313.5 $ & $946663 $ & $ 486076$& $10$\\
\hline
$H_T > 1700$ GeV& $ 9455.1 $ & $75979 $ & $19734 $& $31 $ \\
\hline

\end{tabular}
\captionsetup{font=normalsize, labelfont=bf, labelsep=period}
\caption{Sequence of cuts analyzed in ma5 for $\sin\theta^{L}_u$=0.1.}\label{table4ma5}\end{table}
For the first selection in Table (\ref{table4ma5}) after generated events, we took into account the invariant mass distribution for 
the two more energetic jets as shown in 
Figure~(\ref{selection01_0}). 
To compare the shapes of the curves related to both background and signal datasets, the events were normalized to unity. We select the following cut
\begin{equation}\label{equ-23}
M_{j_1j_2} > 1850.0~\text{GeV}. 
\end{equation}
Figure~(\ref{selection01_1}) shows the distributions of the total transverse energy in jets after the selection (\ref{equ-23}). 
To obtained this plot the events also were normalized to unity. So a good discrimination between signal and background can be obtained by using the 
following cut\newline
\begin{equation}\label{equ-24}
H_T>1700.0~\text{GeV},
\end{equation}
\noindent Now, assuming $70\%$ $b$-tagging efficiency, 
and then considering $4$ $b$ tags the significance will be

\begin{equation}\label{significance0.1}
\frac{S}{\sqrt{B}}\simeq7.
\end{equation}

For more realistic simulation we would like to consider as background the case where the involving final state has 5 jets jointly with $e^- \nu$ or 
$\mu^- \nu$. In addition that, for futures studies we can consider the EW decay modes to study the phenomenology at high energy colliders.\newline

\noindent We do not have different kinematic cuts for each different values of $\sin\theta^{L}_u$, that is the sequence of cuts 
implemented as in Table (\ref{table4ma5}) should be more than adequate.

\noindent The ratio $\dfrac{S}{\sqrt{B}}$ is large enough for $\sin\theta^{L}_u=0.1$ to be promising for larger $m_T$ and smaller values of $\sin\theta^{L}_u$, this work 
is in progress.\newline 

\noindent For more realistic study, we need to simulate more backgrounds. But looks promising.\newline

Also, we need to add pile-up effect, it is still unknown for future machine, we do not know machine parameters, but can guess that pile-up is large.\newline

Ultimately, it looks like $\sqrt{s}$ = 100 TeV pp collider is how we will be able to study a new sector of
vector-like quarks with large masses in detail.

\chapter{\bf Conclusions and Outlook}

Extensions of the SM that solve the hierarchy problem by having the Higgs as a pNGB typically must contain new vector-like quarks. In this thesis are 
studied the phenomenology of vector-like quarks in two such theories: quiver theories and a model-independent effective theory of vector-like quarks. 
In the case of quiver theories we computed the spectrum and couplings of vector-like quarks as well as excited gauge bosons to SM states, 
particularly $W^{\pm}$, $Z$ and the Higgs. In the last case we needed to formulate the Higgs as a pNGB in the quiver theory. Once we obtained 
the spectrum and its couplings we started studying the phenomenology  of vector-like quarks at the LHC. We saw that for quiver theories the production 
cross section are dominated by the single $T$ channel, is too small even for the HL-LHC. The reason was a combination of the small $TWb$ coupling 
and the large value of $m_T$. the vector-like quark mass.\newline

For the model-independent approach we concluded that the LHC could discover the $T\rightarrow Gt$ decay channel with $300~fb^{-1}$ if $c_{TWb}>0.16$, 
and for $3000~fb^{-1}$ (the ultimate HL-LHC luminosity) if $c_{TWb}>0.09$. Thus is clear that for $m_T\gtrsim2$TeV, quiver theories uses 
$c_{TWb}\simeq0.03$, are out of reach of the LHC.\newline

\noindent The results of chapter 4 indicated that in order to study the $T\rightarrow Gt$ in quiver theories for $m_T>2$TeV higher center-of-mass energies are 
needed. In chapter 5 we started a study of this channel for a hypothetical pp collider with $\sqrt{s}$ = 100 TeV. We studied the signal to 
background significance after cuts for various values of the coupling responsible for single $T$ production. As a first pass, we used 
pp$\rightarrow W^+bb\bar{b}\bar{b}j\rightarrow l\nu_lbb\bar{b}\bar{b}j$, and pp$\rightarrow t\bar{b}jb\bar{b}\rightarrow l\nu_lb\bar{b}jb\bar{b}$ as 
background.\newline

\noindent We conclude that for $\sqrt{s}$ = $100$ TeV $5\sigma$ discoveries will be accessible for couplings well below $0.1$, other possible.\newline

\noindent In the future a more complete background study will be needed. Furthermore, there is a need to include the effects of pile-up which were not 
considered here. We can see that if the discoveries of the LHC point to the possibility of heavy vector-like quarks, high energy pp colliders 
could study these sector in detail.

\cleardoublepage
\addappheadtotoc

\appendixpage
\begin{appendices}
\chapter{Wave-Function of Excited States}\label{WF-Excited}
  An specific example for hypothetical fermion with left-handed zero mode and localization parameter $c_{L}=0.2$, 
 in this case the left-handed zero mode is localized toward the N-th site, we show their Wave-Functions 
 in Figures \ref{c_{L}=0.2Nsite5}, \ref{c_{L}=0.2Nsite10} and \ref{c_{L}=0.2Nsite16} for $N=4$, $N=9$ and $N=15$ respectively. 
 
 
 \begin{figure}[h]
   \centering
   \epsfig{file=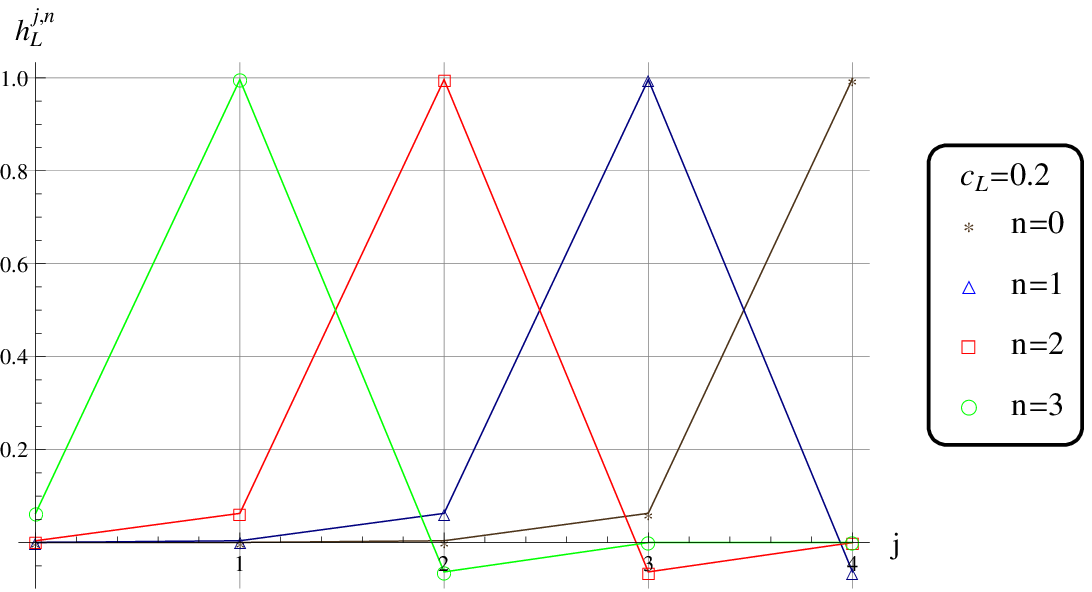}
   \caption{\label{c_{L}=0.2Nsite5}Wave-Functions of hypothetical fermion with left-handed zero mode and localization parameter $c_{L}=0.2$, 
   where j is a index of site and n is a index of Kaluza-Klein mode. In this case $N=4$ and some allowed $n$ are shown. 
   For the visualization, we choose by the opposite signs of $h^{j,1}_{L}$ and $h^{j,2}_{L}$.} 
  
\end{figure} 
  \begin{figure}[!t]
   \centering
   \epsfig{file=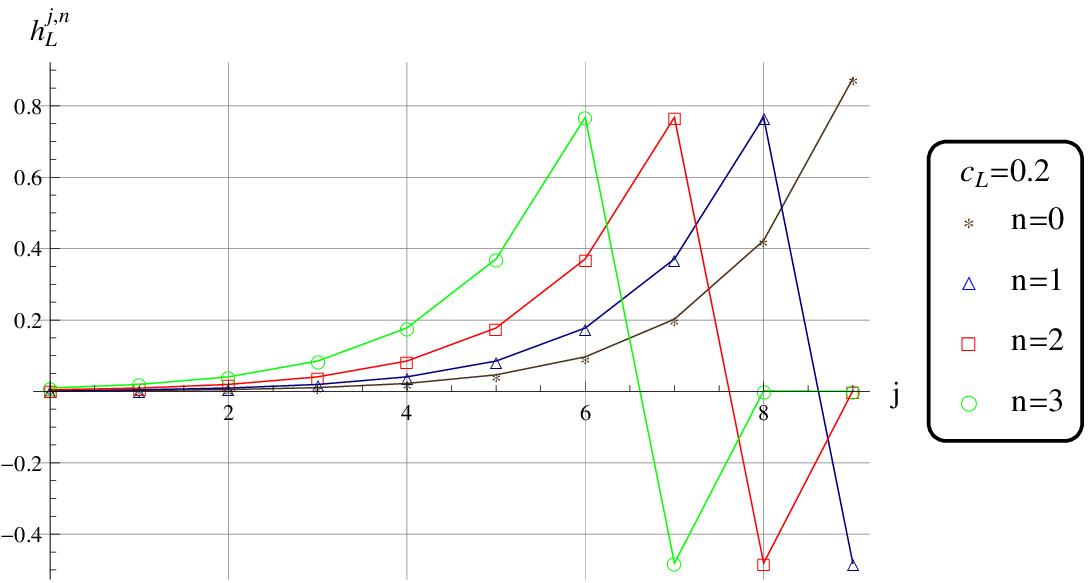}
   \caption{\label{c_{L}=0.2Nsite10}Wave-Functions of hypothetical fermion with left-handed zero mode and localization parameter $c_{L}=0.2$, 
   where j is a index of site and n is a index of Kaluza-Klein mode. In this case $N=9$ and some allowed $n$ are shown. 
   For the visualization, we choose by the opposite sign of $h^{j,3}_{L}$.} 
  \end{figure}
 
  \begin{figure}[H]
   \centering
   \epsfig{file=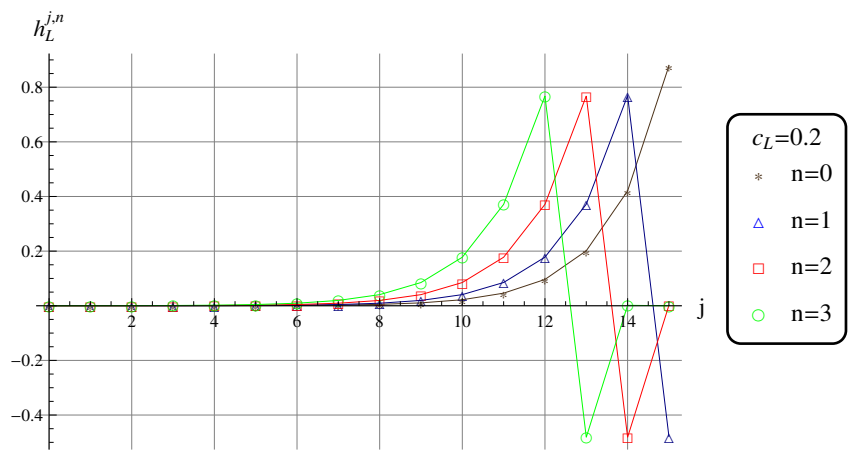}
   \caption{\label{c_{L}=0.2Nsite16}Wave-Functions of hypothetical fermion with left-handed zero mode and localization parameter $c_{L}=0.2$, 
   where j is a index of site and n is a index of Kaluza-Klein mode. In this case $N=15$ and some allowed $n$ are shown.} 
  \end{figure}
  
 \begin{figure}[H]
   \centering
   \epsfig{file=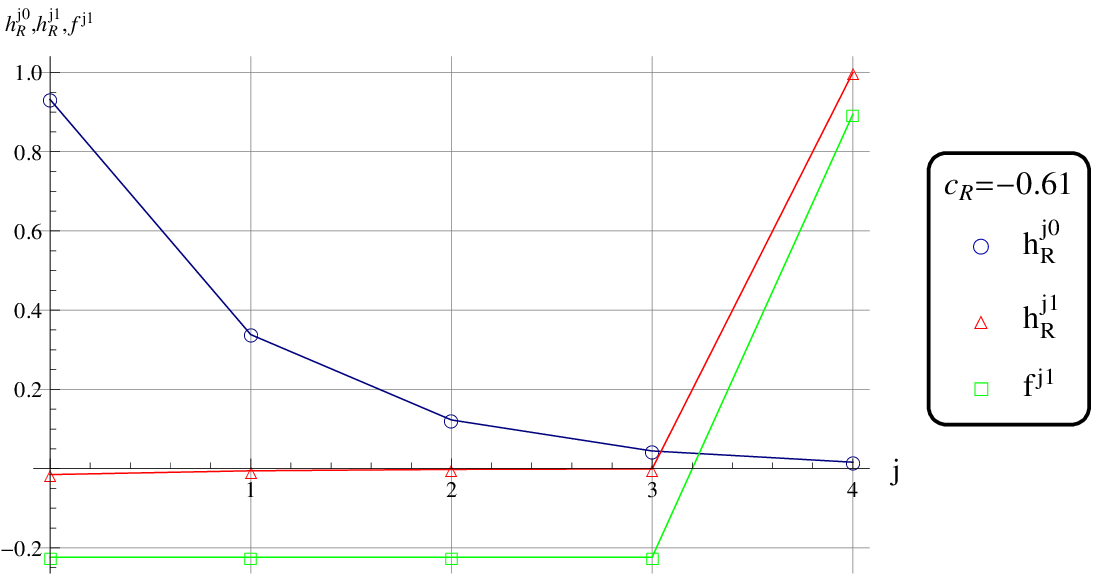}
   \caption{\label{c_{R}=-0.61Nsite4}Wave-Functions zero mode $h^{j,0}_{R}$, first mode $h^{j,1}_{R}$ of bottom fermion with 
   right-handed zero mode and localization parameter $c_{R}=-0.61$, and Wave-Function first mode of gauge boson $f^{j,1}$ where j is a index of site and n is a index of 
   Kaluza-Klein mode. In this case $N=4$ and some allowed $n$ are shown. 
   For the visualization, we choose by the opposite sign of $h^{j,0}_{R}$.} 
  \end{figure}
 
  \begin{figure}[h]
   \centering
   \epsfig{file=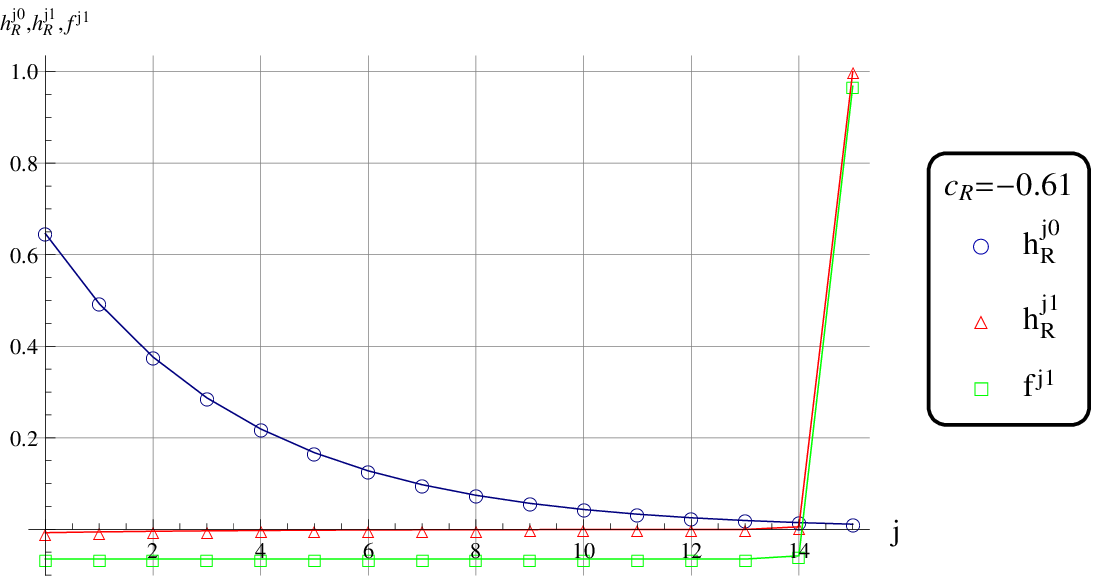}
   \caption{\label{c_{R}=-0.61Nsite4}Wave-Functions zero mode $h^{j,0}_{R}$, first mode $h^{j,1}_{R}$ of bottom fermion with 
   right-handed zero mode and localization parameter $c_{R}=-0.61$, and Wave-Function first mode of gauge boson $f^{j,1}$ where j is a index of site and n is a index of 
   Kaluza-Klein mode. In this case $N=15$ and some allowed $n$ are shown.} 
  \end{figure}
  
  \begin{figure}[t]
   \centering
   \epsfig{file=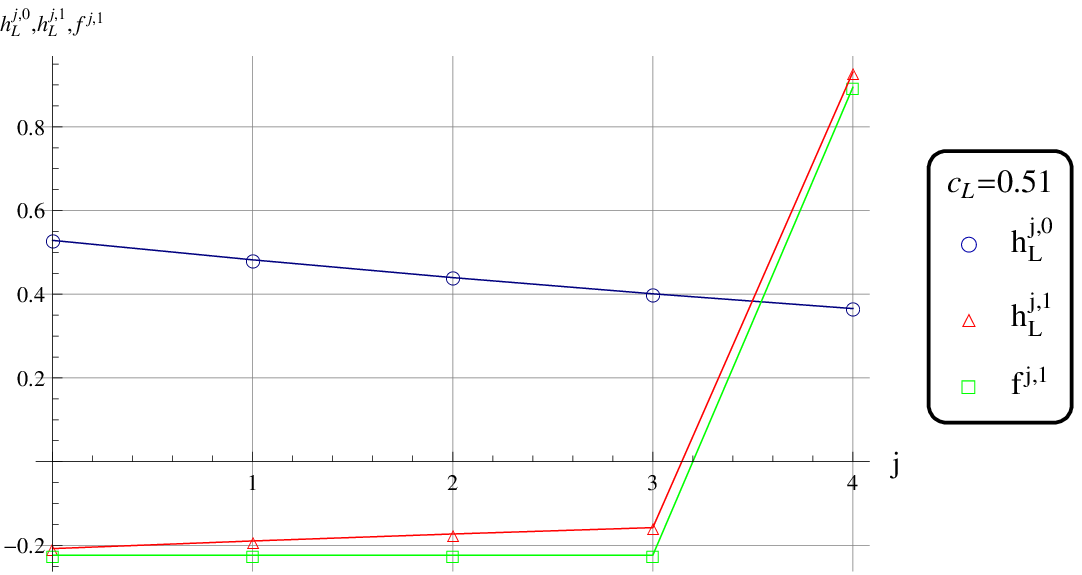}
   \caption{\label{c_{L}=0.51Nsite5}Wave-Functions zero mode $h^{j,0}_{L}$, first mode $h^{j,1}_{L}$ of top fermion with 
   left-handed zero mode and localization parameter $c_{L}=0.51$, and Wave-Function first mode of gauge boson $f^{j,1}$ where j is a index of site and n is a index of 
   Kaluza-Klein mode. In this case $N=4$ and some allowed $n$ are shown.
   For the visualization, we choose by the opposite sign of $h^{j,0}_{L}$.} 
  \end{figure}

  \begin{figure}[H]
   \centering
   \epsfig{file=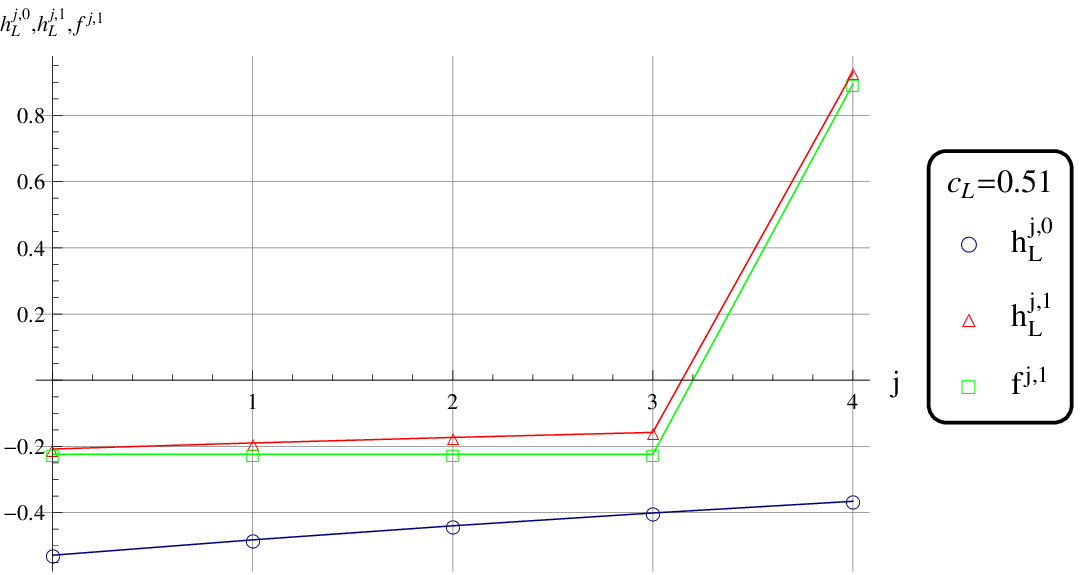}
   \caption{\label{c_{L}=0.51Nsite16}Wave-Functions zero mode $h^{j,0}_{L}$, first mode $h^{j,1}_{L}$ of top fermion with 
   left-handed zero mode and localization parameter $c_{L}=0.51$, and Wave-Function first mode of gauge boson $f^{j,1}$ where j is a index of site and n is a index of 
   Kaluza-Klein mode. In this case $N=15$ and some allowed $n$ are shown.} 
  \end{figure}
  \begin{figure}[t]
   \centering
   \epsfig{file=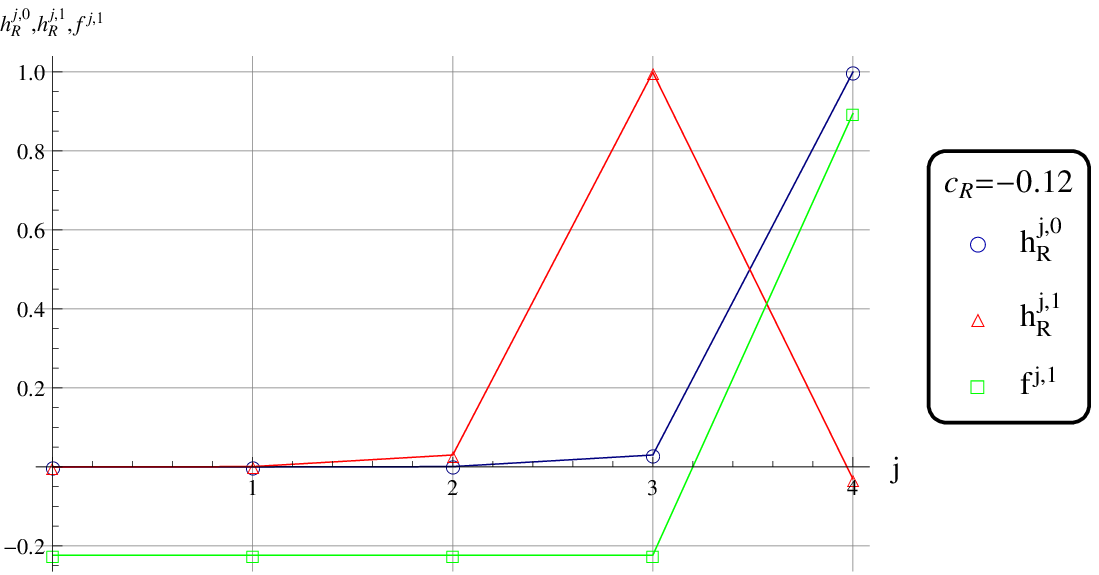}
   \caption{\label{c_{R}=-0.12Nsite5}Wave-Functions zero mode $h^{j,0}_{R}$, first mode $h^{j,1}_{R}$ of top fermion with 
   right-handed zero mode and localization parameter $c_{R}=-0.12$, and Wave-Function first mode of gauge boson $f^{j,1}$ where j is a index of site and n is a index of 
   Kaluza-Klein mode. In this case $N=4$ and some allowed $n$ are shown. 
   For the visualization, we choose by the opposite sign of $h^{j,1}_{R}$. } 
  \end{figure}
  \begin{figure}[H]
   \centering
   \epsfig{file=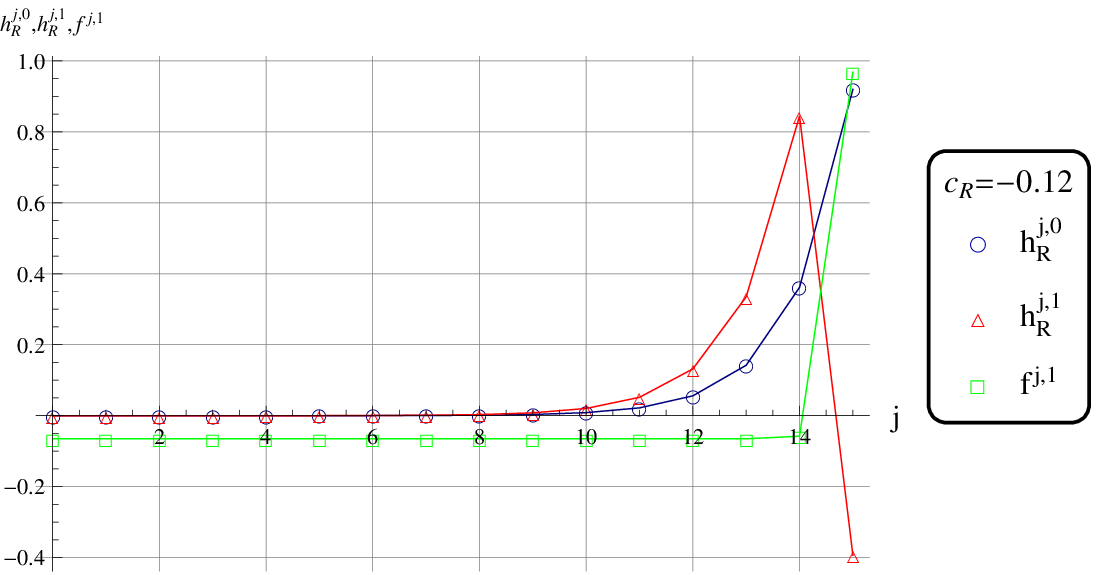}
   \caption{\label{c_{R}=-0.12Nsite16}Wave-Functions zero mode $h^{j,0}_{R}$, first mode $h^{j,1}_{R}$ of top fermion with 
   right-handed zero mode and localization parameter $c_{R}=-0.12$, and Wave-Function first mode of gauge boson $f^{j,1}$ where j is a index of site and n is a index of 
   Kaluza-Klein mode. In this case $N=15$ and some allowed $n$ are shown.} 
  \end{figure}

\chapter{}\label{EW-MassDoubletCase}

For the vector-like quark SM SU(2)-doublet, we pass the interactions between $T'$, $B'$ the SM third generations and the EW gauge bosons from the EW to the mass eigenstate, 

 \begin{align}\label{VLTBdoubletEW1}
&~~\frac{g}{\sqrt{2}}\overline{t}'_L\slashed{W}^+ b'_L+\frac{g}{\sqrt{2}}\overline{b}'_L\slashed{W}^- t'_L\nonumber\\
&+\frac{g}{c_W}\overline{t}'_L\slashed{Z}\left(\frac{1}{2}-\frac{2}{3}s_W^2\right)t'_L+\frac{g}{c_W}\overline{b}'_L\slashed{Z}\left(-\frac{1}{2}+\frac{1}{3}s_W^2\right)b'_L\nonumber\\
&+\frac{2gs_W}{3}\overline{t}'_L\slashed{A}t'_L-\frac{gs_W}{3}\overline{b}'_L\slashed{A}b'_L\nonumber\\
&+\frac{g}{c_W}\overline{t}'_R\slashed{Z}\left(-\frac{2}{3}s_W^2\right)t'_R+\frac{g}{c_W}\overline{b}'_R\slashed{Z}\left(\frac{1}{3}s_W^2\right)b'_R\nonumber\\
&+\frac{2gs_W}{3}\overline{t}'_R\slashed{A}t'_R-\frac{gs_W}{3}\overline{b}'_R\slashed{A}b'_R\nonumber\\
&+\frac{g}{\sqrt{2}}\overline{T}'_L\slashed{W}^+ B'_L+\frac{g}{\sqrt{2}}\overline{B}'_L\slashed{W}^- T'_L\nonumber\\
&+\frac{g}{c_W}\overline{T}'_L\slashed{Z}\left(\frac{1}{2}-\frac{2}{3}s_W^2\right)T'_L+\frac{g}{c_W}\overline{B}'_L\slashed{Z}\left(-\frac{1}{2}+\frac{1}{3}s_W^2\right)B'_L\nonumber\\
&+\frac{2gs_W}{3}\overline{T}'_L\slashed{A}T'_L-\frac{gs_W}{3}\overline{B}'_L\slashed{A}B'_L\nonumber\\
&+\frac{g}{\sqrt{2}}\overline{T}'_R\slashed{W}^+ B'_R+\frac{g}{\sqrt{2}}\overline{B}'_R\slashed{W}^- T'_R\nonumber\\
&+\frac{g}{c_W}\overline{T}'_R\slashed{Z}\left(\frac{1}{2}-\frac{2}{3}s_W^2\right)T'_R+\frac{g}{c_W}\overline{B}'_R\slashed{Z}\left(-\frac{1}{2}+\frac{1}{3}s_W^2\right)B'_R\nonumber\\
&+\frac{2gs_W}{3}\overline{T}'_R\slashed{A}T'_R-\frac{gs_W}{3}\overline{B}'_R\slashed{A}B'_R,
\end{align}
and then with the rotations defined by (\ref{rotationLR-TBdoublet}) into (\ref{VLTBdoubletEW1}) 
we find that
 \begin{align}
&~~\frac{g}{\sqrt{2}}\left(c^{u}_{L}c^{d}_{L}\bar{t}_L\slashed{W}^+ b_L+c^{u}_{L}s^{d}_{L}\bar{t}_L\slashed{W}^+ B_L+s^{u}_{L}c^{d}_{L}\overline{T}_L\slashed{W}^+ b_L+s^{u}_{L}s^{d}_{L}\overline{T}_L\slashed{W}^+ B_L+\mathrm{h.c.}\right)\nonumber\\
&+\frac{g}{c_W}\left(\frac{1}{2}-\frac{2}{3}s_W^2\right)\bigg((c^{u}_{L})^2\bar{t}_L\slashed{Z}t_L+c^{u}_{L}s^{u}_{L}\bar{t}_L\slashed{Z} T_L+s^{u}_{L}c^{u}_{L}\overline{T}_L\slashed{Z} t_L+(s^{u}_{L})^2\overline{T}_L\slashed{Z} T_L\bigg)\nonumber\\
&+\frac{g}{c_W}\left(-\frac{1}{2}+\frac{1}{3}s_W^2\right)\bigg((c^{d}_{L})^2\bar{b}_L\slashed{Z}b_L+c^{d}_{L}s^{d}_{L}\bar{b}_L\slashed{Z} B_L+s^{d}_{L}c^{d}_{L}\overline{B}_L\slashed{Z} b_L+(s^{d}_{L})^2\overline{B}_L\slashed{Z} B_L\bigg)\nonumber\\
&+\frac{2gs_W}{3}\bigg((c^{u}_{L})^2\bar{t}_L\slashed{A}t_L+c^{u}_{L}s^{u}_{L}\bar{t}_L\slashed{A} T_L+s^{u}_{L}c^{u}_{L}\overline{T}_L\slashed{A} t_L+(s^{u}_{L})^2\overline{T}_L\slashed{A} T_L\bigg)\nonumber\\
&-\frac{gs_W}{3}\bigg((c^{d}_{L})^2\bar{b}_L\slashed{A}b_L+c^{d}_{L}s^{d}_{L}\bar{b}_L\slashed{A} B_L+s^{d}_{L}c^{d}_{L}\overline{B}_L\slashed{A} b_L+(s^{d}_{L})^2\overline{B}_L\slashed{A} B_L\bigg)\nonumber\\
&-\frac{2gs_W^2}{3c_W}\bigg((c^{u}_{R})^2\bar{t}_R\slashed{Z}t_R+c^{u}_{R}s^{u}_{R}\bar{t}_R\slashed{Z} T_R+s^{u}_{R}c^{u}_{R}\overline{T}_R\slashed{Z} t_R+(s^{u}_{R})^2\overline{T}_R\slashed{Z} T_R\bigg)\nonumber\\
&+\frac{gs_W^2}{3c_W}\bigg((c^{d}_{R})^2\bar{b}_R\slashed{Z}b_R+c^{d}_{R}s^{d}_{R}\bar{b}_R\slashed{Z} B_R+s^{d}_{R}c^{d}_{R}\overline{B}_R\slashed{Z} b_R+(s^{d}_{R})^2\overline{B}_R\slashed{Z} B_R\bigg)\nonumber\\
&+\frac{2gs_W}{3}\bigg((c^{u}_{R})^2\bar{t}_R\slashed{A}t_R+c^{u}_{R}s^{u}_{R}\bar{t}_R\slashed{A} T_R+s^{u}_{R}c^{u}_{R}\overline{T}_R\slashed{A} t_R+(s^{u}_{R})^2\overline{T}_R\slashed{A} T_R\bigg)\nonumber\\
&-\frac{gs_W}{3}\bigg((c^{d}_{R})^2\bar{b}_R\slashed{A}b_R+c^{d}_{R}s^{d}_{R}\bar{b}_R\slashed{A} B_R+s^{d}_{R}c^{d}_{R}\overline{B}_R\slashed{A} b_R+(s^{d}_{R})^2\overline{B}_R\slashed{A} B_R\bigg)\nonumber\\
&+\frac{g}{\sqrt{2}}\left(s^{u}_{L}s^{d}_{L}\bar{t}_L\slashed{W}^+ b_L-s^{u}_{L}c^{d}_{L}\bar{t}_L\slashed{W}^+ B_L-c^{u}_{L}s^{d}_{L}\overline{T}_L\slashed{W}^+ b_L+c^{u}_{L}c^{d}_{L}\overline{T}_L\slashed{W}^+ B_L+\mathrm{h.c.}\right)\nonumber\\
&+\frac{g}{c_W}\left(\frac{1}{2}-\frac{2}{3}s_W^2\right)\bigg((s^{u}_{L})^2\bar{t}_L\slashed{Z}t_L-s^{u}_{L}c^{u}_{L}\bar{t}_L\slashed{Z} T_L-c^{u}_{L}s^{u}_{L}\overline{T}_L\slashed{Z} t_L+(c^{u}_{L})^2\overline{T}_L\slashed{Z} T_L\bigg)\nonumber\\
&+\frac{g}{c_W}\left(-\frac{1}{2}+\frac{1}{3}s_W^2\right)\bigg((s^{d}_{L})^2\bar{b}_L\slashed{Z}b_L-s^{d}_{L}c^{d}_{L}\bar{b}_L\slashed{Z} B_L-c^{d}_{L}s^{d}_{L}\overline{B}_L\slashed{Z} b_L+(c^{d}_{L})^2\overline{B}_L\slashed{Z} B_L\bigg)\nonumber\\
&+\frac{2gs_W}{3}\bigg((s^{u}_{L})^2\bar{t}_L\slashed{A}t_L-s^{u}_{L}c^{u}_{L}\bar{t}_L\slashed{A} T_L-c^{u}_{L}s^{u}_{L}\overline{T}_L\slashed{A} t_L+(c^{u}_{L})^2\overline{T}_L\slashed{A} T_L\bigg)\nonumber\\
&-\frac{gs_W}{3}\bigg((s^{d}_{L})^2\bar{b}_L\slashed{A}b_L-s^{d}_{L}c^{d}_{L}\bar{b}_L\slashed{A} B_L-c^{d}_{L}s^{d}_{L}\overline{B}_L\slashed{A} b_L+(c^{d}_{L})^2\overline{B}_L\slashed{A} B_L\bigg)\nonumber\\
&+\frac{g}{\sqrt{2}}\left(s^{u}_{R}s^{d}_{R}\bar{t}_R\slashed{W}^+ b_R-s^{u}_{R}c^{d}_{R}\bar{t}_R\slashed{W}^+ B_R-c^{u}_{R}s^{d}_{R}\overline{T}_R\slashed{W}^+ b_R+c^{u}_{R}c^{d}_{R}\overline{T}_R\slashed{W}^+ B_R+\mathrm{h.c.}\right)\nonumber\\
&+\frac{g}{c_W}\left(\frac{1}{2}-\frac{2}{3}s_W^2\right)\bigg((s^{u}_{R})^2\bar{t}_R\slashed{Z}t_R-s^{u}_{R}c^{u}_{R}\bar{t}_R\slashed{Z} T_R-c^{u}_{R}s^{u}_{R}\overline{T}_R\slashed{Z} t_R+(c^{u}_{R})^2\overline{T}_R\slashed{Z} T_R\bigg)\nonumber\\
&+\frac{g}{c_W}\left(-\frac{1}{2}+\frac{1}{3}s_W^2\right)\bigg((s^{d}_{R})^2\bar{b}_R\slashed{Z}b_R-s^{d}_{R}c^{d}_{R}\bar{b}_R\slashed{Z} B_R-c^{d}_{R}s^{d}_{R}\overline{B}_R\slashed{Z} b_R+(c^{d}_{R})^2\overline{B}_R\slashed{Z} B_R\bigg)\nonumber\\
&+\frac{2gs_W}{3}\bigg((s^{u}_{R})^2\bar{t}_R\slashed{A}t_R-s^{u}_{R}c^{u}_{R}\bar{t}_R\slashed{A} T_R-c^{u}_{R}s^{u}_{R}\overline{T}_R\slashed{A} t_R+(c^{u}_{R})^2\overline{T}_R\slashed{A} T_R\bigg)\nonumber\\
&-\frac{gs_W}{3}\bigg((s^{d}_{R})^2\bar{b}_R\slashed{A}b_R-s^{d}_{R}c^{d}_{R}\bar{b}_R\slashed{A} B_R-c^{d}_{R}s^{d}_{R}\overline{B}_R\slashed{A} b_R+(c^{d}_{R})^2\overline{B}_R\slashed{A} B_R\bigg),\label{VLTBdoubletEW2}
\end{align}

  \end{appendices}

\bibliographystyle{unsrt}
\bibliography{./CHAPTERS/BIBLIOGRAPHY/Bibliography}


\end{document}